\documentclass{aa}  

\usepackage{graphicx}
\usepackage{txfonts}
\usepackage[multiple]{footmisc}
\usepackage{hyperref}
\makeatletter
\renewcommand*\aa@pageof{, page \thepage{} of \pageref*{LastPage}}
\makeatother

\newcommand\fnsep{\textsuperscript{,}}
\newcommand{\hii}{H\,{\sc ii}}
\newcommand{\hi}{H\,{\sc i}}
\newcommand{\HI}{H\,{\sc i}}
\newcommand{\Ht}{H$_2$}
\newcommand{\kms}{\,km\,s$^{-1}$}
\newcommand{\msun}{M$_{\odot}$}
\newcommand{\cm}{cm$^{-2}$}

\begin{document}

   \title{CHILES X: Molecular and atomic gas at intermediate redshift}

   \author{Kelley M.~Hess\inst{\ref{oso},\ref{astron}}\thanks{Corresponding author: kelley.hess@chalmers.se}
           \and
           John Hibbard\inst{\ref{cville}} \and
           Jennifer Donovan Meyer\inst{\ref{cville}} \and
           Hansung B.~Gim\inst{\ref{mont}} \and
           Nicholas M.~Luber\inst{\ref{col}} \and
          Min S.~Yun\inst{\ref{umass}} \and
           Julia Blue Bird\inst{\ref{nrao}} \and
           Richard Dodson\inst{\ref{icrar}} \and
           Aeree Chung\inst{\ref{yonsei}} \and
           Danielle Lucero\inst{\ref{vatech}} \and
           Emmanuel Momjian\inst{\ref{nrao}} \and
           D.J.~Pisano\inst{\ref{uct}} \and
           J.~H.~van Gorkom\inst{\ref{col}} 
          }
    \authorrunning{Hess, K.~M.~et al.}

    \institute{Department of Space, Earth and Environment, Chalmers University of Technology, Onsala Space Observatory, 43992 Onsala, Sweden\label{oso}
        \and
        ASTRON, the Netherlands Institute for Radio Astronomy, Postbus 2, 7990 AA, Dwingeloo, The Netherlands\label{astron}
        \and
        National Radio Astronomy Observatory (NRAO), 520 Edgemont Road, Charlottesville, VA 22903, USA\label{cville}
        \and
        Department of Physics, Montana State University, P.O. Box 173840, Bozeman, MT 59717, USA\label{mont}
        \and
        Department of Astronomy, Columbia University, 550 West 120th Street, New York, NY 10027, USA\label{col}
        \and
        Department of Astronomy, University of Massachusetts, Amherst, MA 01003, USA\label{umass}
        \and
        National Radio Astronomy Observatory, P.O. Box O, Socorro, NM 87801, USA\label{nrao}
        \and
        International Centre for Radio Astronomy Research (ICRAR), University of Western Australia, 35 Stirling Hwy, Crawley, WA 6009, Australia\label{icrar}
        \and
        Department of Astronomy, Yonsei University, 50 Yonsei-ro, Seodaemun-gu, Seoul 03722, Republic of Korea\label{yonsei}
        \and
        Department of Physics, Virginia Tech, 850 West Campus Drive, Blacksburg, VA 24061, USA\label{vatech}
        \and
        Department of Astronomy, University of Cape Town, Private Bag X3, Rondebosch 7701, South Africa\label{uct}
             }

   \date{Received XXX; accepted YYY}

 
  \abstract
  {We present ALMA CO observations of 14 \HI-detected galaxies from the COSMOS \hi\ Large Extragalactic Survey (CHILES) found in a cosmic over-density at $z\sim0.12$. This is the largest collection of spatially resolved CO + \HI\ observations beyond the local Universe ($z>0.05$) to date. While the \HI-detected parent sample spans a range of stellar masses, star formation rates (SFRs), and environments, we only directly detect CO in the highest stellar mass galaxies, $\log(M_*/M_{\odot})>10.0$, with SFRs greater than $\sim$2~\msun\ yr$^{-1}$.  The detected CO has the kinematic signature of a rotating disk, consistent with the \hi.  We stacked the CO non-detections and find a mean \Ht\ mass of $\log(M_{H_2}/M_{\odot}) = 8.46$ in galaxies with a mean stellar mass of $\log(M_*/M_{\odot}) = 9.35$.  In addition to high stellar masses and SFRs, the systems detected in CO are spatially larger, have redder overall colors, and exhibit broader (stacked) line widths.  The CO emission is spatially coincident with both the highest stellar mass surface density and star forming region of the galaxies, as revealed by the 1.4 GHz continuum emission from CHILES Con Pol.  We interpret the redder colors as the molecular gas being coincident with dusty regions of obscured star formation.  The 14 \HI\ detections show a range of morphologies, but the \HI\ reservoir is always more extended than the CO.  Finally, we compare with samples in the literature and find mild evidence for evolution in the molecular gas reservoir and \Ht-to-\HI\ gas ratio with redshift in \hi\ flux-limited samples.  We also show that the scatter in the \hi, and \hi-to-stellar mass ratio is too great to conclusively measure evolution below $z=0.2$, and would be even extremely difficult below $z=0.4$.  Detections from CHILES are likely to be the only individual galaxies detected in \hi\ between $0.1<z<0.23$ for the foreseeable future due to the severity of satellite radio frequency interference, and its preferential impact on short baselines which dominate the observations of contemporary \HI\ surveys.
  }


   \maketitle


\section{Introduction}

Despite decades of observations, evolution in the gas reservoirs of galaxies with cosmic time is a poorly understood component of the baryonic cycle.  The abundance of neutral atomic hydrogen (\hi) provides the reservoir out of which molecular hydrogen (\Ht) forms, catalyzed by the presence of dust.  \Ht\ cools efficiently and molecular clouds collapse to form stars, some of which later recycle their material back into the interstellar medium \citep{Tumlinson17}.  In large part, however, the atomic reservoir which provides the initial fuel for star formation must be replenished through accretion of gas from the surrounding environment.

Despite this role, \hi\ has been called a ``pass-through'' phase in the baryon cycle, between the accretion of ionized gas (\hii) on the way to \Ht\ in part because the cosmic mass density of \hi\ appears to only evolve slowly with redshift: declining by approximately a factor of two since Cosmic Noon ($z\sim1.5-2$;~\citealt{Peroux20,Walter20}).  In contrast, \Ht\ has been inferred to evolve rapidly with the cosmic mass density dropping by a factor of roughly six from its peak over the same time.  The precipitous decline in the amount of \Ht\ over the last half lifetime of the Universe has led to the conclusion that molecular gas is the component primarily responsible for the decline in the star formation rate density, which itself falls by a factor of about eight (e.g.,~\citealt{Peroux20,Walter20}, and references therein).

Indeed, a number of studies have argued that star formation is dependent on (only) the molecular gas content \citep{Bigiel08,Bigiel11,Scoville17,Scoville23}, but this is in tension with results from larger studies which have sufficient numbers to look at molecular gas by galaxy properties.  \citet{Saintonge16} show that the amount of star formation is not only correlated with molecular gas content, but by the total cold gas content and the star formation efficiency \citep{Saintonge22}.

In the local Universe, the close connection between star formation and molecular gas was revealed by detailed high resolution studies of the gas surface density in galaxies \citep{Bigiel11,Pessa21,Sun23}.  However, despite detailed studies at $z=0$, and global studies spanning to $z\sim3$ \citep{Tacconi13,Genzel15,Sharon16,Freundlich19}, several scales which are important for understanding the details of the baryonic cycle, and how it has evolved, remain unexplored or are just becoming available with the current and upcoming suite of radio telescopes.  Indeed, the global picture of star formation does not capture how the gas reservoirs evolve over the same timescale where we observe the star formation rate density is shifting from primarily massive galaxies at higher redshift, to lower mass galaxies in the present day \citep{Behroozi13}.

To date, direct observations that measure \hi\ and \Ht\ in the same galaxies are limited to the very local Universe and either to small numbers of resolved sources \citep{Leroy08}, or to larger numbers of unresolved galaxies over a relatively limited stellar mass range.  Targeted studies of resolved \HI\ and CO in specific galaxy Hubble types (spirals, early-types) or specific environments (the field, groups, clusters) amount to of order a couple hundred galaxies within 20 Mpc ($z=0.005$).  For example, the HERACLES survey mapped CO (2-1) in 18 galaxies observed by THINGS in \hi\ \citep{Leroy09,Walter08}.  The PHANGS-ALMA survey mapped 90 ``main sequence'' galaxies in CO (2-1) which are also being observed in \hi\ with the Very Large Array and MeerKAT \citep{Leroy21}.  The VERTICO survey has mapped CO (2-1) in 51 Virgo cluster galaxies observed by the VIVA survey in \hi\ \citep{Brown21,Chung09}.  Indeed, clusters have proven to be interesting targets for CO+HI population studies.  Thirty (15) galaxies in the Fornax Cluster have also been targeted (detected) in CO (1-0) and HI \citep{Zabel19,Loni21,Serra23}.  Meanwhile, the largest sample of CO and HI consists of single-dish measurements of 532 galaxies from the stellar mass-selected xGASS sample \citep{Catinella18} with dedicated CO (1-0) follow-up, xCOLD GASS \citep{Saintonge17}.  Other notable collections of spatially unresolved CO observations with HI measurements of the same galaxies include 273 isolated galaxies in the AMIGA sample \citep{Lisenfeld11,Jones18}; 163 galaxies in filaments around Virgo \citep{Castignani22}; early-type galaxies from ATLAS-3D (56 detections of 260 targeted by \citet{Young11}; and 97 low mass galaxies in the ALLSMOG sample with stellar masses down to $M_*=10^{8.5}$~\msun\ \citep{Cicone17}.

Unfortunately, most higher redshift CO studies lack \HI\ counterparts (e.g.,~EGNoG at $0.05<z<0.5$, \citealt{Bauermeister13}; VALES at $0.02<z<0.35$ \citealt{Villanueva17}; PHIBBS at $z<0.5$, \citealt{Freundlich19}), due to a combination of the weakness of the 21 cm line and, until the last 10-15 years, the lack of receivers with the appropriate frequency coverage on telescopes.

Beyond the local Universe ($z>0.05$), there are a dozen galaxies with combined \hi\ and \Ht\ measurements.  Five are massive isolated galaxies at $z=0.2$ from the HIGHz sample that were detected in CO (1-0) with ALMA.  The COOL BUDHIES sample targeted 23 galaxies in and around two clusters observed by the BUDHIES \hi\ survey at $z~0.2$ with the Large Millimeter Telescope (LMT) \citep{Cybulski16,Gogate20}.  Of these, six have both secure \hi\ and CO detections and six more have CO upper limits.  The last measurement is the highest redshift \hi\ direct detection in emission to date, from the CHILES survey at $z=0.376$ which was also detected in CO (1-0) with the LMT \citep{Fernandez16}.  In all but the CHILES galaxy (Donovan Meyer et al.~in prep), either the \HI\ or the \HI\ and CO detections were unresolved.  But perhaps, more critically, all the galaxies for which both the atomic and stellar mass have been measured are also more massive than $M_* = 10^{10}$ \msun.  Indeed, this also holds true for high redshift samples for which only the molecular content has been pursued: PHIBBS and EGNoG galaxies have masses greater than $\log(M_*/M_{\odot}) = 10.4$ and 10.6, respectively.  Measurements of molecular gas for smaller targeted samples covering another 60 or so galaxies beyond $z=0.1$ all have stellar masses greater than $\log(M_*/M_{\odot}) = 10.0$ (e.g. \citealt{Genzel15} and references therein).  As a result, the available observational data for the gaseous component of baryons in galaxies is limited to the global properties in the most massive systems.  We are missing both the resolved details, and the majority of the galaxy population which is at lower stellar mass (e.g.,~\citealt{Taylor15}).

In this paper, we present results from the full COSMOS HI Legacy Extragalactic Survey (CHILES) and the dedicated CO (1-0) follow-up observations with the Atacama Large Millimeter Array (ALMA) to map the atomic and molecular gas content in the 14 \hi\ detections around $z=0.12$ which were reported in \citet{Hess19}.  These sources were first detected in ``Epoch 1'' (178 hours) of the survey. CHILES was designed to directly detect the most gas-rich galaxies known, with M$_{HI}$ of $\sim$3x10$^{10}$~\msun, out to a redshift of $z<0.5$, with the Karl G.~Jansky Very Large Array \citep[VLA;][]{Fernandez16,Luber25b}.
The data presented here consists of 856 hours of on-source time, on a single pointing in the COSMOS field \citep{Scoville07}, centered at 10h01m24s +02d21m00s \citep{Fernandez13}.

The frequency range in which the 14 galaxies in this paper are detected in \hi\ is among the most heavily impacted by radio frequency interference \citep[RFI;][]{Hess19}. Nonetheless, these galaxies are the first to be resolved in both CO and \hi\ beyond $z=0.1$.  Detections by CHILES are likely to be the only \HI\ measurements in the redshift range between $0.10\lesssim z \lesssim 0.23$ (1160 $\lesssim \nu \lesssim$ 1290 MHz), without advanced RFI mitigation techniques, until L-band observations move to the far side of the moon, due to the worsening RFI environment\footnote{Statistics from MeerKAT in 2019, after the completion of CHILES observing, show 100\% of data flagged on baselines less than 1 km, and 50-100\% of data flagged on baselines greater than 1 km at these frequencies due to RFI from satellites. (See Figure 1: \url{https://skaafrica.atlassian.net/wiki/spaces/ESDKB/pages/305332225/Radio+Frequency+Interference+RFI}).  Currently, this range is being avoided by MeerKAT Key Science Projects in the data processing \citealt{Heywood24,KazemiMoridani25}}.

The paper is organized as follows.  In Section \ref{sect:observations} we describe the ALMA and VLA data reduction and source finding, as well as ancillary data and comparison samples.  In Section \ref{sect:data_stack} we describe our three dimensional stacking technique applied to the CO data cubes.  In Section \ref{sect:results} we present the results for the resolved molecular and atomic gas observations, including derived quantities and resolved \Ht\ and \hi\ maps. Over the course of this paper we found that the CO and \HI\ communities assume different unit conventions, which has not been widely discussed in the literature. In Section \ref{sect:results1}, we present consistent formulae for the mass and column density with redshift corrections in the fundamental observed units of interferometric data. In Section \ref{sect:discussion} we discuss our results compared to existing knowledge of \Ht\ and \hi\ gas ratios, and models of baryon evolution through cosmic time.  Throughout the paper we assume a $\Lambda$CDM cosmology with $\Omega_m = 0.27$, $\Omega_{\Lambda} = 0.73$, and $H_0 = 70$ \kms\ Mpc$^{-1}$.

\section{Data}
\label{sect:observations}

\subsection{Atacama Large Millimeter Array CO (1-0) data}
\label{sect:alma_obs}

The 14 \HI-detected galaxies from the CHILES project in the ``cosmic wall'' from $z\sim 0.11 - 0.13$ were the subject of successful ALMA Cycle 6 \& 7 proposals 2018.1.01852.S and 2019.1.01615.S (P.I.~K.~Hess) using Band 3 \citep{Claude08}. We estimated the expected CO luminosity of each galaxy based on its UV+IR star formation rate (SFR) using the relationship in \citet[see Figure 8]{Solomon05}, and grouped the observations into three different science goals based on the requested sensitivity. Two galaxies were in the low SFR sample ($\sim$0.4 \msun\ yr$^{-1}$), five in the medium SFR sample ($\sim$0.6 \msun\ yr$^{-1}$), and seven in the high SFR sample ($\sim$1.0 \msun\ yr$^{-1}$). We requested a resolution of 1.4\arcsec -- 2.5\arcsec\ ($3.0 - 5.5$ kpc at $z\sim0.123$) corresponding to ALMA 12-m array configurations C43-2 and C43-3. The spectral setup for all observations included three continuum spectral windows (SPWs), each with 128 channels over a 2 GHz bandwidth, and a higher spectral resolution SPW covering the redshifted CO (1-0) line with 1920 channels over a 1.875 GHz bandwidth. A factor of two spectral averaging was used, giving a channel separation of 2.8\kms\ and a velocity resolution of 3.3\kms.

The scheduling block (SB) for the science goal consisting of the seven high SFR systems was executed three times in March and April 2019, amounting to 20 minutes on-source time per galaxy. The SBs for the other two science goals were obtained in January 2020 and consisted of seven executions for the five medium SFR systems, amounting to 60 minutes per source; and three executions for the two low SFR systems, amounting to 75 minutes per source. All observations were taken with 41-49 12-m antennas with baselines ranging from 15 to 500 m. Subsequently it was discovered that the local oscillator (LO) tuning for the high-resolution spectral window missed the redshifted CO frequency for two of the targets, so a new SB was generated for these two systems using the appropriate LO tuning, and this was observed in February 2020 with 38 12-m antennas and baselines from 15 to 784 m, accumulating 50 minutes per source. A summary of this information is given in Table \ref{tab:observations}. 

\begin{sidewaystable}[]
    \caption{ALMA observation \& imaging pipeline data product details}
    \centering
    \begin{tabular}{lcccc}
    \hline
    \hline
Property  & Observation 1 & Observation 2 & Observation 3 & Observation 4 \\
    \hline
Project Code & 2018.1.01852.S & 2019.1.01615.S & 2019.1.01615.S & 2019.1.01615.S \\
Observation Dates 	& 2019 Mar 13, Apr 1, Apr 3	& 2020 Jan 13 & 2020 Jan 14, 16, 25	& 2020 Feb 28 \\
Max baseline length	& 360 -- 500 m	& 178 m	& 167 -- 180 m & 784 m \\
Number of antennas after flagging & 45 -- 48 & 38 -- 41	& 40 -- 42	& 38 \\
Bandpass \& Amplitude Calibrator (flux) & J1058+0133 ($\sim$ 4.6 Jy) & J1058+0133 ($\sim$ 2.6 Jy)	& J1058+0133 ($\sim$ 2.6 Jy) & J1058+0133 ($\sim$ 2.9 Jy) \\
Phase Calibrator (flux)	& J0948+022 ($\sim$ 90 -- 210 mJy) 	& J1008+0029 ($\sim$ 61 mJy) & J1008+0029 ($\sim$ 61 mJy) & J1008+0621 ($\sim$ 140 mJy) \\	
Target Selection & $SFR_{UV+IR} > 1$ M$_{\odot}$ yr$^{-1}$ 	& $SFR_{UV+IR} \sim 0.4$ M$_{\odot}$ yr$^{-1}$ &  $SFR_{UV+IR} \sim 0.6$ M$_{\odot}$ yr$^{-1}$ & replacement for bad LO tunings \\
COSMOS targets	& 0969208, 1189669,  & 1399657, 1440745 & 1008875, 1009969, & 1429536, 1430950 \\
	& 1197519, 1200839,  &  & 1197786, 1440643 &  \\
 	& 1204323,  1411106 &  &  &  \\
Number of Executions 	& 3		& 3		& 7		& 1 \\
Time on source	per target &  20 min & 75 min & 60 min & 50 min \\
Cube beam size	& 2.1\arcsec $\times$ 1.6\arcsec & 1.7\arcsec $\times$ 1.5\arcsec			& 1.7\arcsec $\times$ 1.5\arcsec & 1.8\arcsec $\times$ 1.1\arcsec \\
Cube RMS noise$^{a}$ & 1.1 mJy beam$^{-1}$	& 0.65 mJy beam$^{-1}$	& 0.81 mJy beam$^{-1}$	& 0.69 mJy beam$^{-1}$ \\
Cube RMS noise at 5\arcsec resolution$^{a}$ & 1.5 mJy beam$^{-1}$	& 1.0 mJy beam$^{-1}$ & 1.2 mJy beam$^{-1}$	& 2.2 mJy beam$^{-1}$ \\        
    \hline
    \end{tabular}
    \tablefoot{$^{a}$Noise per channel in the final products from the ALMA Interferometric Pipeline (see Section \ref{sect:alma_obs}).  The value does not account for noise correlation between channels.}
    \label{tab:observations}
\end{sidewaystable}

The resulting data from all four SBs were calibrated and imaged using the standard ALMA Interferometric Pipeline \citep{Hunter23}, with a few minor manual flagging commands added by the ALMA data analysts during the quality assurance (QA) process. The final products all met the ALMA QA criteria, including meeting our desired angular resolution range and exceeding our requested sensitivity (Table \ref{tab:observations}). We performed a thorough review of the pipeline web logs accompanying the final products, finding no problems with the calibration, continuum subtraction, or imaging, and used the delivered products for the analysis reported in this paper. 

The resulting CO (1-0) cubes had Gaussian-like noise with sensitivities $10 - 12\%$ above theoretical. Visual inspection identified four line detections from the high SFR sample, but none from the two lower SFR samples apart from a serendipitous detection of a background galaxy in one of the continuum spectral windows (SPW19) in the field of COSMOS-1430950.  This serendipitous detection was made in the first LO tuning for COSMOS-1430950 before it was discovered to be inappropriate for the targeted CO line.  As a result, the detection appears in the ``Observation 3'' dataset, but later fell outside SPW19 in ``Observation 4'' (Table \ref{tab:observations}). Properties of this source are given in the Appendix. Only one of the systems, COSMOS-1411106, had a 3 mm continuum detection over 4.5$\sigma$ (peak emission of 17 mJy beam$^{-1}$, rms=0.17 mJy beam$^{-1}$).  We note that all visually identified sources were also found by auto-masking in the ALMA pipeline.

In order to search for additional spectral line detections, to optimize moment maps, and to measure source properties, we ran the Source Finding Application, SoFiA-2 \citep{Westmeier21} on the ALMA cubes.  In particular, we used the well-tested ``smooth+clip'' (S+C) algorithm, and the reliability module to reject false positives.  Due to the very high quality (high Gaussianity) of the ALMA data, the results are not especially sensitive to changes in the SoFiA-2 input parameters, but we describe our preferred choices here.  
First, we allowed SoFiA-2 to perform a channel-based \texttt{spectral} noise scaling to the data.  Then the S+C algorithm was run with a pixel threshold of 3.8$\sigma$ and a range of spatial and spectral smoothing kernels.  Given the high spatial and spectral resolution, we started with spatial and spectral parameter settings \texttt{scfind.kernelsXY =  0, 3, 6, 9, 12} and \texttt{scfind.kernelsZ =  0, 3, 7, 15}, respectively, where the the numbers indicate the number of pixels in the smoothing kernel.  For each source, we then used the combination of the set of smallest kernels that gave consistent results in the CO morphology in order to not artificially enlarge the mask and include unnecessary noise.  The linker module was run with spatial and spectral linking lengths of 5 and 3 pixels, respectively, and minimum sizes in both directions of 5 pixels.  Finally, the reliability module was run with a probability threshold of 0.75.

Of the 14 sources observed with ALMA, SoFiA-2 recovered all four visually identified sources, and one additional source: COSMOS-0969208.  We found it was only when significant spectral smoothing was applied to the data that the detection of 0969208 became evident.  This source was also not visible when the data were viewed at the native resolutions in 3D using the iDaVIE virtual reality software \citep{Jarrett21}.  
All five CO detected sources were among those predicted to have the brightest CO emission based on their UV+IR SFRs.  For the sources that were not detected with SoFiA-2 using the above parameters, we also experimented with lowering the S+C noise threshold to 3.3, and the probability threshold in the reliability module down to 0.5, and varying the range of kernel sizes, but this did not result in further detections.  Examining the noise properties of the sources we did detect, as well as comparing our SoFiA-2 results with the auto-masking that is performed by the ALMA pipeline,  we are confident in our choice of SoFiA-2 parameters.  The ability of SoFiA-2 to find the emission with high confidence in the faintest cases (i.e.,~COSMOS-0969208) is both a testament to the success of the SoFiA-2 software and the extremely high quality of the ALMA data.

\subsection{CHILES \HI\ data}

The CHILES \HI\ 21\,cm observations were carried out with the VLA using five consecutive B-configurations (``epochs''; maximum baseline 11.1\,km) from October 2013 to April 2019. The observations utilized the L-band ($1-2$ GHz) receiver and the 8-bit samplers of the VLA. The VLA WIDAR correlator was set up to cover the nominal frequency range $950-1430$ MHz via fifteen 32\,MHz wide subbands. The observations employed frequency dithering, which consisted of using three different frequency settings in each of the five observing epochs to minimize the loss of sensitivity at the edges of the subbands. Both recirculation and baseline board stacking techniques were used in the correlator\footnote{\url{https://science.nrao.edu/facilities/vla/docs/manuals/oss/widar\#section-7}}\fnsep\footnote{\url{https://library.nrao.edu/public/memos/evla/EVLAM_163.pdf}} to obtain 2048 channels in each 32\,MHz subband, resulting in a frequency spacing of 15.6 kHz (3.3 km s$^{-1}$ at $z=0$).  For the imaging described below, the data have been binned to 250 kHz leading to a velocity resolution of 59~\kms\ at $z=0.12$.

Calibration of the CHILES data was done with a custom data reduction pipeline utilizing CASA 5.3 \citep{mcmullin07,casa22} running on the Spruce Knob High Performance Computing facility at West Virginia University.  We followed the standard data reduction procedure: import data, apply online flags, bandpass/flux density scale calibration, complex gain calibration, and application of calibration to target, with a couple of important modifications.  First, we identified frequency ranges with persistent RFI on short baselines and masked these frequencies for the calibration step.  These masks were customized for each epoch and changed over the six years of the survey.  For calibration, we first derived the initial solutions before flagging the calibrator using the {\it rflag} algorithm in CASA's {\it flagdata}, and then re-derived the final calibration.  Once this calibration was applied to the data, we did a final flagging of the target before splitting it off for imaging.  At the end of the pipeline, we produced quality assessment plots that allowed us to determine if there were problems with any calibration steps and to localize the problem to specific sources, times, or antennas.  On the occasions where the QA plots showed a problem, we went back and flagged the appropriate visibilities.  Section \ref{sect:epoch1} provides an overview of the data quality in comparison with previously published Epoch 1 results \citep{Hess19}.  More details on the pipeline and data quality will be provided in Pisano et al.~(in prep).  

We have reported in detail on the imaging pipelines developed for the CHILES dataset in \citet{Dodson22} and \citet{Luber25b}.  Here we summarize both of these approaches, which differ in the details of the continuum subtraction, but commonly separate the processing into domains in which tasks can be done in parallel. These domains are temporal processing which can be applied on the level of individual sessions and observational epochs; and image processing in which the parallelization can be applied across independent frequency channels. The former is handled by treating observing sessions separately when possible and only combining sessions when strictly necessary, and the latter by separating the processing into smaller frequency chunks.

In particular, the imaging pipeline of \citet{Dodson22} generates a global model of the continuum emission from the combined continuum data that is subtracted from the different epochs in parallel. This includes hour angle (HA) variations in the model for continuum sources far from the phase center. The technique accounts for instrumental variation due to the rotation of the primary beam with hour angle, but assumes that the HA variations are constant between observations. Meanwhile, the imaging pipeline of \citep{Luber25b} relies on a multi-step and multi-scale, low spectral resolution model of the sources for each observing session. This accounts for any daily variations, but does not properly account for rotational variation of the primary beam. The two approaches produce very similar outcomes, and are both considered successful, as the corrections for the sources in the CHILES field lie at or below the level of the noise. However, in certain frequency ranges, where artifacts can arise for different reasons, one method can marginally outperform the other. In the redshift range considered here, we found \citet{Dodson22} cubes have channels with, on average, 15.9\% better rms values while maintaining similar values of kurtosis compared to the cubes from \citep{Luber25b}.  These quantitative measures are supported by a qualitative assessment that there are also fewer remaining visual artifacts, thus all results presented are derived from the \citet{Dodson22} data products.

We used SoFiA-2 in a similar manner to that on the ALMA data to generate source masks and parameterize the 14 sources that were previously detected.  First, we extracted 40 pixel $\times$ 40 pixel $\times$ 176 channel (1.3\arcmin\ $\times$ 1.3\arcmin\ $\times$ 10,384~\kms) cubelets from the CHILES data, centered on the spatial position of each known galaxy and spanning the full velocity range of the volume of interest.  The limited spatial extent minimizes how the variation in noise properties across the CHILES field-of-view impacts the mask generation.  We then ran SoFiA-2 with spectral noise scaling and the S+C algorithm with smoothing kernels \texttt{0, 3, 6}, and \texttt{0, 3, 5}, with linking lengths of 2 in both the spatial and spectral dimensions.  In this case, we turned the reliability module off because the relativley small size of the cubelets prevented a sufficient number of negative features from being detected to perform the reliability calculation.  The smoothing kernel and linking length values were further tailored to each individual source to account for local noise properties, which can vary significantly across the field in this redshift range \citep{Luber25b}.

Using the source detection masks output by SoFiA-2, we performed an image-plane H\"ogbom CLEAN \citep{hogbom74} in the cubelets centered on each detection. In this image-based CLEAN, we identify the pixel of maximum emission, subtract off 10\% of the emission, saving this subtracted flux in a model cube, as well as 10\% of the synthesized beam dictated by the CHILES point-spread function. This process is done for all pixels in the mask above one times the local rms noise until no pixels in the mask lie above the noise criteria. Once this is accomplished, we convolve the model with a two dimensional Gaussian fit to the inner peak of the point-spread function and add this to the cube with the subtracted components. This methodology allows us to perform a deep clean on the regions of \HI\ emission from these sources.  The final resolution of the \hi\ cubes are $7.1\arcsec \times5.2\arcsec$, corresponding to $15.6\times12.6$ kpc at $z=0.12$.

\subsection{CHILES Con Pol 1.4 GHz continuum}
\label{sect:con_pol}

In this work we derive 1.4 GHz star formation rates for CHILES galaxies based on the CHILES Continuum and Polarization survey (``CHILES Con Pol'', or CCP) source catalog by \citet{Gim25}.  
CHILES Con Pol consists of commensal observations of the CHILES \HI\ field, utilizing the full stokes capabilities of the VLA \citep{Luber25a}. The CCP observational setup involved the deliberate selection of four additional SPWs to avoid or mitigate the adverse effects of frequency ranges with known strong RFI. Each SPW had a bandwidth of 128 MHz, comprising 64 channels, and recording the full polarization products.

Detailed information regarding the data reduction and imaging processes can be found in the comprehensive survey description paper \citet{Luber25a}. The final CHILES Con Pol image was generated using Briggs weighting \citep{Briggs95} with a robust value of 0.5, as implemented in CASA, resulting in a synthesized beamwidth of 5.5\arcsec$\times$5.0\arcsec and an RMS noise level of 1.09 $\mu$Jy beam$^{-1}$ measured in regions far away from the phase center in the Stokes I image. The RMS noise measured within the central 3\arcmin$\times$3\arcmin region shows a higher value of 1.92 $\mu$Jy beam$^{-1}$, due to the presence of source confusion at this depth and resolution, and residual imaging artifacts \citep{Luber25a,Gim25}.

To calculate the 1.4 GHz star formation rates, we use Eqn 15 from \cite{Murphy11}, assuming values of $T_e = 10^4$~K and $\alpha_{nt}=0.8$ derived from the same work, and the 1.4 GHz flux in the rest frame of the source.  $T_e$ is the electron temperature taking into account the thermal bremsstrahlung (free–free) emission around massive star-forming region, and $\alpha_{nt}$ is the spectral index of the non-thermal (synchrotron) emission from cosmic ray electrons moving in the galaxy's magnetic field--predominantly from core collapse supernovae (see \citealt{Murphy11} for details).  The rest-frame 1.4~GHz luminosities were estimated using its measured spectral indices \citep{Gim25}.

\subsection{Stellar counterparts with Spitzer, Hubble, and DECaLS}

In addition of the radio data described above, we retrieved archival 3.6 $\mu$m \textit{Spitzer Space Telescope} \citep{Werner04} Infrared Array Camera (IRAC; \citealt{Fazio04}) images for the CO detected galaxies from the COSMOS cutout server hosted by NASA/IPAC Infrared Science Archive (IRSA)\footnote{\url{https://irsa.ipac.caltech.edu/data/COSMOS/index_cutouts.html}}. The imaging was done as part of the S-COSMOS survey to map the full COSMOS field in all seven \textit{Spitzer} bands \citep{Sanders07}. The native units of the calibrated images are in Jy steradian$^{-1}$ which we convert to \msun\ kpc$^{-2}$ assuming a mass-to-light ratio of 0.47 \citep{McGaugh14} to estimate the stellar mass surface density, $\Sigma_*$, of the galaxies.

In addition, we retrieved COSMOS Hubble Space Telescope Advanced Camera for Surveys (HST-ACS) F814W mosaics \citep{Koekemoer07,Massey10} from IRSA, and DECaLS \emph{griz} false-color images from Legacy Survey\footnote{\url{https://www.legacysurvey.org/}} for a qualitative assessment of the CO and \HI\ detections relative to their stellar counterparts.

\subsection{xCOLD GASS, COOL BUDHIES, and HIGHz comparison samples}
\label{sect:samples}

To investigate the potential for evolution in the gas content of galaxies with redshift, we compare our CHILES detections with three stellar mass, \hi\ mass, and color-matched samples from the literature for which both \HI\ and CO measurements exist.  First, xCOLD GASS \citep{Saintonge17} is an IRAM 30m and APEX CO survey of a large (N=532) sample of low redshift ($0.01 < z < 0.05$) galaxies with \hi\ measurements from the GALEX Arecibo SDSS Survey (GASS, \citealt{Catinella18}) and represents the largest sample of CO-detected galaxies to date.  Second, COOL BUDHIES \citep{Cybulski16} is an LMT CO-survey of the WSRT intermediate redshift ($z\sim0.2$) Blind Ultra-Deep \HI\ Environment Survey (BUDIES, Verheijen et al. 2007) and represents the largest sample (N=15) of CO-detected galaxies beyond the local Universe with \HI\ observations of the same field.  Finally, there are five ALMA CO-detected galaxies which are a subset of the \hi-detected HIGHz Arecibo survey of galaxies at intermediate redshift ($0.17 < z <0.25$; \citealt{Cortese17}).  Given the different motivation and observing depths for each sample, we attempt to homogenize the samples with our subset of CHILES galaxies by applying mass- and color-selections, when necessary, before making a comparison.  We describe this process below.

The CHILES galaxies presented in this paper comprise an \HI\ flux-limited sample in a narrow redshift range ($0.11 < z < 0.13$), which corresponds to a log \hi\ mass limit between 9.0 and 9.2.  Therefore, we apply an \HI\ mass cut of $\log(M_{HI}/M_{\odot})>9.2$ to the xCOLD GASS sample from the outset.  In addition, xCOLD GASS is the combination of two different samples: one stellar mass selected sample with $\log(M_*/M_{\odot})>10.0$ and a second at lower redshift with $9.0<\log(M_*/M_{\odot})<10.0$.  As described in Section \ref{sect:results}, this mass break in the two xCOLD GASS samples conveniently corresponds to the same stellar mass at which CO is or is not detected in the CHILES galaxies and so we consider the two mass ranges separately in our analysis.  A stellar mass-selected sample will tend to be redder in color than an \hi\ selected sample (e.g.,~\citealt{Durbala20}), so to ensure the most fair comparison, we only consider xCOLD GASS galaxies in the same k-corrected NUV-r color range as the CHILES detections $\pm0.1$ magnitudes, for the two stellar mass ranges.  For the high stellar mass galaxies, this color range is $2.36<NUV-r<4.14$.  For the low stellar mass galaxies, the color range spans $0.4<NUV-r<2.54$.  We have taken the NUV-r colors for the CHILES galaxies from \citet{Hess19}.

From the COOL BUDHIES survey, we only consider galaxies that are detected in \HI: these all lie above $\log(M_{HI}/M_{\odot})=9.2$ and are bluer than NUV-r=4.0 \citep{Jaffe16}, so we do not apply any further \hi, or color sample selection.  In addition, all galaxies have stellar masses greater than $\log(M_*/M_{\odot})>10.0$.  As a result, the COOL BUDHIES galaxies are most directly comparable to the CHILES CO direct-detections, and the high stellar mass xGASS sample.  It is interesting to note that COOL BUDHIES targeted cluster galaxies, whereas xCOLD GASS and CHILES have not targeted specific environments. We take this into consideration in the discussion of gas evolution with redshift (Section \ref{sect:redshift}).  By comparison, five of the CHILES \hi\ detections are in a relatively massive group (31 confirmed group members), while about a third are not in groups, and the rest are in small loose groups of 2-4 members \citep{Knobel12,Hess19}.  We also note that the calibration of the RSR on Large Millimeter Telescope (LMT; \citealt{Hughes20}) has been updated using the entire historical calibration data in 2019 (Yun, M., private communication).  The updated calibration at 97 GHz is 6.0 Jy/K: 15\% smaller than the value used by \citet{Cybulski16} for their COOL BUDHIES sample.  We apply this correction in our analysis.

The last comparison sample is the ALMA follow-up of a subset of five HIGHz galaxies \citep{Cortese17}.  These galaxies are selected to be extremely massive, both in stars ($\log(M_*/M_{\odot})>10.3$) and atomic gas ($\log(M_{HI}/M_{\odot})>10.3$) and are chosen to be isolated.  In fact, only one of the CHILES galaxies would make the cut to be included in the original HIGHz sample \citep{Catinella15}, and none of the CHILES galaxies would have been included in the ALMA follow-up, where the lowest stellar mass galaxy has $\log(M_*/M_{\odot})>10.8$, and the lowest \hi\ mass galaxy is $\log(M_{HI}/M_{\odot})>10.35$.  As a result, we include this sample for completeness, but since it is not representative of the average galaxy population, we hesitate to draw strong conclusions from it.  This HIGHz subset belongs to the most massive $\sim$1\% of all \hi\ detected galaxies, and the 15\% most \HI\ massive systems at their stellar mass (e.g.,~\citealt{Maddox15}).

\section{Image and ALMA CO spectral line stacking}
\label{sect:data_stack}

In addition to source finding in the images generated by the ALMA pipeline, we performed 3D stacking \citep[e.g.][]{Chen21} of the CO data in order to achieve greater sensitivity and to measure the average molecular gas properties of the non-detections.  For validation of the method, we performed this stacking separately on the set of five CO-detected galaxies, and the set of nine CO non-detected galaxies.  The non-detections include data from a range of array configurations (Table \ref{tab:observations}), so we first smooth them all to a common 2\arcsec$\times$2\arcsec\ beam.  We extracted cubelets centered on the optical position and optical spectroscopic redshift\footnote{The optical redshifts are used instead of the \hi\ because the SNR of the \hi\ detections are relatively low and the redshift relies on a single line integrated over the entire galaxy.} of each galaxy, spanning 36\arcsec $\times$36\arcsec\ and $\sim$1000~\kms\ ($121\times121$ pixels and 700 channels), and co-add them on a pixel-by-pixel basis with equal weights.  This results in a single stacked cube of the same dimensions each for the detections, and for the non-detections.  We then binned the stacked data cubes by 20 channels in frequency corresponding to a spectral resolution of 56.5~\kms.

To estimate the best aperture from which to extract the CO spectra, and to test that our stacking algorithm was correctly centering the galaxies, we also stacked the HST images, and CHILES \HI\ total intensity maps.  In this case, the individual HST (CHILES) images were extracted over the same area corresponding to $1201\times1201$ ($18\times18$) pixels, and the images are co-added with equal weights--effectively averaged together on a pixel-by-pixel basis.  We then considered three apertures at integer units of the smoothed ALMA beam: [2, 4, 6] arcsec radius, and compared the extracted CO spectra from the stacked cubes of detections and non-detections.  The analysis of the apertures and overall CO stacking results are further discussed in Section \ref{sect:stacking}.

\begin{figure*}
    \centering
    \includegraphics[width=0.208\textwidth]{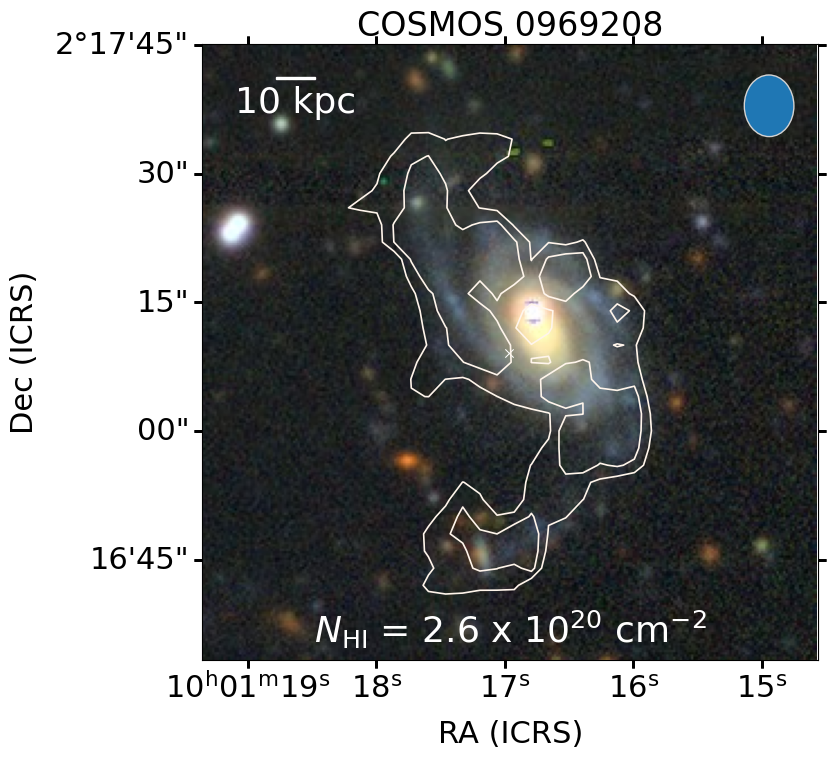}
    \includegraphics[width=0.192\textwidth,trim=35 0 0 0,clip]{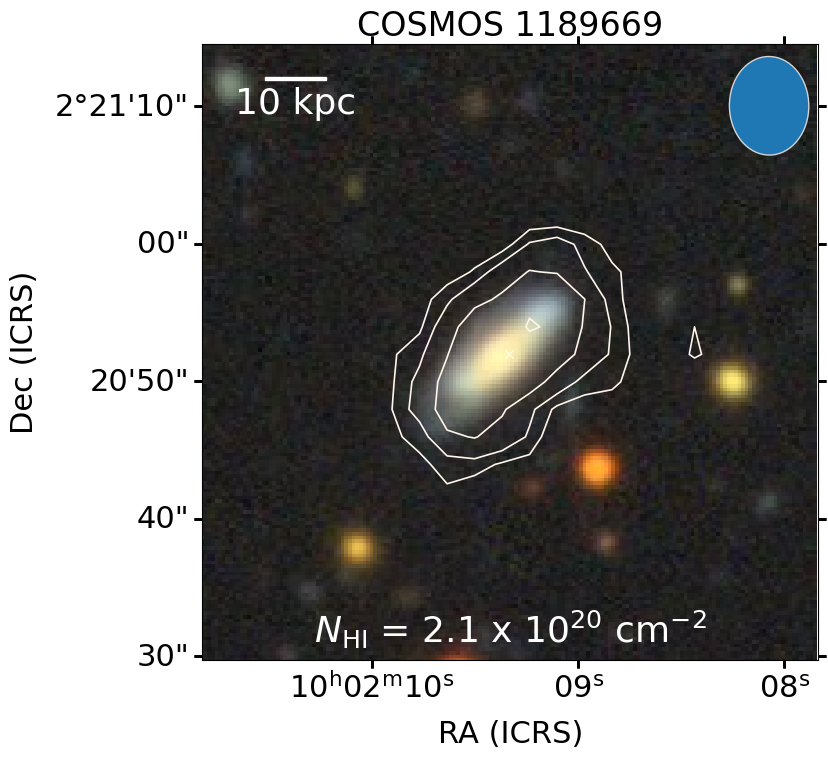}
    \includegraphics[width=0.192\textwidth,trim=35 0 0 0,clip]{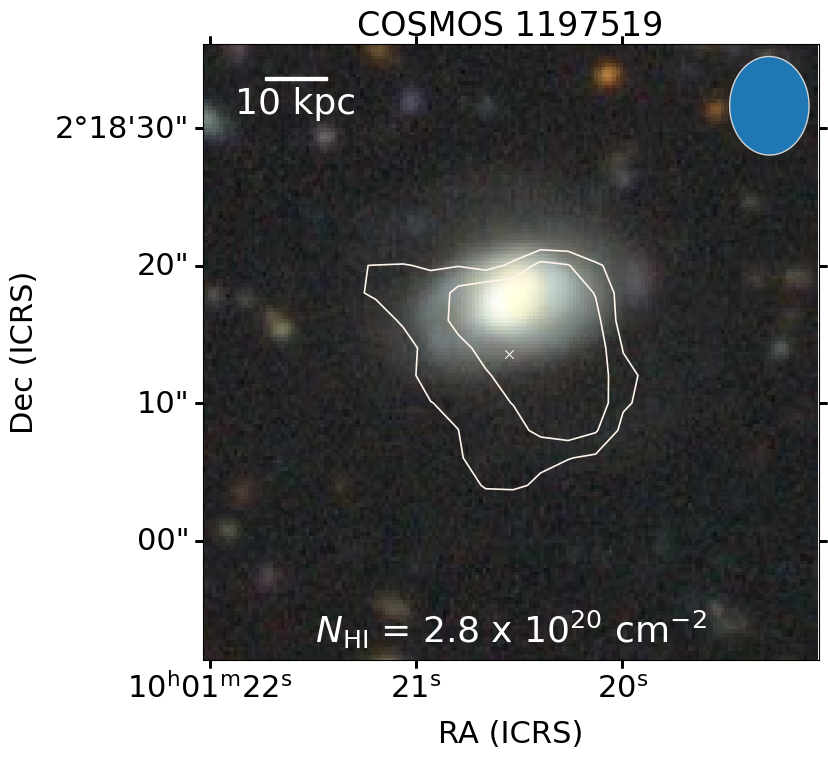}
    \includegraphics[width=0.197\textwidth,trim=35 0 6 0,clip]{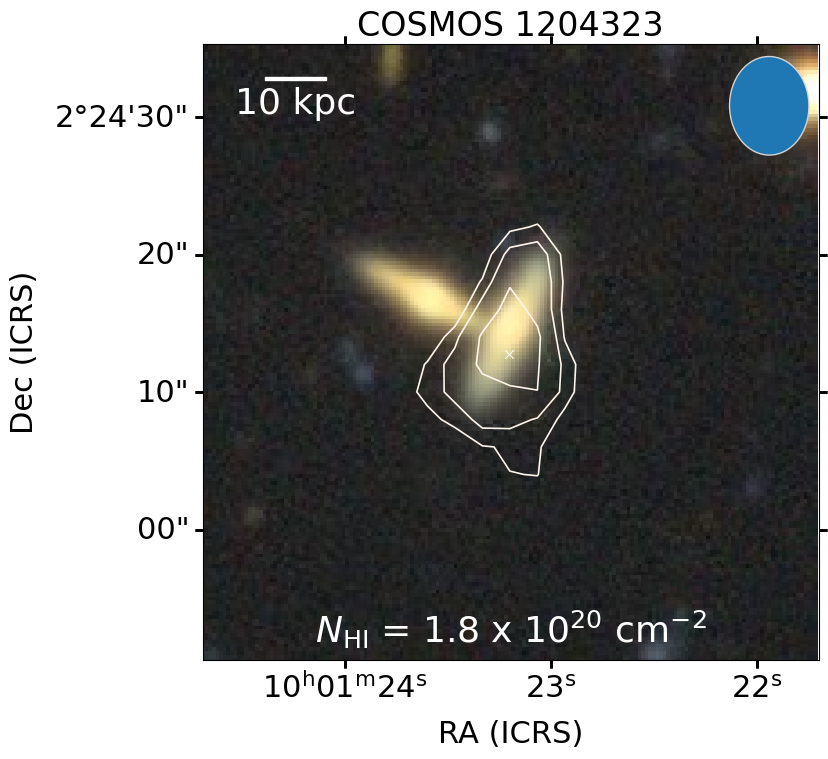}
    \includegraphics[width=0.191\textwidth,trim=35 0 0 0,clip]{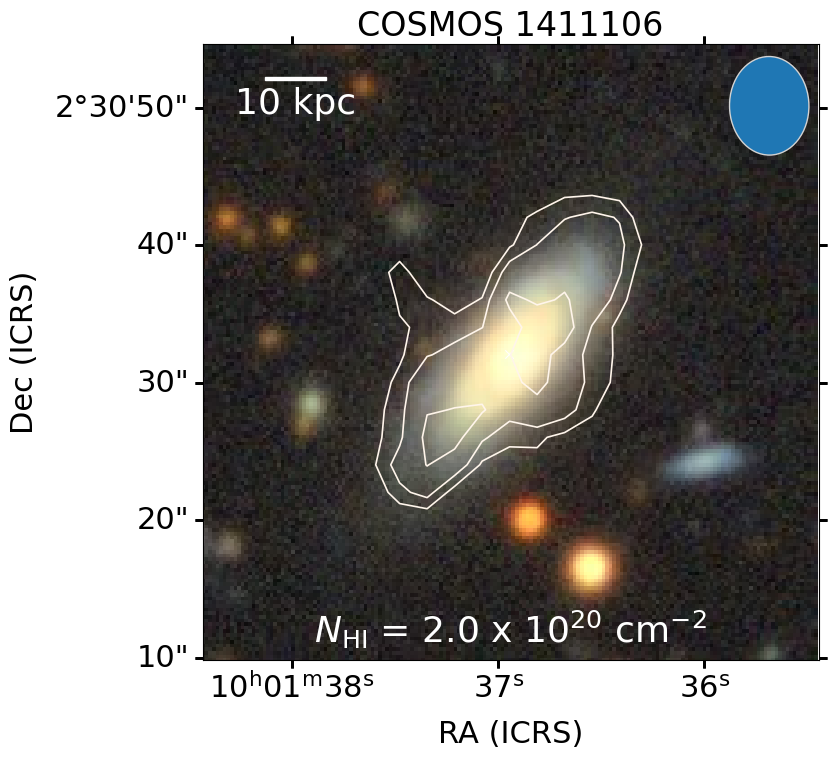} \\
    \vspace{0.6cm}
    \includegraphics[width=0.208\textwidth]{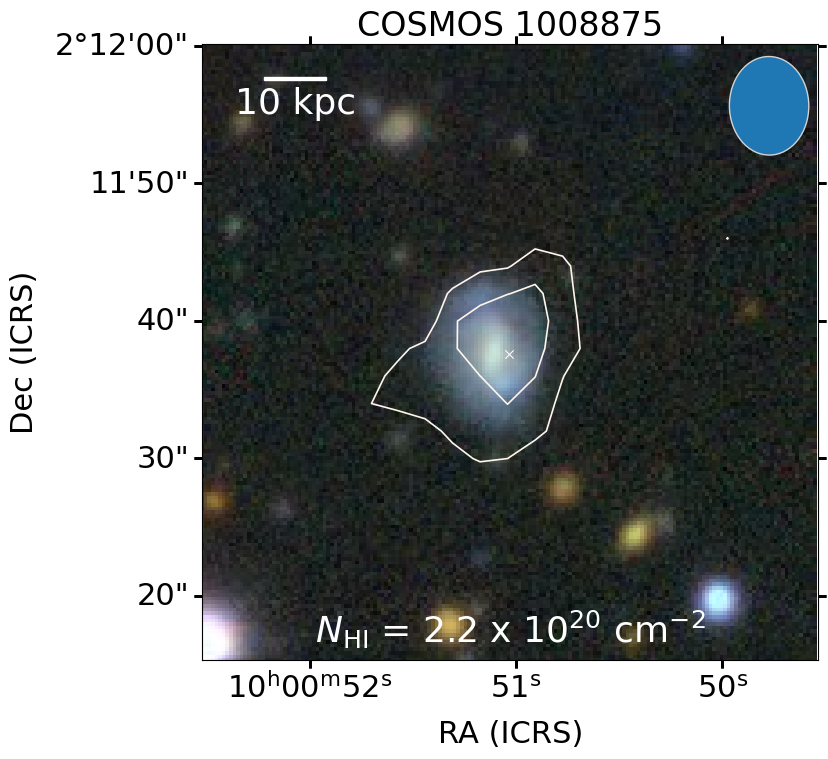}
    \includegraphics[width=0.195\textwidth,trim=35 0 0 0,clip]{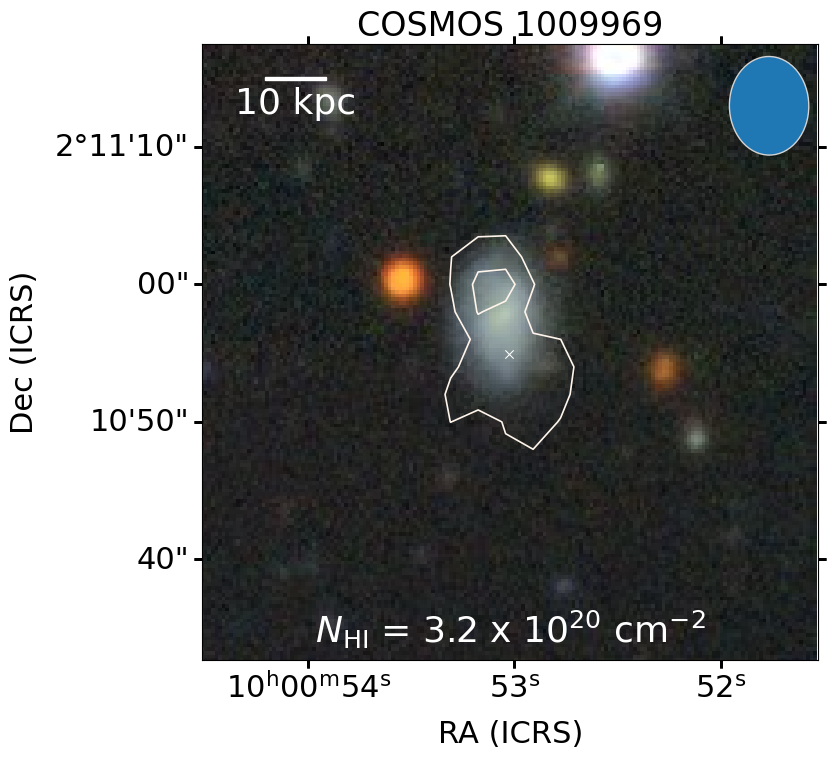}
    \includegraphics[width=0.195\textwidth,trim=35 0 0 0,clip]{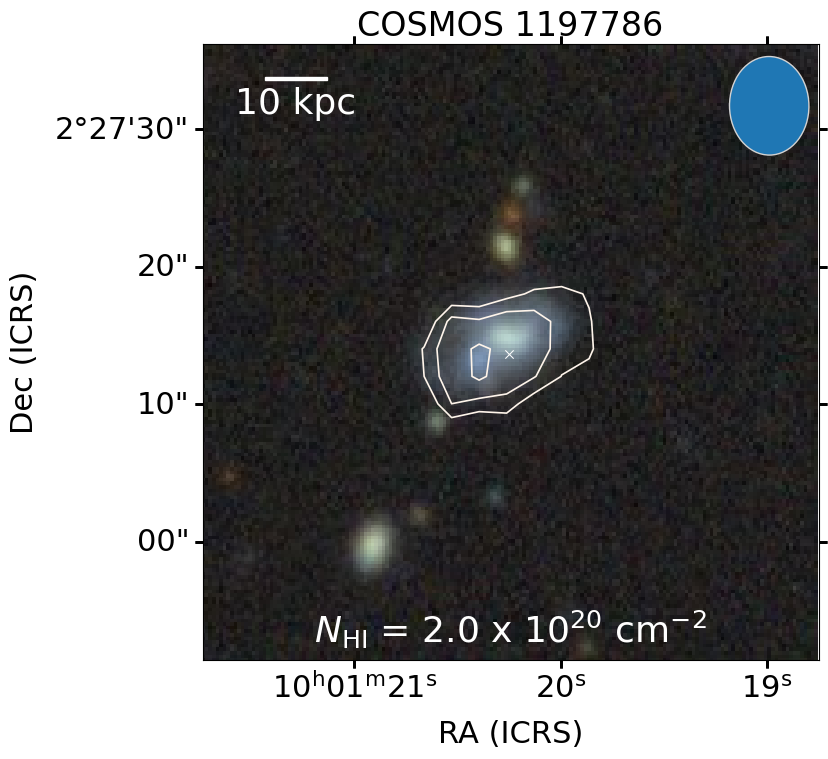}
    \includegraphics[width=0.202\textwidth,trim=35 0 0 0,clip]{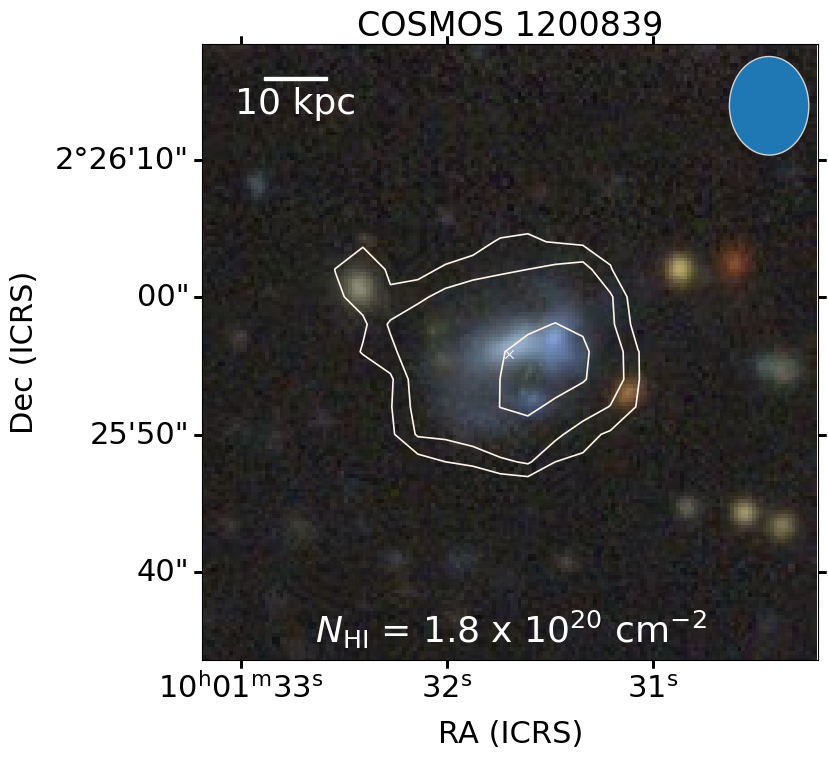}
    \includegraphics[width=0.207\textwidth]{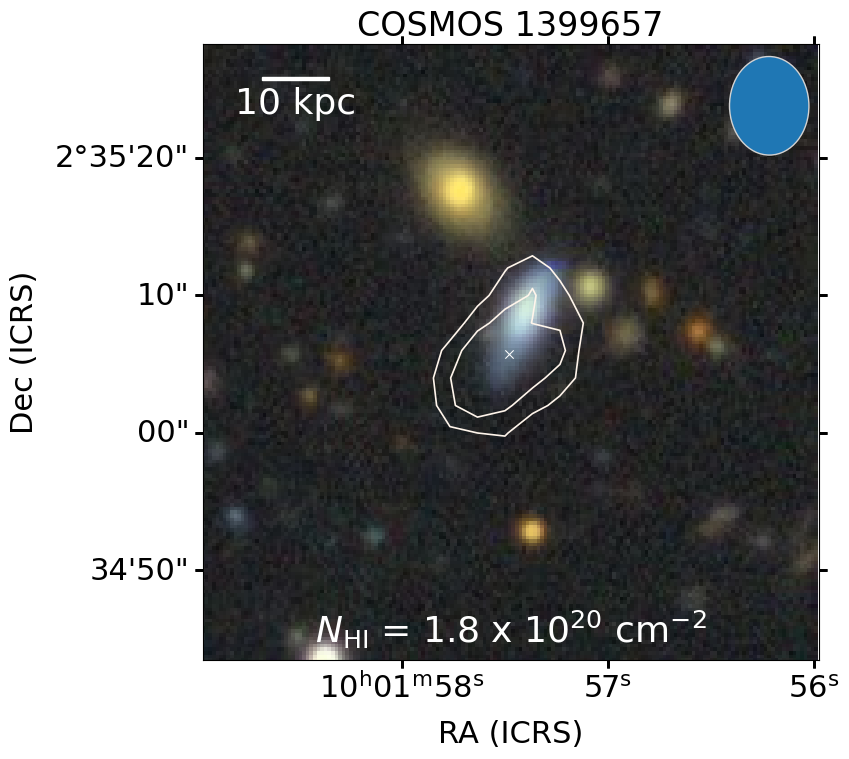}
    \includegraphics[width=0.193\textwidth,trim=35 0 0 0,clip]{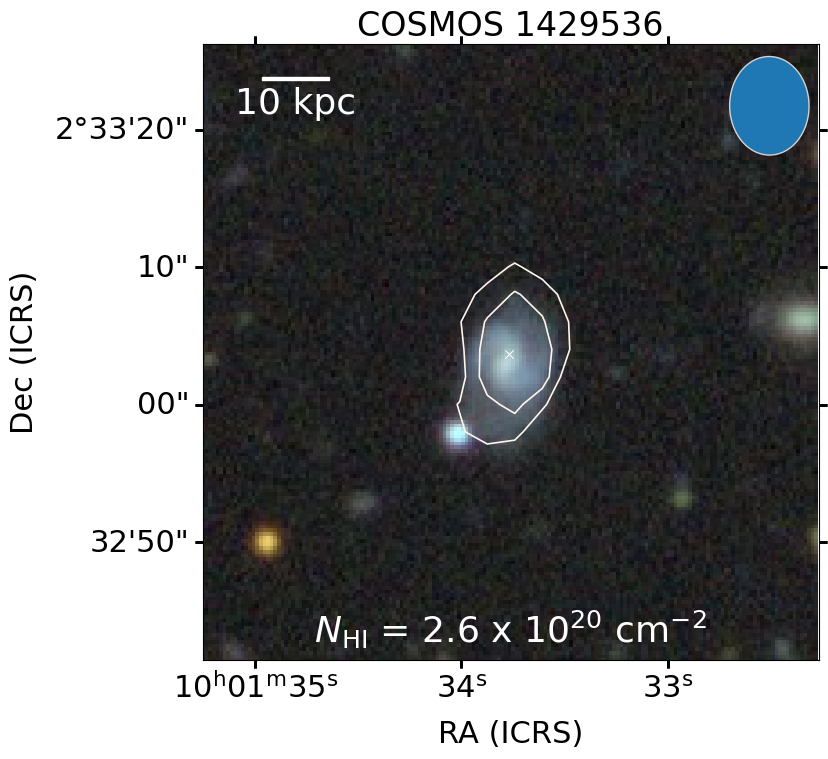}
    \includegraphics[width=0.193\textwidth,trim=35 0 0 0,clip]{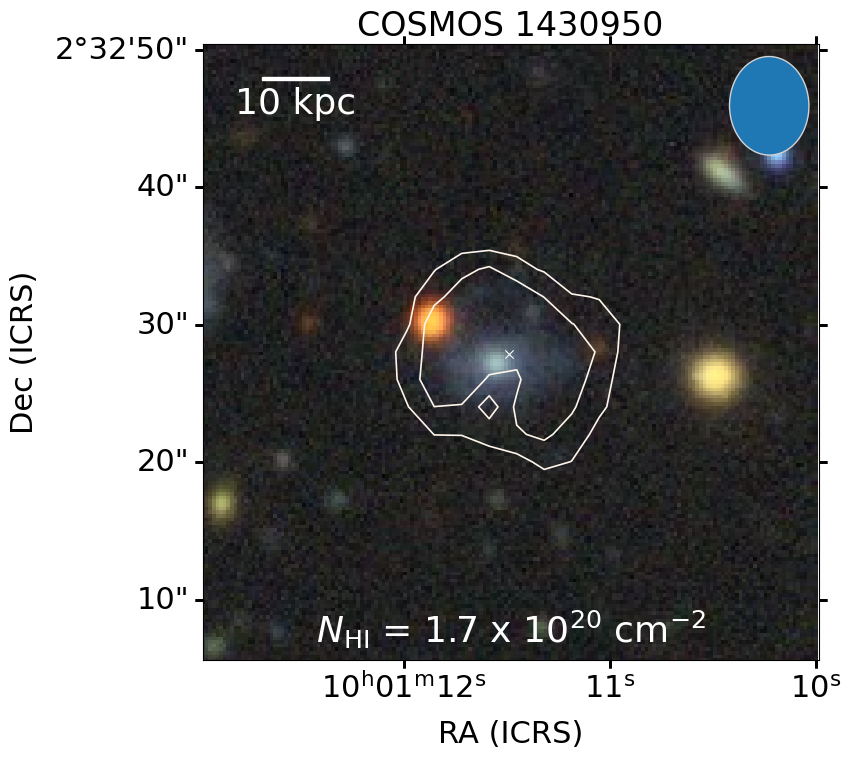}
    \includegraphics[width=0.193\textwidth,trim=35 0 0 0,clip]{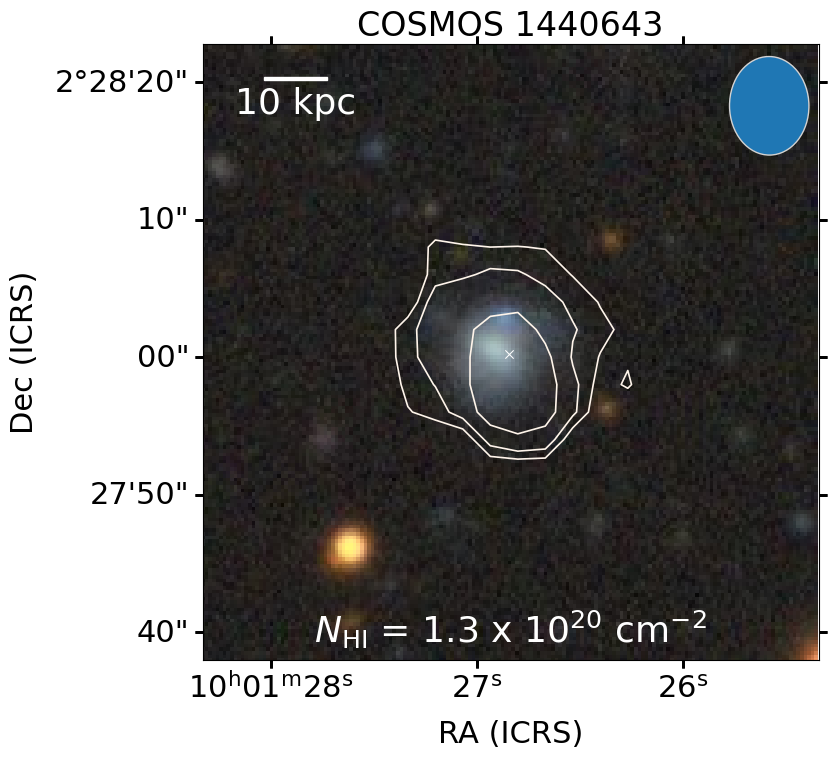}
    \includegraphics[width=0.193\textwidth,trim=35 0 0 0,clip]{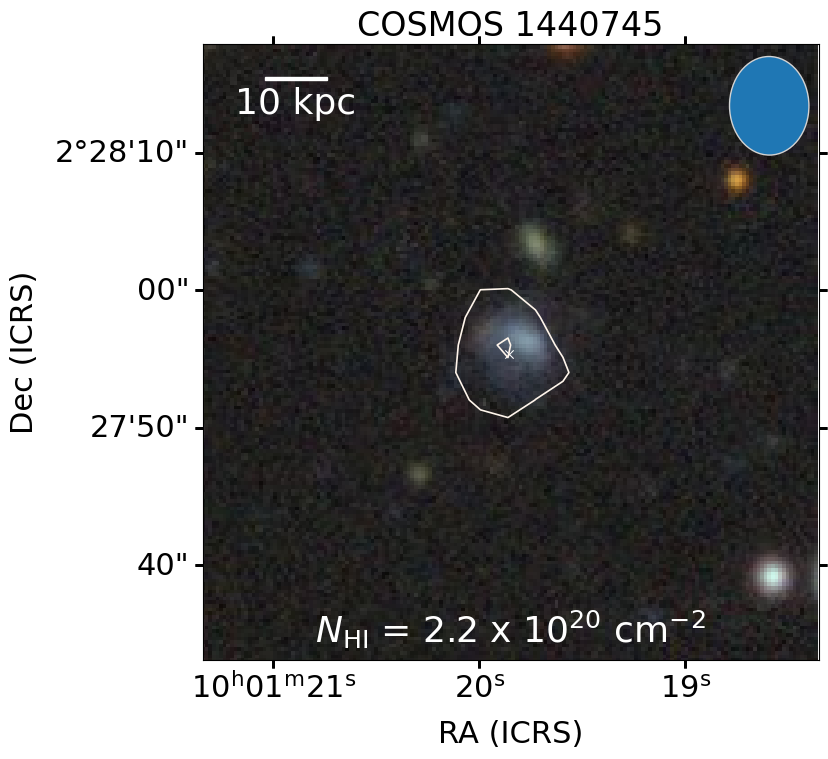}
    \caption{\HI\ contours overlaid on DECaLS \emph{griz} false-color images.  Galaxies are ordered top-to-bottom, left-to-right as they appear in Table \ref{tab:observations}.  Top row: CHILES galaxies detected in CO (1-0) by ALMA.  \HI\ column densities are $2^n\times[2.6, 2.1, 2.8, 1.8, 2.0]\times10^{20}$\cm\ $(n=0,1,2...)$. Middle and bottom row: CHILES galaxies that are undetected in CO (1-0). \HI\ column densities are as above: $2^n\times[2.1, 3.2, 2.0, 1.8]$, and $[1.8, 2.6, 1.7, 1.3]\times10^{20}$\cm\ $(n=0,1,2...)$. There is a clear dichotomy between the two sets of galaxies: CO detections are noticeably redder, while CO non-detections are significantly bluer.  See text for discussion.  Note that the top left-most galaxy (0969208) has a star superimposed to the north of the nucleus which appears white.  The red area to the south of the star is from the background galaxy.}
    \label{fig:decals}
\end{figure*}

\section{Results}
\label{sect:results}

All 14 of the sources identified in the CHILES 178 hour Epoch 1 data at $z=0.12$ and reported by \citet{Hess19} are confirmed in the combined 856 hour CHILES data cube.  Figure \ref{fig:decals} shows the CHILES \hi\ contours on DeCALS false color images.  

\begin{figure*}
    \centering
    \includegraphics[width=0.84\textwidth]{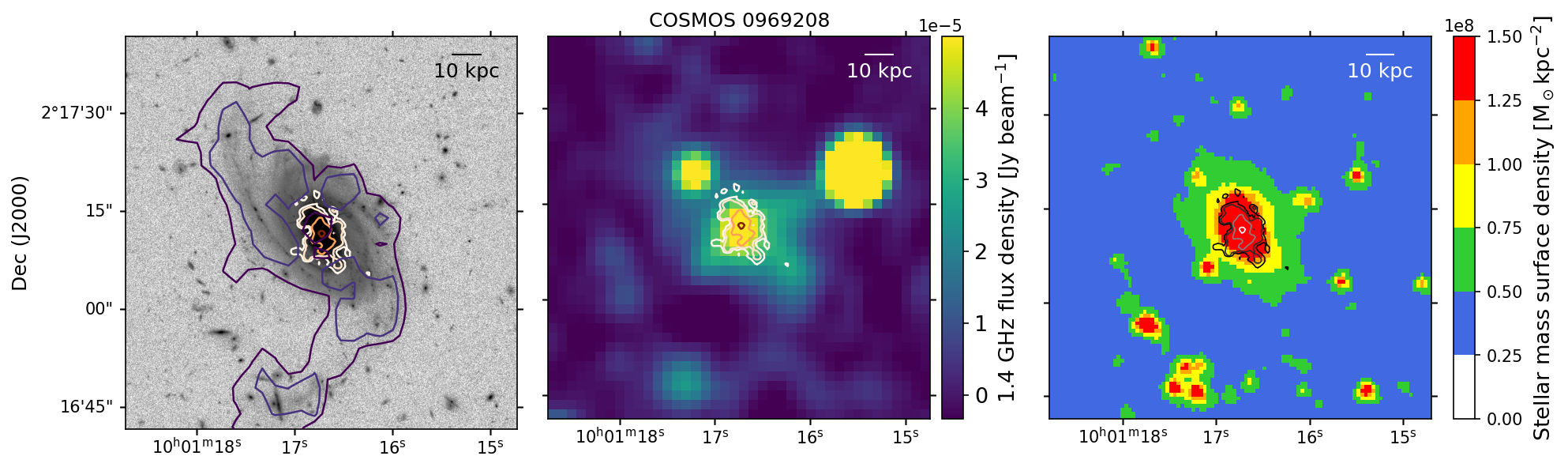}
    \includegraphics[width=0.84\textwidth]{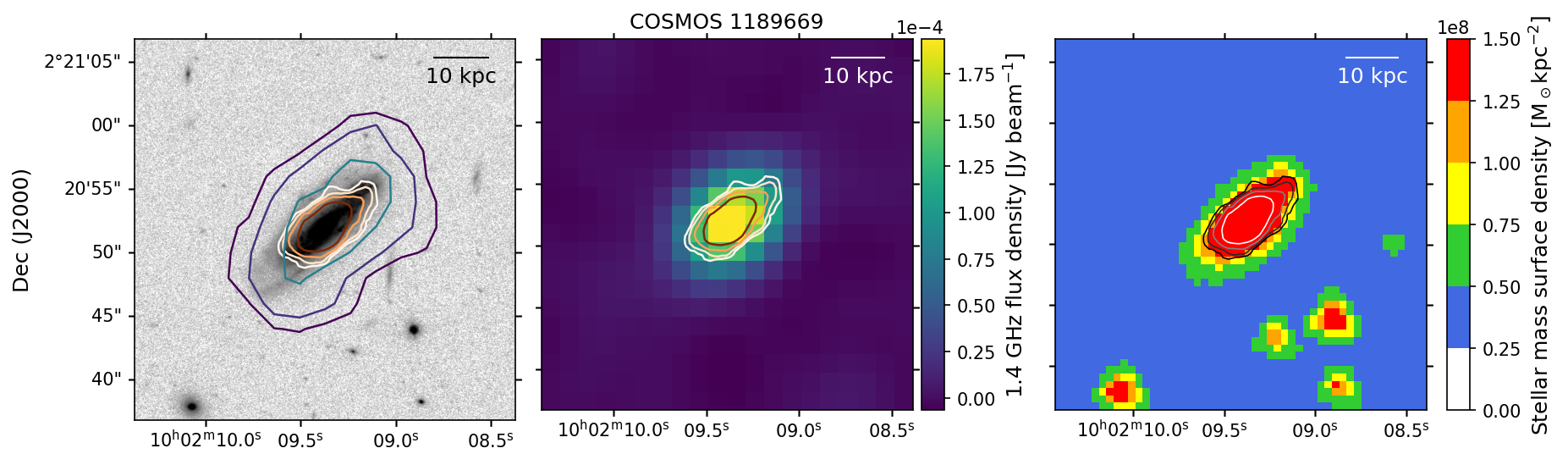}
    \includegraphics[width=0.84\textwidth]{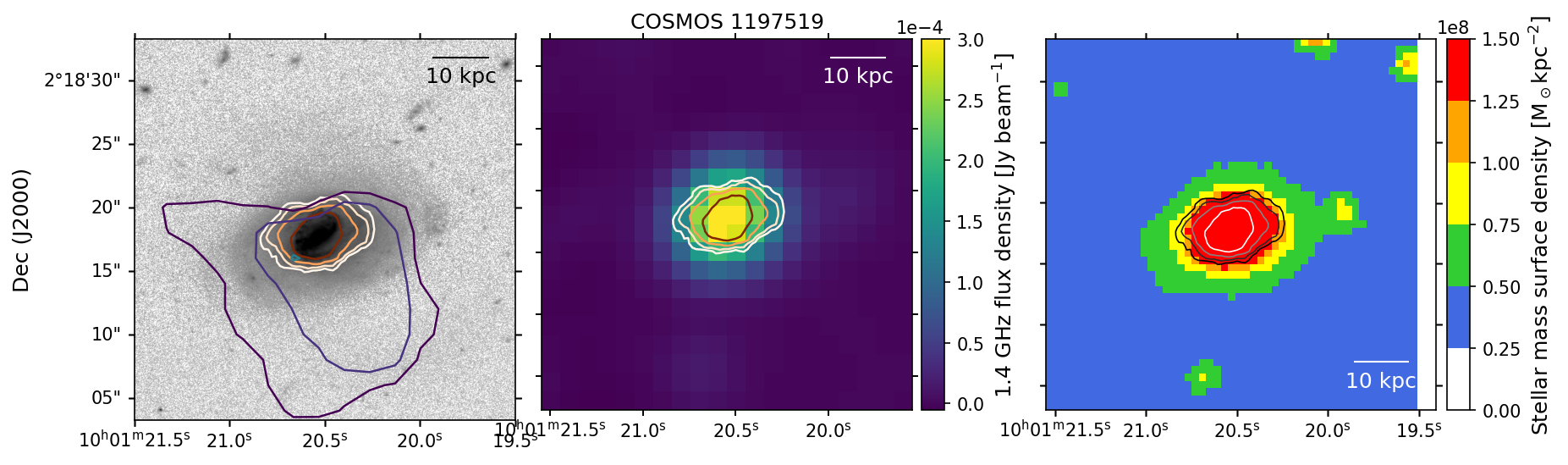}
    \includegraphics[width=0.84\textwidth]{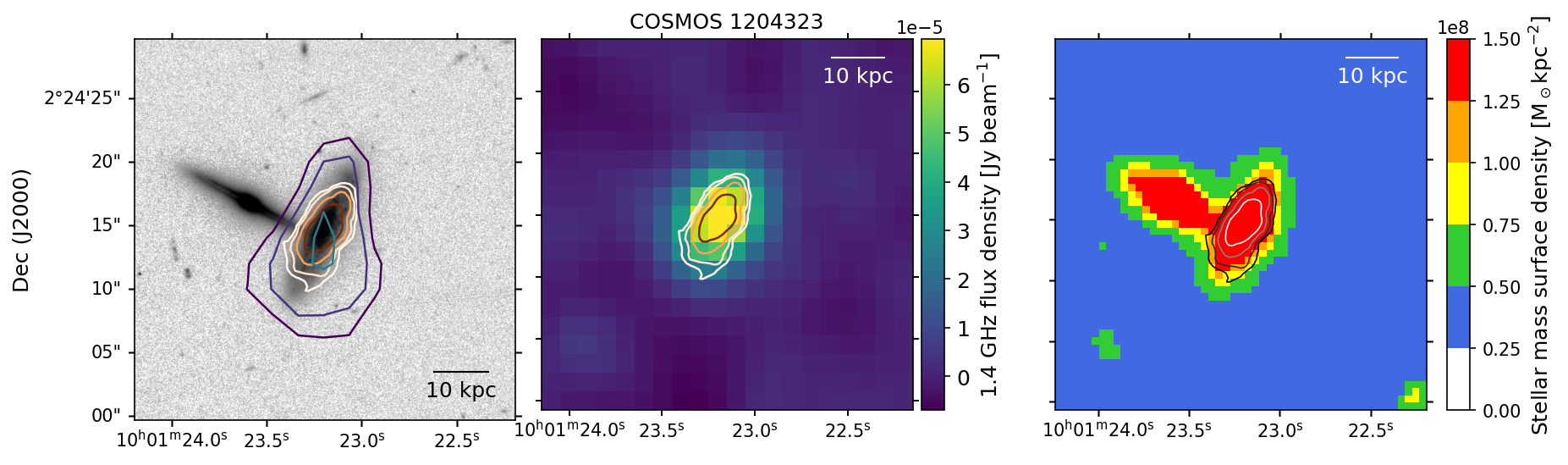}
    \includegraphics[width=0.84\textwidth]{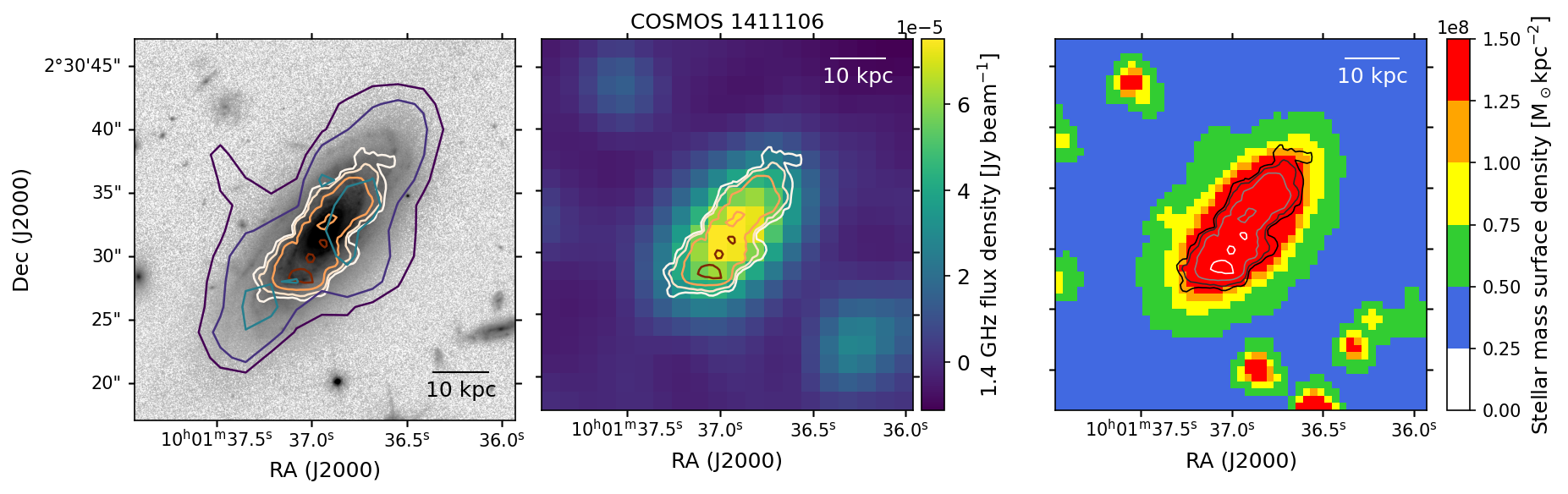}
    \caption{Left: HST/ACS F814W images overlaid with \HI\ contours (purple to teal) as in Fig \ref{fig:decals} and CO (1-0) contours (white to red).  CO contours correspond to \Ht\ column densities $2^n\times[3.6, 2.6, 2.3, 2.3, 3.9]\times10^{20}$\cm\ $(n=0,1,2,...)$  Center: CHILES Con Pol 1.4 GHz continuum images, overlaid with CO contours (white to dark red). Right: Spitzer 3.6 $\mu$m images, converted to bins of stellar mass surface density overlaid with CO contours (black to white).  Galaxies are presented in the same order top-to-bottom that they are left-to-right in Figure \ref{fig:decals}.}
    \label{fig:hst_gas}
\end{figure*}

We detect and spatially resolve CO (1-0) in five of the 14 targeted systems (top row of Figure \ref{fig:decals} and left column of Figure \ref{fig:hst_gas}).  These correspond to the galaxies with the highest stellar masses ($\log(M_*/M_{\odot})>10.0$) and star formation rates (SFR > 2.0 $M_{\odot}$ yr$^{-1}$) in the sample.  The CO detected galaxies are also qualitatively the reddest in the DeCALS imaging, in contrast to those not directly detected in CO which are significantly bluer in color (bottom two rows of Figure \ref{fig:decals}).  Our ALMA detections increase the number of known galaxies with spatially and spectrally resolved atomic and molecular gas maps beyond $z=0.06$ by 500\%.  The lone other system to be resolved in both CO+\HI\ above these redshifts is the CHILES $z=0.376$ galaxy reported by \citet{Fernandez16} which has been resolved with ALMA in CO (3-2) and will be reported on in a future paper (\citealt{DonovanMeyer23}; Donovan Meyer et al.~in prep).  

Based on the CCP images and available multi-wavelength COSMOS data, we find only limited evidence for AGN activity in one of the CHILES galaxies in this sample. 
AGN activity was assessed based on three diagnostic criteria: X-ray luminosity thresholds \citep{Szokoly04}, excess of X-ray luminosity relative to that expected from star formation (Gim et al.~in prep.), and mid- and far-infrared color-color diagnostics \citep{Kirkpatrick13}. While no X-ray selected AGN were identified in our sample, one galaxy (COSMOS 1008875) exhibits signatures of AGN activity based on its position in the mid- and far-infrared color-color diagram. We estimate that AGN contributes $\approx 30$\% to the radio luminosity of this source, derived from its $q_{\rm FIR}$ value of 2.184 compared to the canonical value of $q_{\rm FIR} = 2.34$ for local star-forming galaxies \citep{Yun01}.  This is noted in Table \ref{tab:gal_properties}.  For the rest, we assume that all the 1.4 GHz continuum emission is due to star formation.  

\begin{sidewaystable}[]
    \caption{Molecular and atomic gas properties of CHILES $z=0.12$ galaxies}
    \centering
    \begin{tabular}{cccccccccccc}
    \hline
    \hline
        COSMOS  & RA Dec  & $z_{HI\odot}$$^a$ & $D_L$ & $S_{CO}\delta v$~$^b$ & $\log(M_{H_2})$~$^c$ & S$_{HI}\delta v$ & $\log(M_{HI})$ & $\log(M_*)$~$^d$ & $S_{1.4GHz}$     & SFR$_{1.4 GHz}$ & SFR$^d_{UV+IR}$  \\
        ID      & J2000   &                   & Mpc   & Jy~\kms               & $M_{\odot}$     & Jy Hz            & $M_{\odot}$    & $M_{\odot}$ & $\mu$Jy          & $M_{\odot}$ yr$^{-1}$ & $M_{\odot}$ yr$^{-1}$ \\
        (1)     & (2)     & (3)               & (4)   & (5)                   & (6)             & (7)              & (8)            &  (9)        & (10)             & (11)                          & (12)     \\ 
    \hline
        0969208 & $10^h01^m16.7^s 2^d17^m11.9^s$ & 0.12163 & 569 & 1.648      &   9.71          & 14028.9          & 10.42          & 10.50       & $470.3\pm21.4$ & 13.05              & 5.87 \\
        1189669 & $10^h02^m09.4^s 2^d20^m52.1^s$ & 0.12688 & 595 & 2.226      &   9.88          &  5138.1          & 10.02          & 10.20       & $338.7\pm17.2$ &  10.27             & 6.14 \\
        1197519 & $10^h01^m20.5^s 2^d18^m17.8^s$ & 0.123\phantom{00} & 576 & 1.973   &   9.80   &  4583.6          &  9.94          & 10.30       & $562.7\pm9.8$  & 16.04              & 7.20 \\
        1204323 & $10^h01^m23.2^s 2^d24^m15.0^s$ & 0.12745 & 596 & 1.552      &   9.72          &  1997.0          &  9.61          & 10.10       & $122.0\pm6.5$  &  3.71              & 3.89 \\
        1411106 & $10^h01^m36.9^s 2^d30^m32.1^s$ & 0.12353 & 579 & 2.320      &   9.88          &  4367.7          &  9.92          & 10.50       & $209.0\pm10.9$ &  6.01             & 5.87 \\
        1008875 & $10^h00^m51.1^s 2^d11^m37.7^s$ & 0.12289 & 575 & < 0.107    & < 8.53          &  2204.8          &  9.62          &  9.60       &   $62.7\pm8.5$ &  1.25$^{\dagger}$ & 0.33 \\
        1009969 & $10^h00^m53.1^s 2^d10^m57.7^s$ & 0.12389 & 580 & < 0.107    & < 8.55          &  1764.8          &  9.53          &  9.56       &      --        &  --\,             & 0.62 \\
        1197786 & $10^h01^m31.5^s 2^d25^m57.7^s$ & 0.12337 & 578 & < 0.107    & < 8.53          &  4087.5          &  9.39          &  9.45       &  $49.1\pm3.1$  &  1.41             & 0.57 \\
        1200839 & $10^h01^m20.3^s 2^d27^m14.9^s$ & 0.12133 & 567 & < 0.150    & < 8.66          &  1272.9          &  9.88          &  9.35       &  $21.6\pm2.5$  &  0.60              & 1.28 \\
        1399657 & $10^h01^m57.4^s 2^d35^m08.8^s$ & 0.11358 & 528 & < 0.085    & < 8.35          &  1196.7          &  9.28          &  9.54       & $32.9\pm10.7$  &  0.79             & 0.32 \\
        1429536 & $10^h01^m33.8^s 2^d33^m03.0^s$ & 0.11269 & 524 & < 0.146    & < 8.59          &  1597.5          &  9.40          &  9.39       &      --        &  --\,             & 0.56 \\
        1430950 & $10^h01^m11.6^s 2^d32^m27.1^s$ & 0.11159 & 518 & < 0.107    & < 8.44          &  2263.2          &  9.54          &  8.92       &      --        &  --\,             & 0.51 \\
        1440643 & $10^h01^m26.9^s 2^d28^m02.9^s$ & 0.12197 & 571 & < 0.107    & < 8.52          &  2295.6          &  9.63          &  8.48       &  $36.6\pm2.8$  &  1.02             & 0.59 \\
        1440745 & $10^h01^m19.8^s 2^d27^m56.5^s$ & 0.12309 & 576 & < 0.085    & < 8.43          &   725.9          &  9.14          &  8.79       &  $22.8\pm2.7$  &  0.65             & 0.24 \\
    \hline
    \end{tabular}
    \tablefoot{The table is sorted first by CO detections (top) vs non-detections (bottom), and then by COSMOS ID number. $^a$The uncertainty on the \hi\ redshift is $\sim$0.00008. $^b$The CO flux is calculated from CASA.  The limiting CO luminosity and molecular mass are calculated from 3$\sigma$ the 5\arcsec\ smoothed data, over 50~\kms.  $^c$We assume $\alpha_{CO}=4.35$ \citep{Saintonge22} to convert flux to molecular mass.  $^d$We reproduce the stellar mass and SFR from \citep{Hess19}; the uncertainties were reported of order 25\% and 35\%, respectively. $^{\dagger}$Star formation for 1008875 is calculated from 70\% of the total $S_{1.4GHz}$ flux, where the other 30\% is due to AGN activity.  See text for details.
    }
    \label{tab:gal_properties}
\end{sidewaystable}

\subsection{Molecular and atomic gas masses}
\label{sect:results1}
Over the course of this paper it became apparent that the \HI\ and CO communities have different assumptions about velocity conventions.  In particular, within the \HI\ community it is historically common to convert frequency to velocities defined by the optical convention when referring to recessional velocities because of the natural comparison with optical redshifts for extragalactic objects.  \citet{Meyer17} went so far as to comment that radio velocities were deprecated.  Meanwhile, it is common in the CO community to use radio velocities, which have the advantage that they are linear with frequency, due to the historic connection with Galactic observations.  Unfortunately, at non-zero redshift, these velocity conventions diverge, giving rise to some ambiguity as to what ``velocity'' refers to in each context.  In practice, these problems are removed if equations that are dependent on velocity are in the rest frame of the galaxy.  However, in all three of these cases, the different velocity conventions require different correction factors of (1+z).

For \hi, we perform all calculations of mass and column density in frequency space following \citealt{Meyer17}.  For CO mass and column density calculations, the common equations use CO flux integrated over (presumably) the rest frame velocity width (e.g.,~\citealt{Solomon05}).  
To remove ambiguity, we derive the equivalent mass and column density equations for CO flux integrated over frequency, analogous to the \HI.  For both the \hi\ and CO, this has the advantage that the calculations are then performed in the native units of the data.  The different CO lines these equations are described in Appendix \ref{sect:serendip}.  For the presentation of spectra, line widths, and intensity-weighted velocity (moment 1) maps we convert frequency to the galaxy rest frame for direct comparison between the \hi\ and CO.  This also has the advantage that the velocities and velocity widths reported are independent of the redshift of the spectral line (i.e.,~for unknown lines and serendipitous detections such as reported in Appendix \ref{sect:serendip}).

Table \ref{tab:gal_properties} summarizes the measured molecular and atomic gas, stellar, and star forming properties of the CHILES galaxies, including detection upper limits where appropriate.  The table is ordered by CO detection vs non-detection, and then by COSMOS ID number.  The columns are as follows: (1) COSMOS 2008 ID; (2) optical RA and Dec in J2000 coordinates; (3) \HI\ redshift calculated from SoFiA-derived central frequency; (4) luminosity distance in Mpc; (5) CO flux or 4.5-sigma flux limits for an unresolved source over 300~\kms; (6) total molecular gas mass; (7) SoFiA-measured \HI\ flux; (8) \HI\ mass; (9) stellar mass as reported in \citet{Hess19}; (10) 1.4 GHz continuum flux density reported by CHILES Con Pol; (11) SFR derived from the 1.4 GHz flux density as described in Section \ref{sect:con_pol}; (12) SFR from COSMOS FIR+UV photometry as reported in \citet{Hess19}. 

The \hi\ mass has been calculated from the following equation, where $M_{HI}$ is in solar masses, $S_{HI}$ is the integrated flux density in Jy Hz, and $D_L$ is the luminosity distance in Mpc (e.g.,~\citealt{Meyer17}):
\begin{equation}
    M_{HI} = 49.7\, S_{HI}\, D_L^2 \,.
\label{eqn:mhi}
\end{equation}

The \Ht\ mass can be calculated from the following equation where $M_{H_2}$ is in solar masses, $\alpha_{CO}$ is the CO-to-H$_2$ conversion factor, $S^V_{CO}$ is the integrated flux density in Jy~\kms\ in the restframe of the galaxy, $\nu_{rest}$ is the rest frequency of CO (1-0) in GHz, and $D_L$ is in Mpc (e.g.,~\citealt{Solomon05}):
\begin{equation}
    M_{H_2} = \alpha_{CO} \times 3.25\times10^7\, S^V_{CO}\, \nu_{rest}^{-2}\, (1+z)^{-1}\, D_L^2 \,.
\label{eqn:mh2vel}
\end{equation}
This can be further simplified to remove confusion about the velocity frame by expressing the integrated flux density, $S_{CO}$, in units of Jy Hz, analogous to Equation \ref{eqn:mhi}.  We use this forumla to calculate the \Ht\ mass, which is particularly useful as it expresses the flux in the native units of the data:
\begin{equation}
    M_{H_2} = \alpha_{CO} \times 6.36\times10^{-3}\, S_{CO}\, D_L^2 \,.
\label{eqn:mh2}
\end{equation}
We assume $\alpha_{CO}=4.35$ \msun\ (K \kms pc$^{-2}$)$^{-1}$ , which takes into account elements heavier than hydrogen to get the total molecular gas mass \citep{Saintonge22}.  As a result the \Ht\ subscript in fact refers to the total molecular mass rather than pure molecular hydrogen, but this is consistent with the conventions widely used in the literature and in our comparison samples.

In Appendix \ref{sect:sip_co} and \ref{sect:sip_hi} we present atlas pages for the full complement of CO and \HI\ detections including moment maps, spectra, and position-velocity slices along the kinematic major and minor axes.
In all figures where surface brightness has been converted into column density we have been careful to include the corrections for redshift.  The equation for column density can be derived by dividing the equations for the mass by the beam area, $\Omega_{bm}$ in physical units, and recalling that $\Omega_{bm} = \pi a b / (4\ln(2))$ and that the angular distance is related to the luminosity distance by $D_A(z)=D_L(z)(1+z)^2$.  For \HI\ we recover the relation as described in \citet{Meyer17}:
\begin{equation}
    \left(\frac{N_{HI}}{cm^{-2}}\right) = 2.33\times10^{20}\, (1+z)^4\, 
    \left(\frac{S_{HI}}{Jy\,Hz}\right)\, 
    \left(\frac{ab}{arcsec^2}\right)^{-1},
    \label{eqn:nhi}
\end{equation}
where $a$ and $b$ are the synthesized beam major and minor axes, respectively, measured at the half power point.

For the \Ht\ column density we derive the following analogous equation which removes the ambiguity of velocity convention:
\begin{equation}
    \left(\frac{N_{H_2}}{cm^{-2}}\right) = \alpha_{CO} \times  
    1.49\times10^{16}\, (1+z)^4\, 
    \left(\frac{S_{CO}}{Jy\,Hz}\right)\, 
    \left(\frac{ab}{arcsec^2}\right)^{-1}.
    \label{eqn:nh2}
\end{equation}
As above, Jy Hz are convenient as they correspond to the native units of the data.

\subsection{Molecular and atomic gas morphologies}

Figure \ref{fig:hst_gas} shows the resolved CO and \HI\ contours on HST images; and CO contours on radio continuum images from CHILES Con Pol, and Spitzer 3.6 $\mu$m.  The left panels of Figure \ref{fig:hst_gas} show that the \hi\ morphology is more extended than the CO.  In many cases the \hi\ is also more extended than the stellar disk, but the column densities achieved are modest: a few $\times10^{20}$ cm$^{-2}$.  In general, the \HI\ at this depth is well confined to the stellar disk.  Only two of the five CO-detected galaxies exhibit an \hi\ hole, or \hi\ depression at the center, coincident with the highest density \Ht\ (first and last galaxies in Figure \ref{fig:hst_gas}; see also Appendix \ref{sect:sip_hi}), although this may in part be due to insufficient spatial resolution to identify other central \hi\ holes.

\begin{figure*}
    \centering
    \includegraphics[width=0.33\linewidth,trim=0 0 29 0,clip]{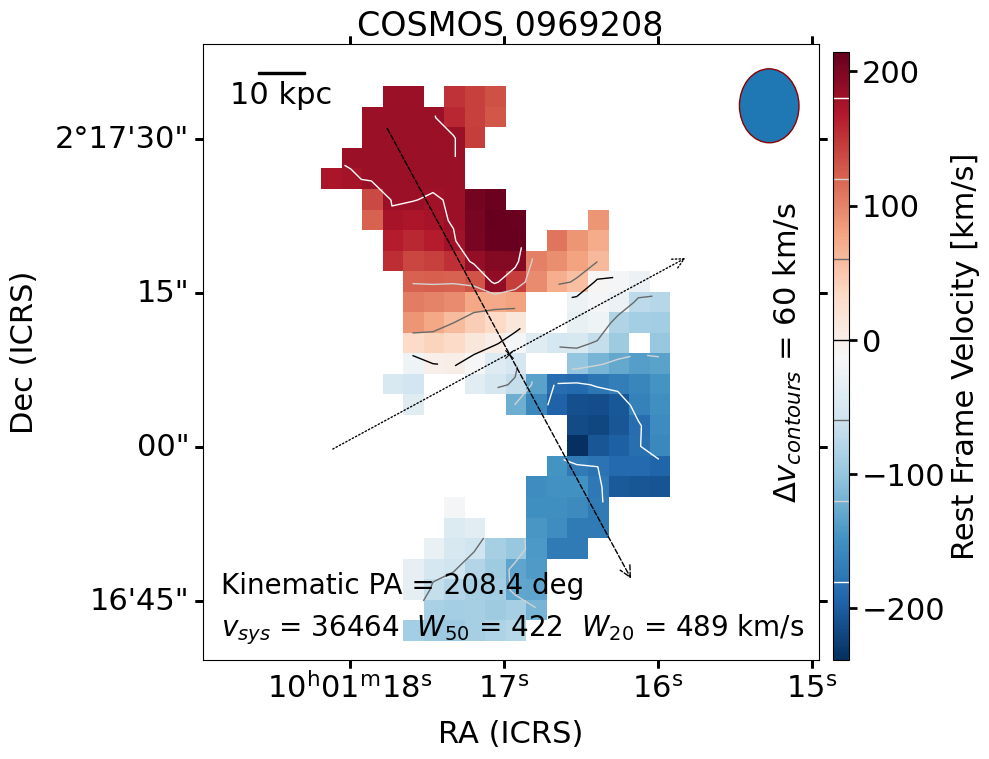}
    \includegraphics[width=0.32\linewidth,trim=26 0 29 0,clip]{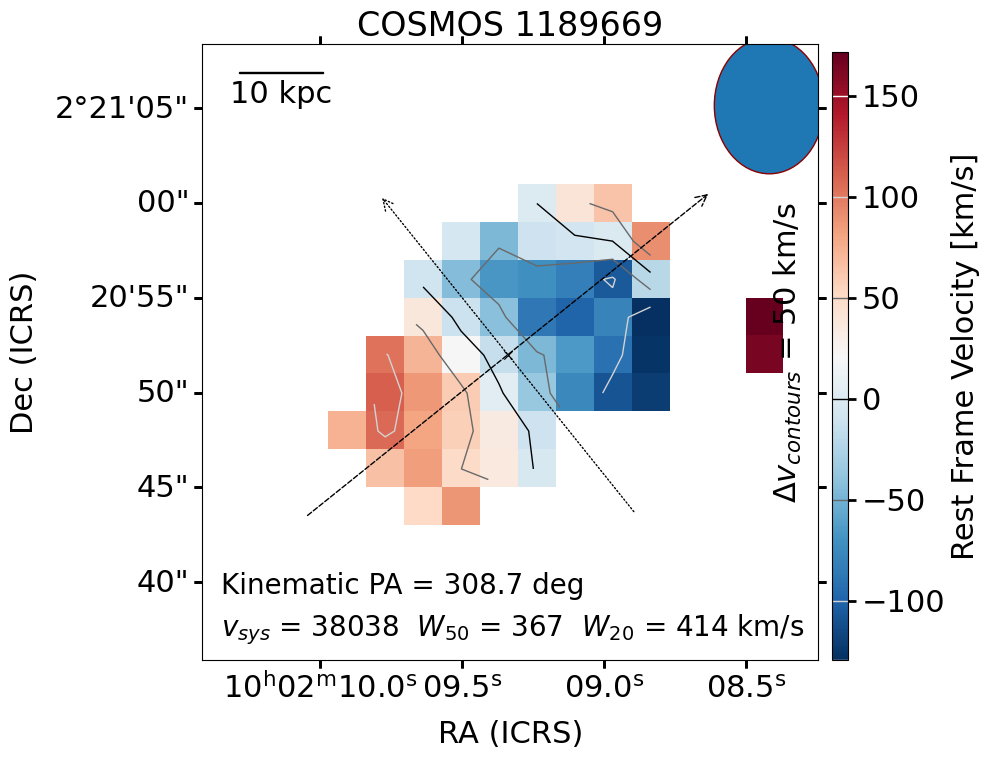}
    \includegraphics[width=0.33\linewidth,trim=26 0 0 0,clip]{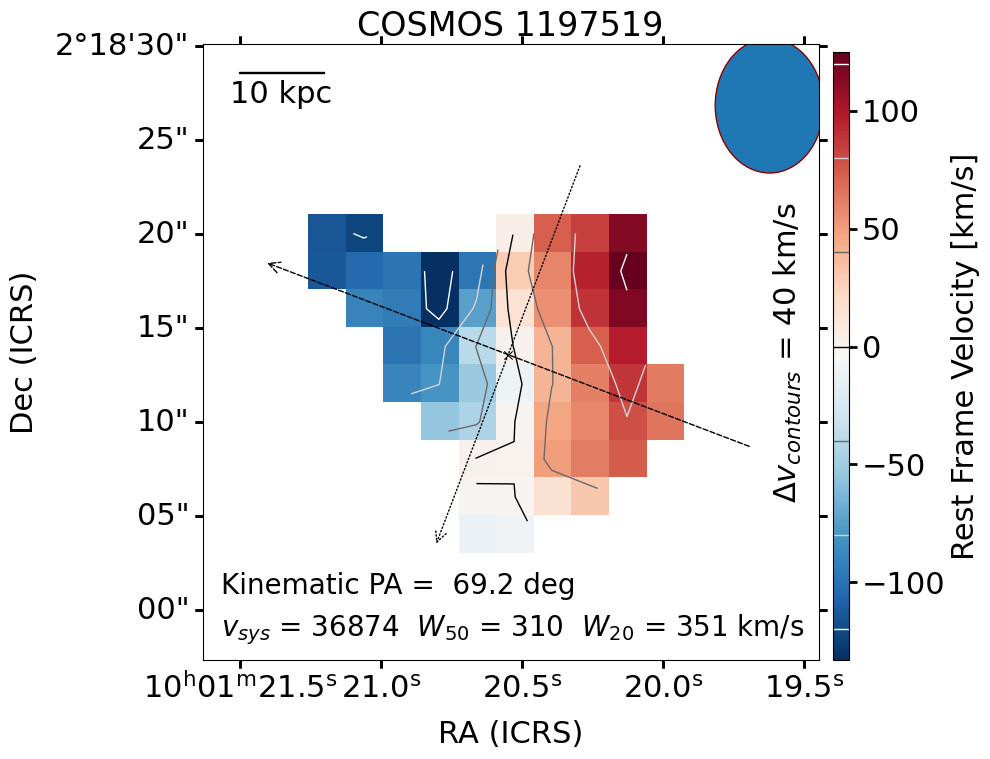} \\
    \includegraphics[width=0.33\linewidth,trim=0 0 29 0,clip]{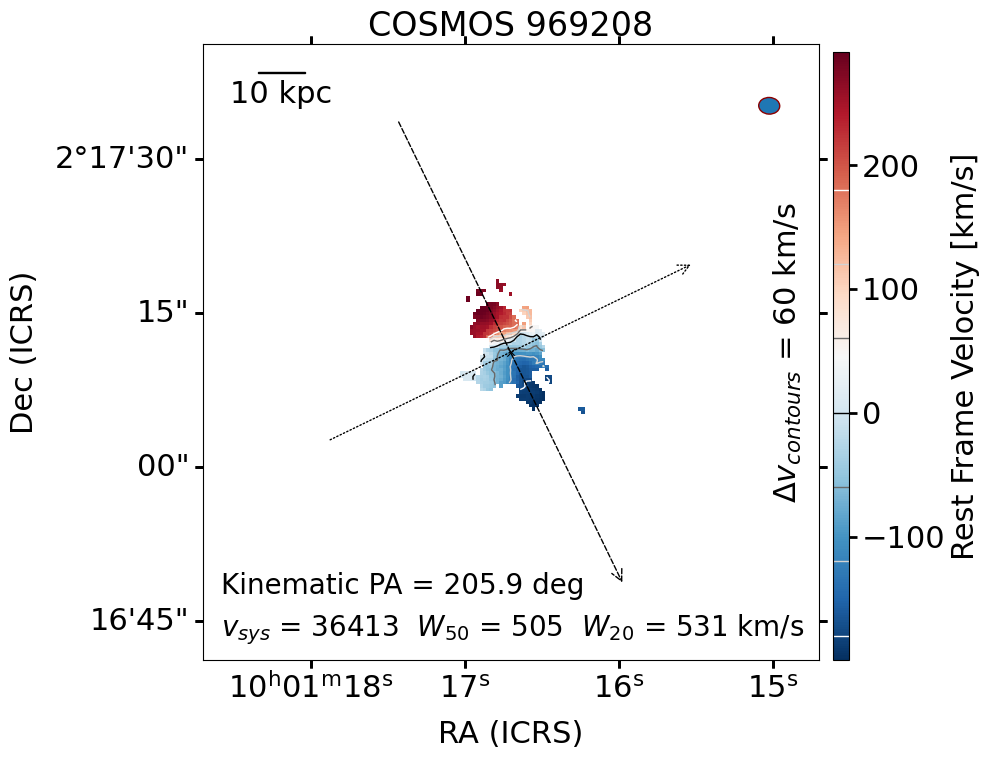}
    \includegraphics[width=0.32\linewidth,trim=26 0 29 0,clip]{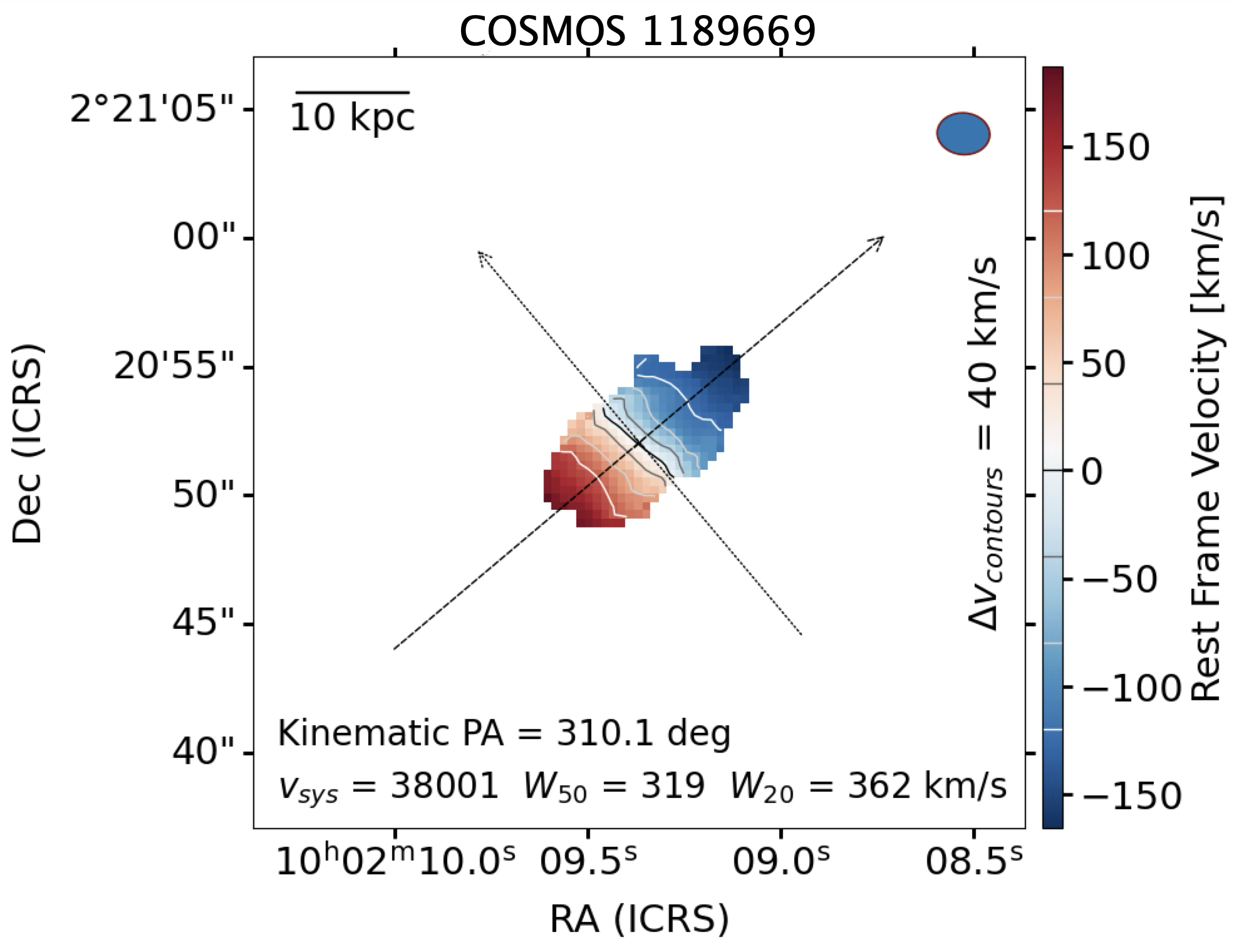}
    \includegraphics[width=0.33\linewidth,trim=26 0 0 0,clip]{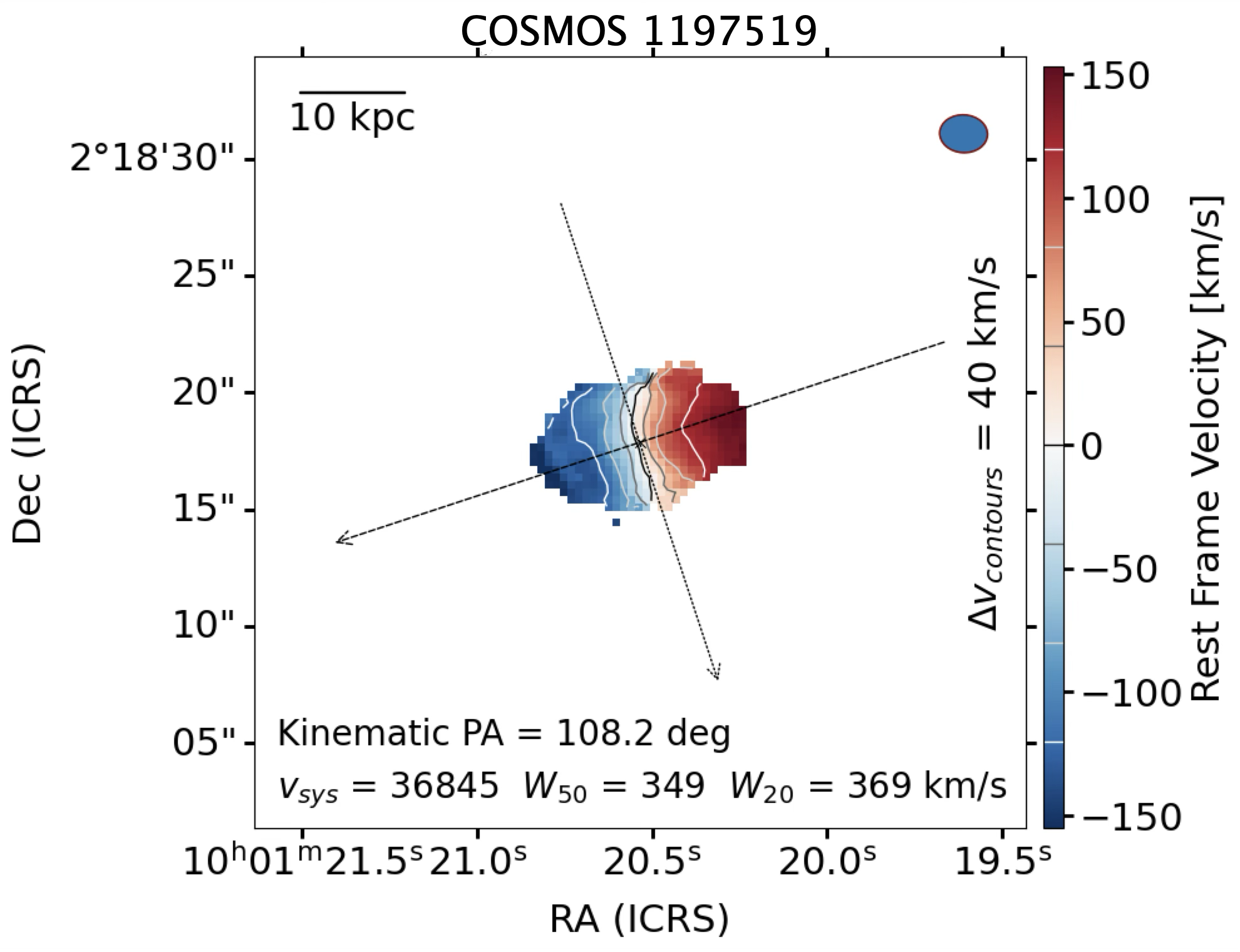} \\
    \includegraphics[width=0.33\linewidth,trim=0 0 29 0,clip]{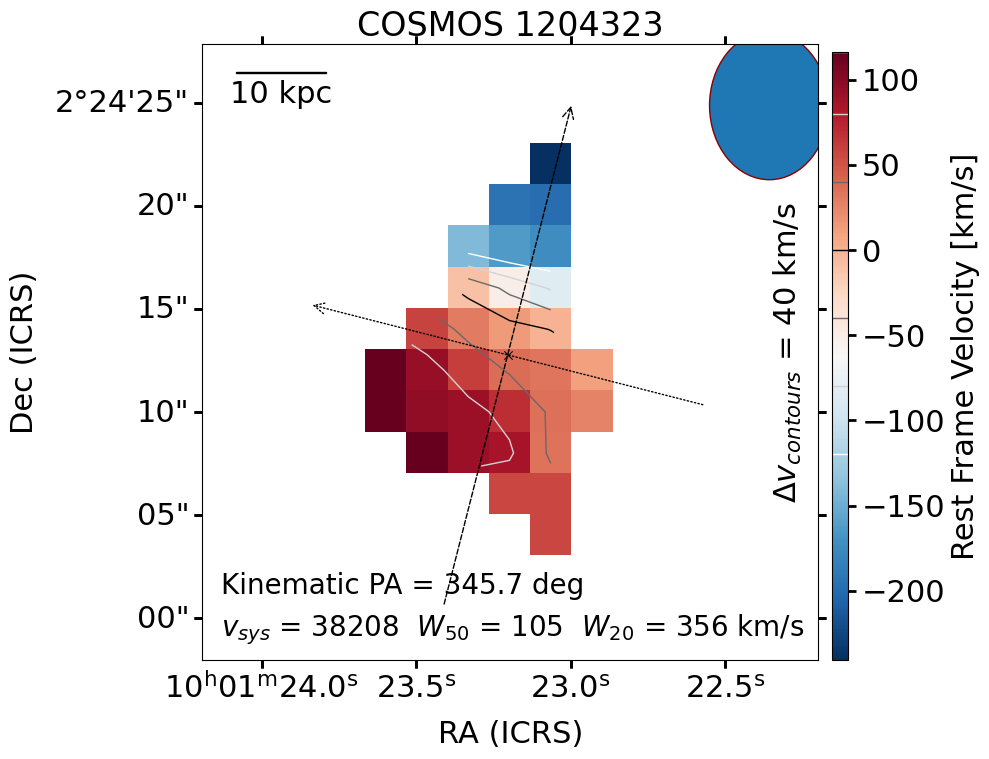}
    \includegraphics[width=0.33\linewidth,trim=26 0 0 0,clip]{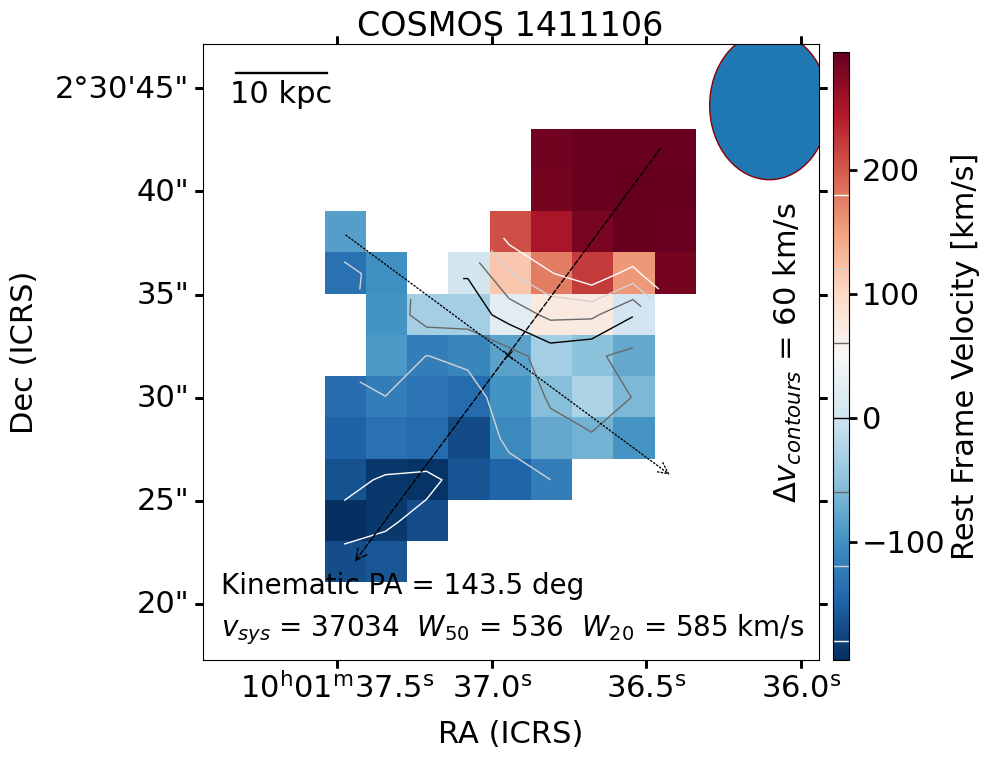} \\
    \includegraphics[width=0.33\linewidth,trim=0 0 29 0,clip]{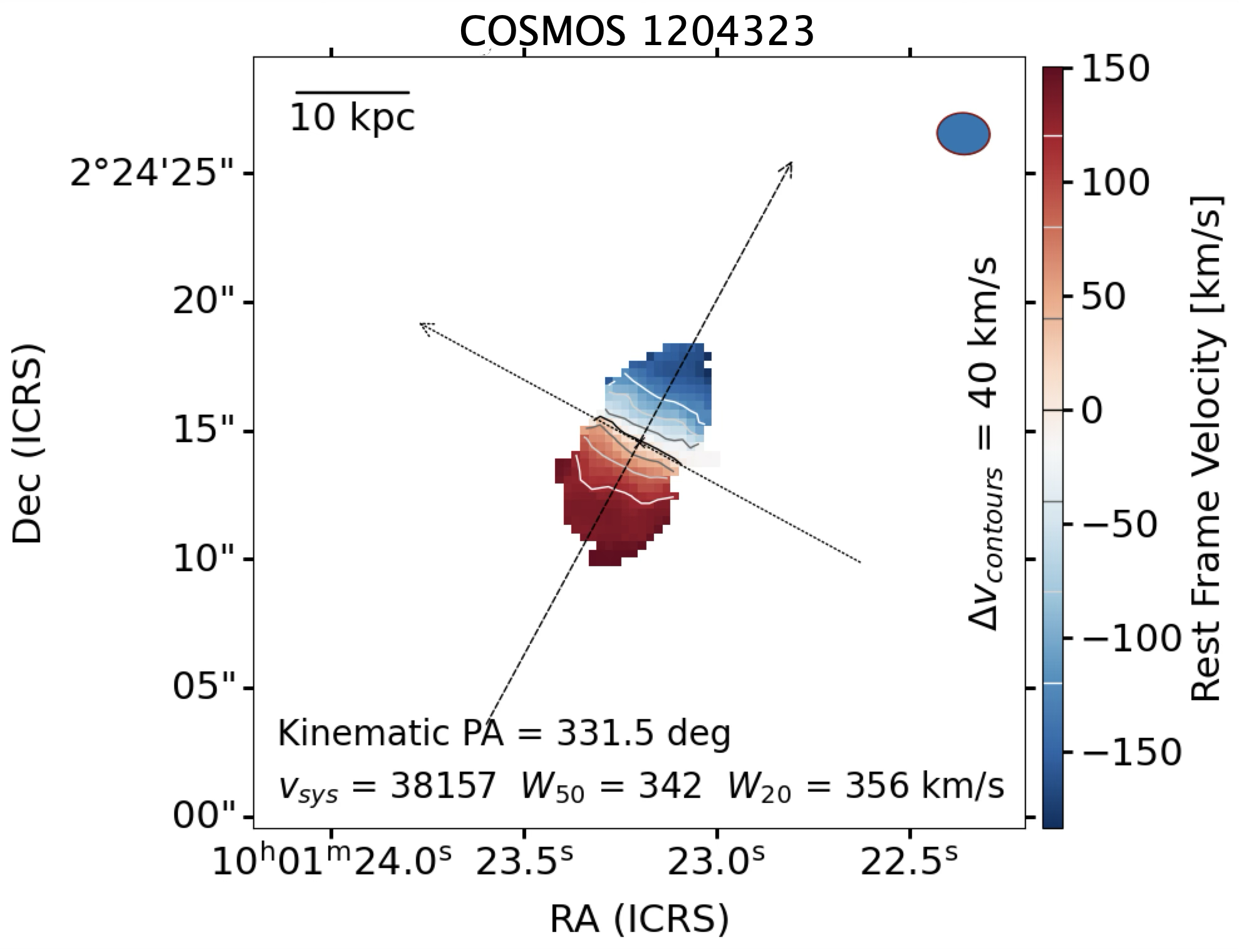}
    \includegraphics[width=0.33\linewidth,trim=26 0 0 0,clip]{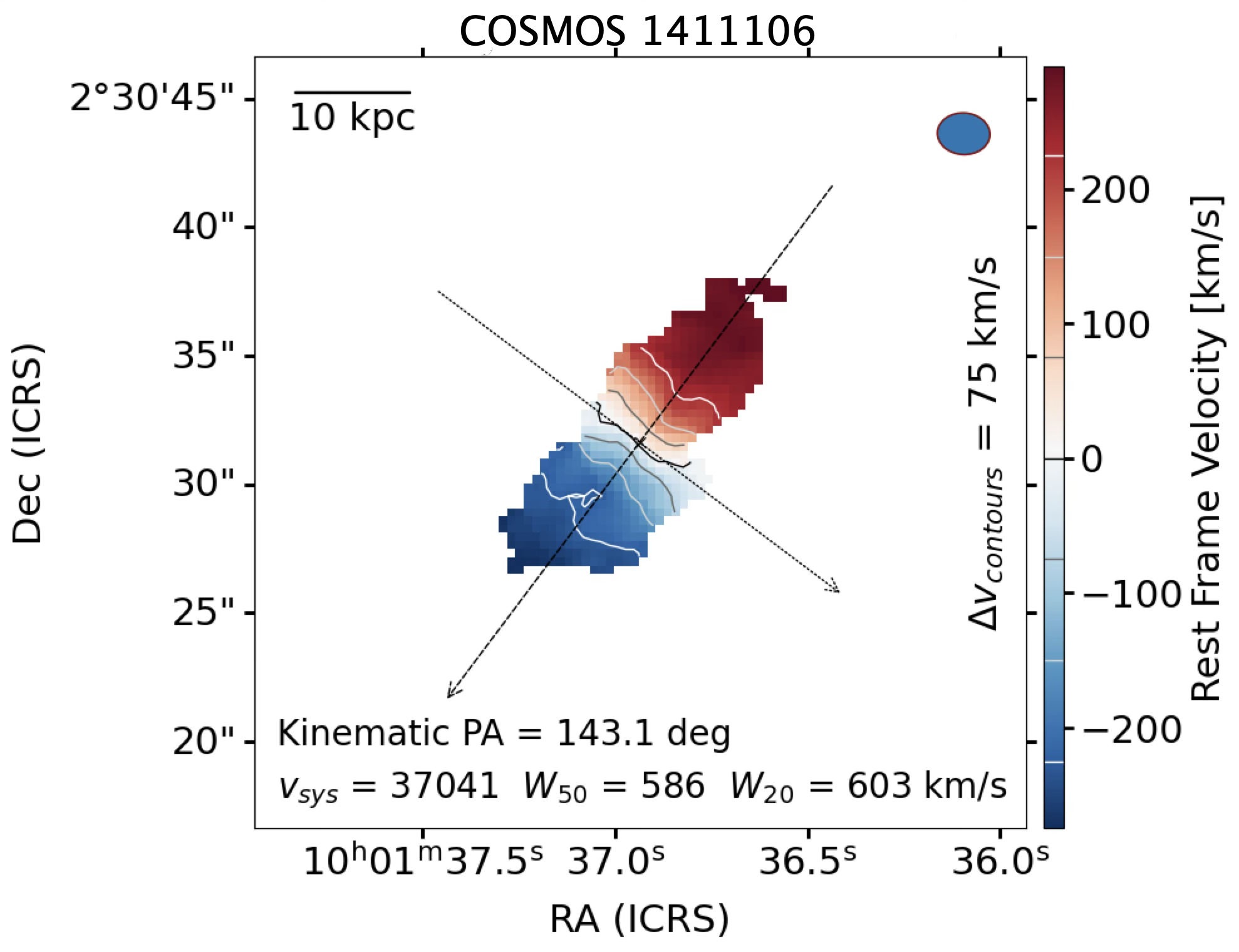}
\caption{\hi\ (top) and CO (bottom) intensity weighted velocity (moment 1) maps for the five CO detections. The kinematic major axis is indicated by the dashed line.  The position angle, systemic velocity, profile width at 20\% and 50\% of the peak ($W_{20}$ and $W_{50}$), as calculated by SoFiA-2, are written inset at the bottom of the figures.  The \HI\ and CO for each galaxy are plotted on the same scale.}
    \label{fig:mom1}
\end{figure*}

Considering the CO morphology, we find that the molecular gas emission in four of the five galaxies is centrally peaked.  In one galaxy the CO peak appears to be offset from the center of the galaxy (COSMOS 1411106; last row of Figure \ref{fig:hst_gas}).  Interestingly, this is the only galaxy of the five CO detections to have been identified as belonging to a COSMOS-identified galaxy group, and it belongs to the most massive group in the CHILES volume \citep{Knobel12,Hess19}.  However, three other CO-detected galaxies (COSMOS-0969208, COSMOS-1197519, and COSMOS-1204323 of Figure \ref{fig:hst_gas}) were noted as candidate interacting pairs \citep{Hess19}.  The central and left panels of Fig.~\ref{fig:hst_gas} show that the CO contours are also coincident with the regions of highest star formation as traced by the 1.4 GHz emission and regions of highest stellar mass surface density as traced in the infrared.  This is discussed in more detail in the following section.

Figure \ref{fig:mom1} shows the \hi\ and CO intensity-weighted velocity maps.  In all cases, the CO shows the signature of a rotating disk.
In four of the five systems, the CO disk is aligned with the atomic gas disk within a few degrees.  The exception is COSMOS 1197519, in which the \hi\ and CO kinematic position angle differ by 39 degrees.  This galaxy may be undergoing an interaction with COSMOS 969208 to the south \citep{Hess19}, which may explain its disturbed \hi\ morphology relative to the more tightly bound CO.  With the exception of COSMOS~1189669 (second column), the CO line-widths are always greater than the HI line-width.  However, the differences between the HI and CO line-widths generally amount to less than one \HI\ channel, suggesting that the difference may be due to resolution and sensitivity rather than a difference in the maximum rotational velocity of the different gas phases.

\subsection{CO traces star formation in dusty regions of galaxies}

A comparison between Figures \ref{fig:decals} and \ref{fig:hst_gas} shows that the CO emission is coincident with the red central regions of the most massive galaxies.  Naively this suggests it is either coincident with an old stellar population, or with dusty regions within the galaxies.  In order to test this, we compare the CO emission with the 1.4 GHz radio continuum images from CHILES Con Pol.  Figure \ref{fig:hst_gas} (center) shows that the CO emission is tightly correlated with the radio continuum.  As stated above, based on the CCP images and available multi-wavelength COSMOS data, we find no evidence for AGN activity in the CO-detected galaxies, and so we assume that all the 1.4 GHz continuum emission in these galaxies is due to star formation.  

The CO-detected galaxies also host the highest total star formation rates in the \hi-detected sample (Table \ref{tab:gal_properties}).  Taken together with the color, it appears that the brightest molecular gas emission is tracing regions of dusty, obscured star formation.  This is supported by the fact that four out of the five CO-detected galaxies have significantly lower UV+IR SFRs as compared to the SFR inferred from the 1.4 GHz emission which should be unimpacted by dust. The two remaining CO-detected galaxies have similar SFR values within their respective uncertainties.  Meanwhile, the CO (1-0) is systematically undetected in blue regions of the galaxies, which one would typically associate with young stars.

We also note that the CO emission is confined to the central regions with the highest stellar mass surface density.  To validate this assessment, we plot CO contours on binned maps of stellar mass surface density derived from Spitzer 3.6 $\mu$m images, and find that most of the CO detected gas is within regions greater than approximately $\sim$$1.25\times10^8$~\msun\ kpc$^{-2}$ (Figure \ref{fig:hst_gas}, right).  The CO extent declines where the gradient in stellar mass surface density is rapidly changing, confirming this observation.  A similar correlation with stellar mass surface density has been noted at various resolutions in EDGE-CALIFA, PHANGS and VERTICO galaxies \citep{Bolatto17,Pessa21,Villanueva21,Villanueva22}.

We attribute the observed trends to the presence of dust as a prerequisite for forming molecular gas, or for forming it more efficiently.  In hindsight this is perhaps unsurprising: \citet{Whitaker17}, find that the fraction of obscured star formation (defined as $f_{obsc} = SFR_{IR}/SFR_{UV+IR}$) is strongly dependent on stellar mass, with >80\% of star formation being obscured for galaxies with $\log(M{_*/M_{\odot}}) > 10.0$.  The difference we observe between $SFR_{1.4GHz}$ and $SFR_{UV+IR}$, suggests they may be under counting both the total and obscured SFR. 

\begin{figure*}
    \centering
    \includegraphics[width=0.92\textwidth]{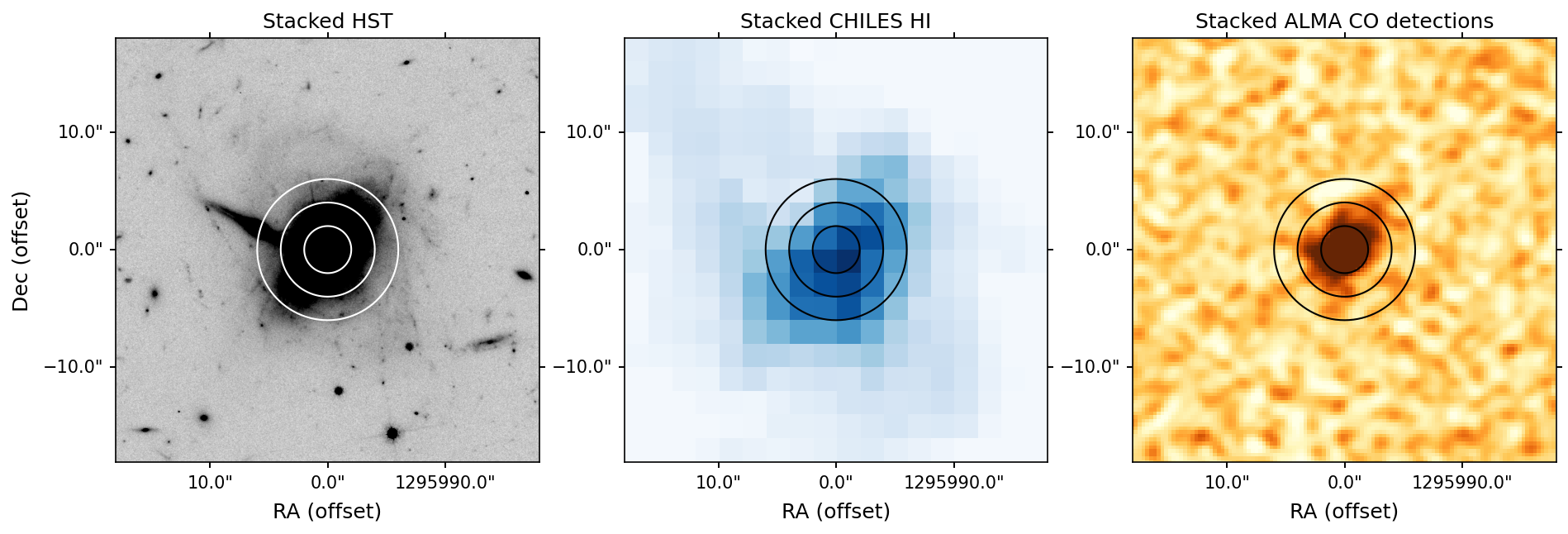}
    \includegraphics[width=0.96\textwidth]{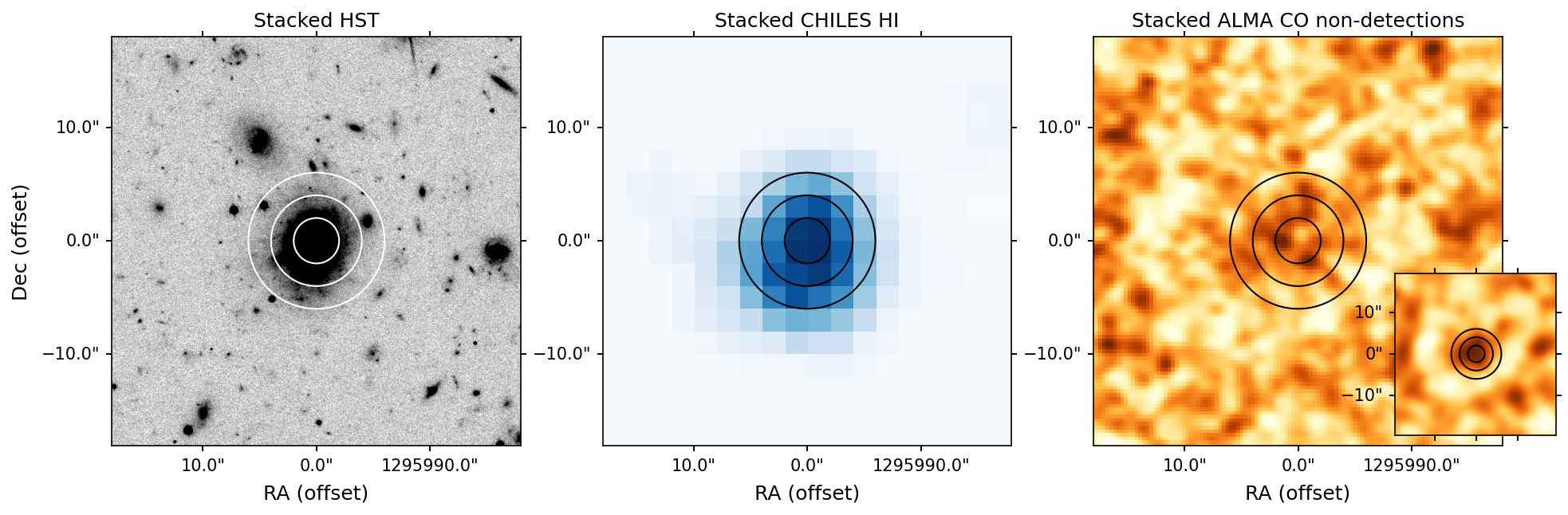}
    \caption{Stacked images for the CO detections (top row) and CO non-detections (bottom row).  From left to right, we stacked HST ACS mosaic images in gray scale, HI total intensity (moment 0) maps in blue, and CO maps collapsed over the channels indicated in Figure \ref{fig:stack_spectra} in orange. In each image the concentric circles represent [2, 4, 6] arcsec radius apertures over which we extracted the CO spectra.  Inset in the lower right corresponds to the stacked non-detections, smoothed to 4 arcsec resolution (see text).  Consistently across all images, the CO-detected galaxies in the top row appear spatially larger on the sky.}
    \label{fig:stack_images}
\end{figure*}

\begin{figure}
    \centering
    \includegraphics[width=0.48\textwidth]{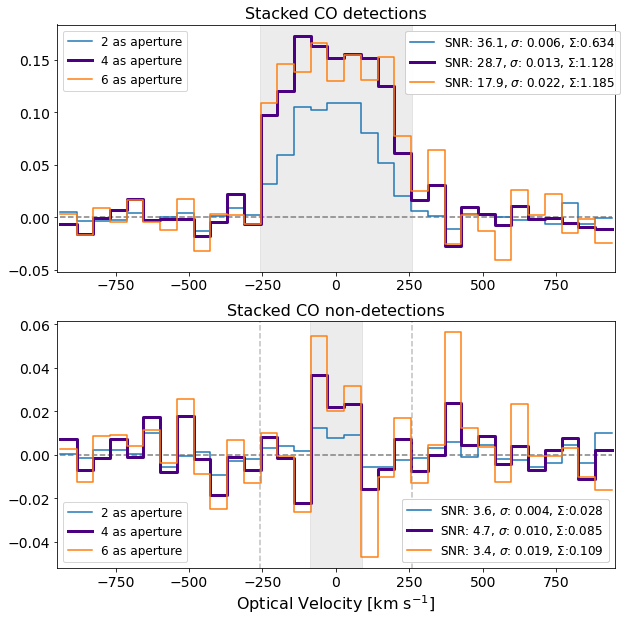}
    \caption{Spectra extracted from the [2, 4, 6] arcsec apertures for the CO detections (top) and over the 4 arcsec aperture for the CO non-detections (bottom).  Light gray regions indicate channels over which emission is detected and the cubes are collapsed make the CO maps in Figure \ref{fig:stack_images}.  See text for further details.}
    \label{fig:stack_spectra}
\end{figure}

\subsection{Image and CO spectral line stacking}
\label{sect:stacking}

Figure \ref{fig:stack_images} shows the results of image and spectral line stacking for the five CO-detected galaxies (top row) and for the nine CO non-detected galaxies (bottom row; Section \ref{sect:data_stack}).  In each panel, we over-plot the three different apertures from which the CO data were extracted at [2, 4, 6] arcsec radii.  The rightmost panel shows the CO moment map integrated over the best channel range estimated from the spectra in Figure \ref{fig:stack_spectra}. Based on the individual direct detections, we expect the CO (1-0) emission to be confined to the stellar disk, and the \HI\ to be more extended, placing upper limits on the aperture.  The stacks of the direct detections show this to hold true even as the detections are averaged.
While the stack of the CO non-detections (bottom right, Figure \ref{fig:stack_images}), shows clumpy emission at the center, when we smooth this image to 4 arcsecond resolution (inset), we see a clear peak at the center of the image.  

Figure \ref{fig:stack_spectra} shows the extracted CO (1-0) spectra for each aperture for CO-detections (top) and CO-non-detections (bottom).  The channel ranges over which the ALMA CO (1-0) moment maps are made for Figure \ref{fig:stack_images} are indicated in gray.   The stacking shows that, in addition to the CO-detected galaxies being more massive, redder, and on average more extended in both the optical and \hi; they also have a broader CO line width, corresponding to 509 \kms\ (9 channels) versus 170 \kms\ (3 channels).  The widths of the emission profiles are consistent regardless of what aperture we extract over.  However, we find an optimal circular aperture of 4 arcsecs for both the CO detections and non-detections.  This is evaluated based on which profile captures the greatest flux over the line width ($\Sigma$), without adding significant noise ($\sigma$).  In the case of the CO detections, the 4 arcsec aperture captures essentially all the flux (missing at most 5\%), while the 6 arcsec aperture adds marginally more flux but is significantly more noisy resulting in a lower signal-to-noise ratio (SNR) detection.  The pattern is similar for the CO non-detections: the 4 arcsec aperture has the highest SNR (4.7$\sigma$), while the 6 arcsec aperture has significantly higher noise and emission is only detected at the 3.4$\sigma$ level.

Ultimately, we measure a 4.7$\sigma$ signal after co-adding the signal from the 9 CO non-detections corresponding to an average CO luminosity of 97.2 Jy\kms\ and an \Ht\ mass of $\log(M_{H_2}/M_{\odot})=8.46$. For later discussion, the nine CO non-detections have an average stellar mass of $\log(M_*/M_{\odot})=9.35$ and average \HI\ mass of $\log(M_{HI}/M_{\odot})=9.52$.  These average values put the stacked galaxies precisely on the molecular gas-stellar mass and \hi-stellar mass relations of $z=0$ low mass galaxies of the ALLSMOG survey \citep{Hagedorn24}.

\section{Discussion}
\label{sect:discussion}

Until recently, estimates of the \hi\ content of the Universe beyond $z=0.1$ have been based on indirect or average global measurements, and suggested that the cosmic volume density of \hi\ varies little as a function of redshift \citep[e.g.][and sources therein]{Rhee18}.  The contributing data include absorption line measurements \citep[e.g.][]{Rao06,Rao17,Bird17,Parks18} and \hi\ stacking experiments \citep[e.g.][]{Rhee18,Bera22,Chowdhury22a,Chowdhury22c,Bianchetti25,Luber25b,Luber25c}.  
Perhaps the most complete analytic prescription for the evolution of baryons in galaxies to date has been summarized by \citet{Walter20} who fit functional forms to the observationally derived gas and stellar mass densities as a function of redshift.  \citet{Peroux20} also provide an overview of the observed variation in gas content with redshift albeit with different parameterization of the same \hi\ data points.  Based on these collected works, it has broadly been supposed that the decline in the star formation rate density within galaxies since a redshift of $z=1-2$ is most directly related to changes in the molecular gas reservoir.  However, the underlying data upon which both of these analyses were conducted provide an incomplete picture: either because they lack atomic and molecular gas measurements in the same galaxies, or because the gas measurements are derived from only the most massive galaxies.  

The redshift of this CHILES \hi\ flux-limited sample of galaxies at $z\sim0.12$, combined with the fact that the galaxies come from an untargeted survey, make them interesting for comparative studies of gas evolution as a function of redshift to begin testing the current paradigm.  Further, this CHILES sample spans a range of galaxy properties including two orders of magnitude in stellar mass and star formation rate, permitting a multi-dimensional view of variations across gas and stellar properties.  In Section \ref{sect:redshift} we compare our CHILES measurements of the atomic and molecular gas to other galaxy samples from the literature which contain both measurements; to analytic fits of the cosmic density of atomic and molecular gas as a function of redshift \citep{Walter20,Peroux20}; and to empirically motivated models for the gas content of a typical galaxy on and above the main sequence.

To close, in Section \ref{sect:epoch1} we discuss how results derived from Epoch 1 of CHILES compare with the present data where we have co-added 856 hours.  We provide a qualitative assessment of the reliability and reproducibility of the \HI\ mass measurements of sources at the detection limit, which may be a useful comparison for other ongoing deep spectral line surveys.  

\begin{figure*}
    \centering
    \includegraphics[width=\textwidth]{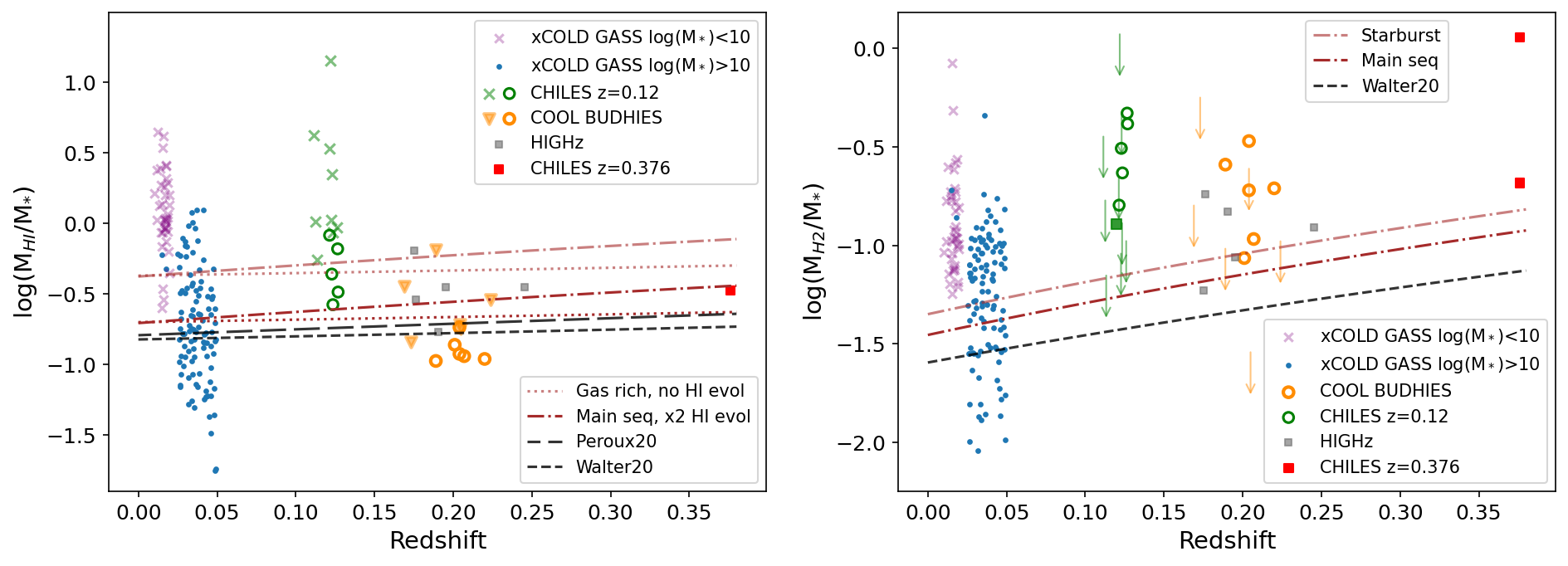}
    \caption{Left: \HI\ mass divided by stellar mass as a function of redshift for CHILES, xCOLD GASS, COOL BUDHIES, and HIGHz.  The dark brown lines are the typical gas ratio for a gas-rich galaxy ($\log(M_*/M_{\odot})=10.3$ from a stellar mass-selected sample of \citet{Catinella18}.  The light brown lines are the gas ratio for a gas-rich galaxy of the same mass based on \citet{Maddox15}.  Right: \Ht\ mass divided by stellar mass for the same samples.  Here the dark brown line is the molecular gas-to-stellar ratio for a $\log(M_*/M_{\odot})=10.3$ main sequence galaxy, and the light brown line is for a starburst galaxy, both based on the empirical evolutionary models of \citet{Scoville23}.  The dotted brown lines are for no \hi\ evolution; the dot-dashed brown lines are for a ``fast'' factor of two evolution between $z=0$ to $z=1$.  See the text for details.  The black short-dashed lines indicate the functional fits to the evolution of the cosmic density from \citet{Walter20}. In the left plot, the black long-dashed lines are the fit to the same \hi\ data by \citet{Peroux20}.  The rest of the symbols are the same for both plots: purple `x's and blue dots are detections in the xCOLD GASS sample below and above $\log(M_*/M_{\odot})=10$, respectively.  Green symbols are CHILES $z=0.12$ galaxies (this work); orange symbols are COOL BUDHIES galaxies.  In both CHILES and COOL BUDHIES the open circles are galaxies with both CO and \HI\ detections, the downward arrows and `x'/downward triangles are galaxies with CO upper limits. In CHILES, the green `x's/downward arrows have the same mass range as the xCOLD GASS `x's.  The green square is the stacked CO non-detections.  Gray squares are HIGHz galaxies. Red symbols are the CHILES $z=0.376$ detection corresponding to conversion factors for interacting galaxies ($\alpha_{CO}=0.8$, bottom) and the Galactic value ($\alpha_{CO}=4.3$, top) \citep{Fernandez16}. The uncertainties for the CHILES galaxies in the left panel are $\pm0.15-0.25$ dex, and $\pm0.1-0.15$ dex in the left panel. See text for further details.}
    \label{fig:gas}
\end{figure*}

\begin{figure*}
    \centering
    \includegraphics[width=\textwidth]{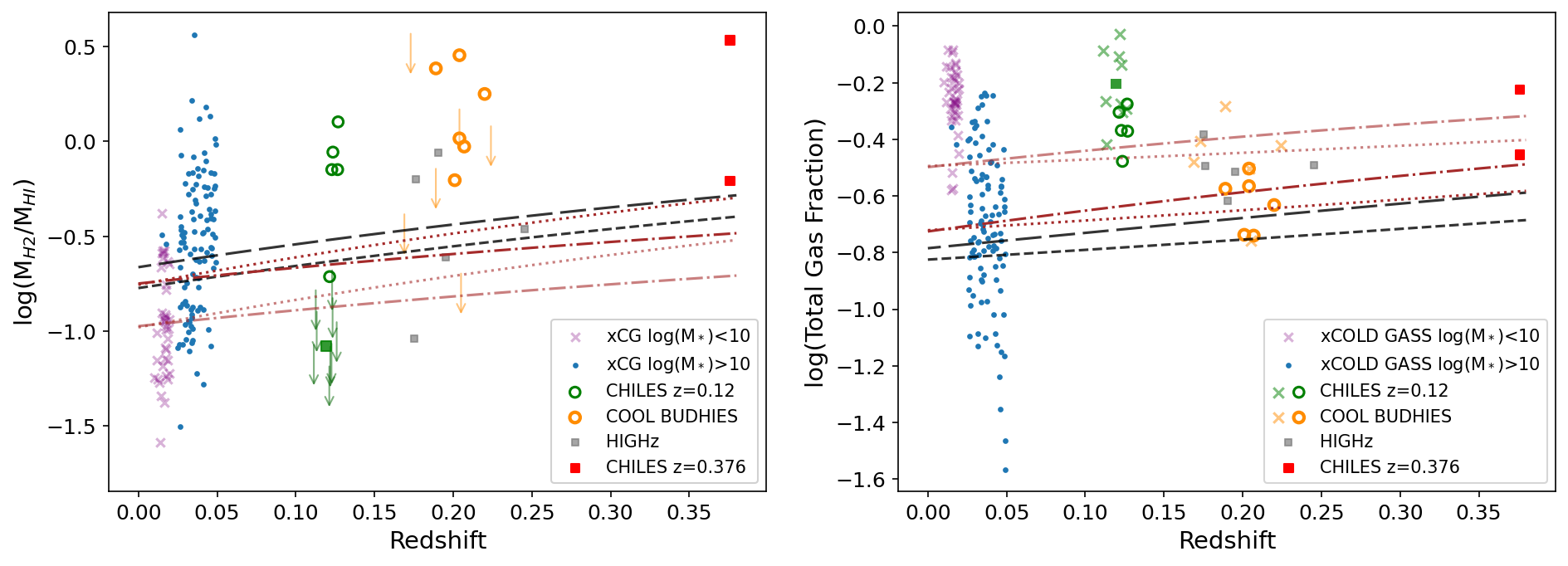}
    \caption{Comparison of molecular and atomic gas samples. Left: the ratio of molecular versus atomic gas as a function of redshift. Right: the total gas fraction ($(M_{HI}+M_{H_2})/(M_*+M_{HI}+M_{H_2})$) as a function of redshift. The symbols are the same as Figure \ref{fig:gas}.  In both plots, the lower red symbol corresponds to the CHILES $z=0.376$ detection assuming $\alpha_{CO}=0.8$ and the higher red symbol corresponds to the same galaxy assuming $\alpha_{CO}=4.3$. We have omitted the second legend for clarity.  To plot the dark brown lines, we combine the ``typical'' \HI\ model from 6a with the ``main sequence'' model for molecular gas from 6b.  To plot the light brown lines, we combine the \HI\ ``gas-rich'' model from Fig 6a with the molecular gas ``starburst'' model from 6b.  See the text for more details.}
    \label{fig:atom_mol_surveys}
\end{figure*}

\subsection{Evolution of the molecular and atomic gas content of galaxies with redshift}
\label{sect:redshift}

Figures \ref{fig:gas} and \ref{fig:atom_mol_surveys} show the gas properties of CHILES galaxies compared to samples in the literature, as a function of redshift. We remind the reader that, as described in Section \ref{sect:samples}, we have ensured to the best of our ability that the samples have the same \HI\ mass, stellar mass, and color selection.  

At low redshift, we plot the xCOLD GASS sample which is separated into high (blue) and low (purple) stellar mass objects.  As mentioned previously, the stellar mass separation in xCOLD GASS coincidentally corresponds to the same mass above and below which CO is detected (green open circles) or is not detected (green `x's or arrows) in the CHILES galaxies.  The CHILES CO-detected galaxies all have stellar masses in excess of $\log(M_*/M_{\odot})=10$, making it convenient throughout these plots to compare CHILES green open circles directly to xCOLD GASS blue dots, and CHILES green `x's or arrows to xCOLD GASS purple `x's.  The green `x's in the left plots correspond to green CO upper limits in the right plots.  The stacked molecular gas measurement is represented by the green squares.  In fact, the stacked measurement and the ``upper limits'' may be considered conservative values because we have used a fixed CO-to-\Ht\ conversion factor.  In reality, $\alpha_{CO}$ may be higher due to the lower metallicity in the low mass galaxies, resulting in a higher \Ht\ mass \citep{Sandstrom13,Accurso17,Bolatto13}.

At $z\sim0.2$, we plot the COOL BUDHIES galaxies in orange.  The open circles correspond to galaxies detected in both \hi\ and CO, while the orange triangles in the left plot correspond to CO upper limits in the right.  All COOL BUDHIES galaxies have stellar masses greater than $\log(M_*/M_{\odot})=10$, and so may be compared with the open green circles and blue points at low redshift.  However, the COOL BUDHIES live in a relatively high density cluster environment \citep{Jaffe13,Gogate20}. This likely accounts for why they have a relatively lower distribution of \hi\ masses than the CHILES galaxies, but comparable \Ht\ masses: the cluster environment impacts the atomic gas more readily than the molecular gas component \citep{Cortese21,Villanueva22,Zabel22}.  
At a similar redshift range, the gray squares correspond to the subset of HIGHz galaxies.  These all have stellar masses greater than $\log(M_*/M_{\odot})=10.8$, and \hi\ masses greater than $\log(M_{HI}/M_{\odot})=10.35$, but they live in relatively isolated environments.  Finally, the highest redshift direct \hi\ detection is plotted as a red square. It has a stellar mass of $\log(M_*/M_{\odot})=10.9$, and \hi\ mass of $\log(M_{HI}/M_{\odot})=10.46$, making it similar to the HIGHz galaxies.  The upper and lower red square correspond to \Ht\ mass estimates assuming an $\alpha_{CO} = 4.35$ or $\alpha_{CO}=0.8$ \msun\ (K \kms pc$^{-2}$)$^{-1}$ for a star forming galaxy, respectively \citep{Fernandez16}.

The uncertainties for the mass ratios of the CHILES galaxies in the left panel of Figure \ref{fig:gas} and both panels of Figure \ref{fig:atom_mol_surveys} are dominated by the \HI\ uncertainties and span $\sim$0.3-0.5 dex. In the right panel of Figure \ref{fig:gas} the uncertainties for the CHILES galaxies are dominated by the stellar mass uncertainties and span $\sim$0.2-0.3 dex. However, these estimates only account for the uncertainties on the integrated flux measurements and the stellar masses.  They do not take into account the uncertainty in the CO-to-\Ht\ conversion factor which may be an even larger contributor to the uncertainty, as evidence by the range in the two values for the CHILES $z=0.376$ galaxy.  In preparation for the discussion to come, we note that the uncertainties in the mass are assumed to be symmetrically distributed, which cannot account for systematic offsets between a given dataset and models or other datasets.

To provide additional context, we compare the data points with the evolutionary models for the stellar, atomic, and molecular gas cosmic densities from \citet[W20]{Walter20}, as well as the atomic gas cosmic density from \citet[P20]{Peroux20}.  These are plotted as a black short-dashed line in Figures~\ref{fig:gas} and \ref{fig:atom_mol_surveys} to indicate the evolutionary trends from W20 (see their Eqns.~1-2 and Table 1), and as a black long-dashed line for P20.  In particular, the W20 parameterizations are derived by fitting functional forms to the results from volumetric surveys covering the redshift range from 0 -- 4, and arguably offer the most complete understanding of how the stellar and gas mass density evolve with redshift.  Interestingly, W20 fit stellar and molecular gas evolution with power law functions, but use a tanh function for the \hi\ cosmic density.  In contrast P20 fit the same \HI\ data, but using a power law parameterization which results in noticeable differences between the \hi\ curves over the range $z=0$ to $z=2$ (e.g.,~Fig 5 of \citealt{Oyarzun25}): where the W20 curve rises slowly at low redshifts, and the P20 curve rises more rapidly.  The limitations of these analytic models will be discussed in Section \ref{sect:summary}.

Finally, we also plot empirically motivated models for how two $\log(M_*/M_{\odot})=10.3$ galaxies would evolve over our redshift range: (1) a galaxy that may be considered to have an \hi-to-stellar mass ratio at $z=0$ of a ``typical'' galaxy, as estimated from the stellar mass-selected GASS sample (\citealt{Catinella18}; -0.7, dark brown lines); and (2) one that may be considered to be ``HI-rich'' having an \hi\ gas fraction at $z=0$ similar to the ALFALFA HI-selected sample from \citet{Maddox15} (-0.37, light brown lines).  The canonical $\log(M_*/M_{\odot})=10.3$ is chosen because it corresponds to the average stellar mass of our CHILES CO-detected galaxies, as well as the typical stellar mass of high redshift molecular gas studies. Given the apparent flexibility in fitting existing \hi\ data as a function of redshift that is discussed above, we consider two extreme evolutionary scenarios for each of these empirical models: (1) one in which there is no \hi\ evolution (brown dotted lines), and (2) one in which the evolution is described by a power law that increases by a factor of two between $z=0$ and $z=1$ (brown dot-dashed lines), which is faster than either the W20 or P20 parameterizations.

For the molecular mass of the ``typical'' galaxy, we assume that it would sit on the galaxy main sequence at z=0.12 and use the methods from \citet{Scoville23} to predict the molecular gas evolution based on the main sequence star formation rate as a function of stellar mass \citep{Lee15}, and redshift \citep{Speagle14}.  For the ``gas-rich'' galaxy, we use the same formulae to predict the molecular gas, but assume it is a ``starbursting'' galaxy with an SFR comparable to the average SFR derived from the 1.4 GHz measurements of our CO-detected CHILES galaxies reported in Table \ref{tab:gal_properties}: roughly a factor of 2 higher than the main sequence.

\subsubsection{\hi\ gas fraction}
Figure \ref{fig:gas} (left) shows that the low stellar mass objects of both CHILES and xCOLD GASS have higher \HI\ gas fractions ($M_{HI}/M_*$) than their high stellar mass counterparts, and are consistent between the two samples.  This is not surprising based on \HI\ flux-limited studies in the local Universe \citep{Maddox15}.  The high stellar mass objects in CHILES share the same range of \hi\ gas fraction with the highest \hi\ gas fraction of the companion xCOLD GASS subset.  This may indicate that although we have attempted to make an equal comparison between xCOLD GASS and CHILES, xCOLD GASS is still fundamentally a stellar mass-selected sample.  By comparison, COOL BUDHIES has a lower \hi\ gas fraction on average, which we attribute to \hi\ deficiency due to gas loss in the cluster environment (e.g.,~\citealt{Giovanelli85,BravoAlfaro00,Chung09,Hess15,Jaffe16,Yoon17}).  Overall there is no obvious evolution in the global \hi\ content of individual galaxies between $z<0.05$ and $z=0.12$ (or $z=0.376$).  

However, we note that the functional fits from W20 and P20 only agree with the gas fractions of the highest stellar mass galaxies.  If galaxies at $z=0.376$ follow the same \hi\ trends as at lower stellar mass, then the vast majority of galaxies at the same redshift would have higher \hi\ gas fractions and sit above the single red point in the left hand plot of Figure \ref{fig:gas}, suggesting that the functional fits to the cosmic baryon densities are systematically under-representing or missing the \hi\ population below $\log(M_*/M_{\odot})=10$, at all redshifts from $z=0$ to $z=0.376$.  On the other hand, the cosmic baryon densities are global measurements, and therefore also include \hi\ non-detections.  Thus the possibility is open for greater evolution in the \hi\ content of galaxies than previously recognized--an idea consistent with recent results from \hi\ stacking experiments with the GMRT \citep{Chowdhury22a,Chowdhury22b,Chowdhury22c}, although any conclusions require greater statistics and the interpretation requires additional nuance (e.g.,~\citealt{Bera19,Bera22}).

Of the two empirical models, we find that the CHILES CO-detected galaxies at $z=0.12$ are most consistent with the ``gas-rich'' galaxy model (light brown lines).  However, perhaps what is the most striking take-away is that the data points from each of the literature samples exhibit significantly greater internal scatter than the differences produced by any assumed redshift evolution over the redshift range between $z=0$ and $z<0.38$ (e.g.,~the differences between the dotted and dot-dashed lines for a given model galaxy).  We suggest that, even with large statistical samples, it would be nearly impossible to detect evolution in the \hi\ content of galaxies below at least $z=0.2$, where the difference between ``no evolution'' (dotted lines) vs ''fast evolution'' (dot-dashed lines), is of order of the difference between W20 and P20 fitting the same \hi\ data with different functional forms.  To be quantitative, the scatter in \hi-to-stellar mass fraction in the massive CHILES galaxies alone is $\sim$7.5 times larger than the difference between the two evolutionary models at a redshift of $z=0.12$.  The scatter in HIGHz and COOL BUDHIES galaxies is $\sim$7.3 times the difference between the two evolutionary models at a redshift of $z=0.2$.  At $z=0.4$, the same scatters are reduced to $\sim$1.6-2 times the difference in the two evolutionary models.  Nonetheless, it may still be challenging to claim evolution over this range, as our models represent two extreme cases and therefore a best-case scenario.  Large statistical samples which allow us to examine evolution for galaxies of fixed stellar mass, or fixed position relative to the star forming main sequence may be key to teasing out trends with redshift.

\subsubsection{\Ht\ gas fraction}
In Figure \ref{fig:gas} (right) we plot the \Ht\ gas fractions ($M_{H_2}/M_*$) as a function of redshift.  The upper limits on the non-detections, plotted at 4.5$\sigma$, are not particularly stringent as they depend on the choice of CO line width and assumed $\alpha_{CO}$. We have chosen a conservative value for $\alpha_{CO}$ of 4.35 that is independent of metallicity, and of 300\kms\ for the line width.  By comparison, xCOLD GASS uses a metallicity dependent $\alpha_{CO}$ based on the calibration from \citet{Accurso17}.  For low mass galaxies, the metallicity dependent $\alpha_{CO}$ may be as high as $\sim$10-20, while their line widths may be a factor of two or three times narrower, suggesting that using a different combination of values will not change the upper limits significantly in the plot.
Since the stacked \Ht\ mass of the CHILES non-detections is at the relatively high end of values observed in low mass galaxies by xCOLD GASS in the local universe, our conservative $\alpha_{CO}$ value is either comparable, or under estimating the amount of molecular gas in our low stellar mass detections.  

Similarly, or to an even greater degree to what was found for the \hi\ gas fraction, the CHILES CO detections have systematically higher \Ht\ gas fractions than galaxies in the local Universe of the same mass.  The COOL BUDHIES CO detections have lower \Ht\ gas fractions than CHILES on average, but are still higher than the xCOLD GASS high mass galaxies.  The global molecular gas content may be less impacted by the cluster environment as it is held more tightly in the stellar disk and so is harder to strip (e.g.,~\citealt{Cortese21,Villanueva22,Zabel22}).  Taken together, this suggests that the molecular gas content is trending upwards with redshift, for galaxies of all stellar masses, even over this relatively short redshift range. The molecular gas trend continues for the single CHILES detection at $z=0.376$, even if the conversion factor is assumed to be low ($\alpha_{CO}=0.8$ for starburst galaxies).  Further, we note that the W20 functional fits also systematically under-predict the amount of molecular gas in an \hi-selected sample for galaxies of all stellar masses.

Interestingly, both of the empirical models, based on \citet{Scoville23} for a main sequence galaxy (dark brown, dot-dashed line) or a starburst galaxy (light brown, dot-dashed line) also under-predict the amount of \Ht\ in these \hi\ selected galaxies.  In fact, the \citet{Scoville23} parameterization does even worse for low mass galaxies.  While a full discussion of the parameter space is outside the scope of this paper, we will note that the same lines move upward and parallel to the existing lines if either the stellar mass or star formation rate of the galaxy goes up, but lines go down (in a parallel fashion) when the stellar mass or star formation rate decreases.  As has been stated by others (e.g. \citealt{Saintonge22}), it is clear that some of the greatest strides in our understanding of the gas evolution of galaxies can come from pushing our observations to lower mass galaxies at higher redshift.

\subsubsection{\Ht\ to \hi\ gas ratio}
In Figure \ref{fig:atom_mol_surveys} (left), we compare the ratio of \Ht\ to \HI\ in galaxies as a function of redshift.  Massive galaxies in CHILES have a higher \Ht/\hi\ ratio than the low stellar mass CHILES galaxies, as found in studies of nearby galaxies \citep{Boselli14}.  We cautiously find that the ratio of molecular-to-atomic gas is increasing with redshift for the most massive galaxies, although note that the \Ht/\hi\ ratio for COOL BUDHIES is likely enhanced in the cluster environment (e.g.,~\citealt{Cortese21,Loni21}).  If the \hi\ mass of the CHILES galaxy at $z=0.376$ is taken as an upper limit rather than a detection, its \Ht/\hi\ ratio is even higher than shown here.  On the other hand, there is no apparent evolution for galaxies below $\log(M_*/M_{\odot})=10$ over the narrow range $z=0$ to $z=0.12$, although the individual CHILES \Ht\ masses are lower limits.

Several simulations have suggested that a decline in the molecular gas mass relative to atomic gas mass may be responsible for the cosmic decline in star formation rate since $z=1$ \citep{Obreschkow09}, but the data presented here is the first direct measurement to test the evolution of \Ht/\HI\ with redshift. Previous studies have used indirect methods to infer the molecular gas content, for example based on star formation rate \citep{Chowdhury22a} or dust \citep{Scoville17}.  The fact that the molecular versus atomic gas ratio increases in the cluster environment is consistent with what has been found in Virgo \citep{Cortese21,Villanueva22}.  However, for non-cluster galaxies, atomic gas dominates over molecular gas--a trend that is observed to $z=1.0-1.3$ \citep{Cortese17,Chowdhury22a,Chowdhury22b,Chowdhury22c}.  

\subsubsection{Total gas fraction}
Finally, in Figure \ref{fig:atom_mol_surveys} (right) we plot the total gas fraction, defined as $(M_{HI} + M_{H_2})/(M_* + M_{HI} + M_{H_2})$, as a function of redshift.  As suggested by the previous figures, the total gas fraction measured in the \hi-selected samples is higher than expected from the functional fits of W20 and P20.  The empirical model galaxies span the range of the data, primarily due to the \hi\ contribution to their estimated gas content.  With the exception of the COOL BUDHIES sample, the trend in total gas fraction is primarily driven by the \hi\ content rather than the \Ht\ content.  In the cluster environment, the total gas fraction is dominated by molecular gas.  Meanwhile, between $0<z<0.12$, low stellar mass galaxies are even more gas-rich than their high stellar mass counterparts at the same redshift.  Nonetheless, the functional fits to the cosmic baryon densities systematically underestimate the gas content for all \HI-selected galaxy populations shown.

\subsubsection{Summary of gas fractions with redshift}
\label{sect:summary}
In summary, we find evidence that the \Ht\ content of galaxies, in an \hi-selected sample, may be evolving and increasing with redshift, relative to the \hi\ and stellar components (Fig \ref{fig:gas} right, Fig \ref{fig:atom_mol_surveys} left).  Galaxies at all stellar masses seem to be more gas rich in terms of their total gas reservoir per unit stellar mass than that predicted by their estimated cosmic volume density (Fig \ref{fig:atom_mol_surveys} right).  For high stellar mass galaxies, this is driven by their relatively high \Ht\ content (Fig \ref{fig:gas} right, \ref{fig:atom_mol_surveys} left), and to a lesser degree by their \hi\ content (Fig \ref{fig:atom_mol_surveys} left).   For low stellar mass galaxies, this is driven by their high \hi\ content (Fig \ref{fig:gas} left).  

For the COOL BUDHIES sample, the cluster environment can remove the less bound atomic gas from galaxies, lowering the total gas content (Fig \ref{fig:atom_mol_surveys} right), and increasing the \Ht\ to \hi\ ratio (Fig. \ref{fig:atom_mol_surveys} left).  By comparison, the HIGHz galaxies reside in more isolated environments and have higher total gas content, driven by the \hi\ (Figs. \ref{fig:gas} left, \ref{fig:atom_mol_surveys} right).  Interestingly, in COOL BUDHIES, which are known to be Butcher-Oemler clusters \citep{Butcher84}, the \hi\ reservoir at $z \sim 0.2$ appears to be on par with the average gas reservoir for galaxies in the local Universe not selected to be in a particular environment (and therefore most likely dominated by non-cluster galaxies), even though they have lower \hi\ than the CHILES detections of similar stellar mass.

We have investigated the mass ratios of various galaxy components rather than their absolute values.  In this way, we probe the relative scaling relations of gas to stars.  However, the underlying analytic prescriptions as a function of redshift from W20 and P20 are derived from heterogeneous samples which may have little to no overlap, and each of which suffer from their own selection effects.  For example, most samples which go into the prescriptions are limited to stellar masses above $\log(M_*/M_{\odot})=10$.  We suppose that, as a comparison of samples constructed to be \HI\ flux-limited, and the fact that \hi\ is the dominant gas component of galaxies at all redshifts, three of the four plots which contain \hi\ on the y-axis in Fig \ref{fig:gas} and \ref{fig:atom_mol_surveys}, are indeed suggesting that the \hi\ component has been systematically under-counted to date.  When it comes to the molecular vs stellar mass ratio for an \hi-selected sample (Fig \ref{fig:gas}, right), it is possible that this plot is simply showing that the most \Ht-rich galaxies are also \hi-rich.

\subsection{Comparison of Full CHILES with Epoch 1 data}
\label{sect:epoch1}

Epoch 1 of CHILES consists of 178 hours of observations, making up 20\% of the combined 856 hours from the CHILES survey presented here.  In theory, if the noise is Gaussian and the RFI does not get worse with later CHILES epochs, the noise should have improved by a factor of $\sim$2.2, and the signal-to-noise of our detections improved by a similar value.  In practice, after correcting for different clean beams and channel widths, we find that the noise in the 856 hour cubes at this frequency range only improves by a factor of $\sim$1.4.  As discussed in \citet{Hess19}, the frequency range in which these galaxies are found is one of those hardest hit by RFI.  Unfortunately, we have only seen it get worse with time, potentially explaining the less than expected noise improvement.

Nonetheless, morphologically, we find that the \HI\ detections in the ``full CHILES'' maps are more well localized around the optical galaxies than they were in the Epoch 1 data.  One notable exception is COSMOS 1197519 (third row of Figure \ref{fig:hst_gas}), whose \hi\ morphology has changed, but whose \hi\ mass is consistent with the previously measured value.  The marginal detections from \citet{Hess19} are also not only confirmed, but seem to be well detected. 

\begin{figure}
    \centering
    \includegraphics[width=0.48\textwidth]{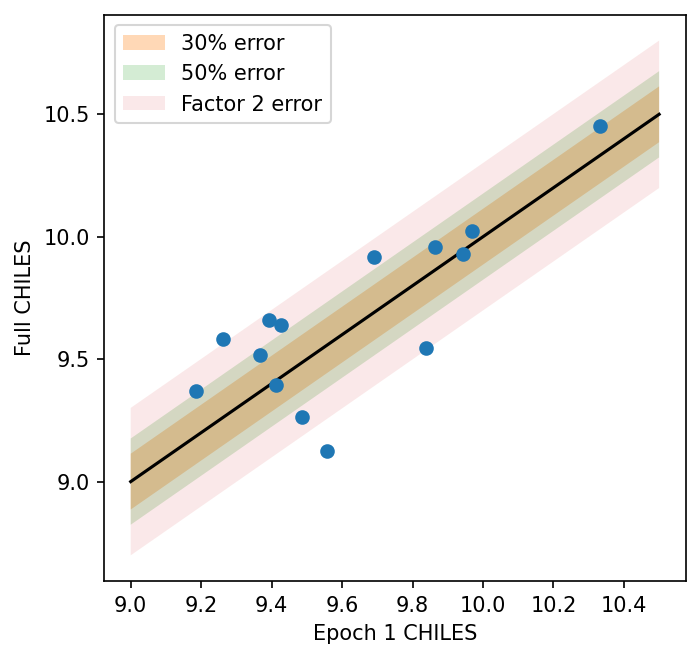}
    \caption{Comparison between Epoch 1 HI masses and HI masses from the "full survey."  The black line is the one-to-one line.  In Epoch 1 data, \citep{Hess19} predicted uncertainties up to 50\%.  With one exception, the HI masses are all within a factor of 2. See Section \ref{sect:epoch1} for further discussion.}
    \label{fig:epoch1}
\end{figure}

In Figure \ref{fig:epoch1}, we show the comparison between Epoch 1 \HI\ masses and \HI\ masses from the full survey.  The black line is the one-to-one line.  We note that more than half of the detections sit above the line suggesting that on average we may be detecting slightly more \hi\ in the deeper data.  The scatter about the line is larger at lower \hi\ masses where one detects objects at lower signal-to-noise, and the scatter is generally smaller at high \hi\ masses, as would be expected.

A more direct comparison between Epoch 1 and Full CHILES is challenging: the data were processed with different pipelines (Pisano et al in prep); but additionally, the RFI environment got worse in this frequency range over the course of the survey.  In Epoch 1 data, \citet{Hess19} predicted uncertainties up to 50\%, which in hindsight appears to be an underestimate.  Instead, about half the galaxies are within 50\% uncertainty, and all of them--with the exception of COSMOS 1440745--are within a factor of two.  For COSMOS 1440745 the spectrum in Figure \ref{fig:hi_atlas3} suggests that the galaxy is sitting in a local minimum which, if corrected for, may raise the \HI\ mass by ~30\% putting it within a factor of two error. 

\section{Conclusions}

We present the first study of resolved molecular and atomic gas in galaxies beyond the local Universe, at $z=0.12$.  The resolved \HI\ emission maps show that at column densities of $1-2\times10^{20}$ cm$^{-2}$ the atomic gas extends modestly beyond the stellar disk of the galaxies.  The resolved CO emission maps show that the molecular gas is exclusively detected in the red, high stellar mass surface density regions of galaxies above $\sim$$1.25\times10^8$~\msun\ kpc$^{-2}$, and that the CO kinematics are consistent with disk rotation aligned with the \hi\ kinematics observed at lower spatial and spectral resolution.  The CO emission is coincident and well matched to the morphology of the 1.4 GHz continuum emission, suggesting that the molecular gas is tracing obscured star formation in these massive galaxies.  In addition, we stacked the CO non-detections to get an average molecular gas mass of $\log(M_{H_2}/M_{\odot}) = 8.46$ in galaxies with a mean stellar mass of $\log(M_*/M_{\odot}) = 9.35$.

We compare our atomic and molecular gas and stellar mass measurements with those from the literature, as well as the evolutionary fits from \citet{Walter20}, and an alternative \hi\ fit from \citet{Peroux20}, to show that the gas reservoir in \hi-selected samples is systematically under-predicted by the parameterization of the cosmic baryon density at all redshifts.  This under-prediction is worse for low stellar mass galaxies ($\log(M_*/M_{\odot})<10$) than for high stellar mass galaxies.  In the \hi-selected samples, the \hi\ and \Ht\ can contribute equally to the gas mass budget in massive galaxies, but when galaxies below $\log(M_*/M_{\odot}) = 10.0$ are included, the atomic gas always dominates.

In addition, we examine the behavior of two empirically modeled galaxies with a canonical mass of $\log(M_*/M_{\odot})=10.3$: one assuming no \hi\ evolution, and one assuming moderately faster evolution than currently suggested by the sum of our knowledge in the literature.  We show that the scatter in \hi-to-stellar mass ratio in \hi-detected galaxies is significantly larger than the difference between the two models: by a factor of $\sim$7.5 at a redshift of $z=0.12$, and $\sim$7.3 at a redshift of $z=0.2$.  This suggests it is almost impossible to measure redshift evolution over this range.  It will likely be difficult even out to $z=0.4$, but here the same scatter is reduced to $\sim$1.6-2 times the difference in the two models.

Due to the increasingly severe RFI environment between $\sim$1290-1150 MHz, the CHILES data set is likely to be the only measure of \hi\ at this redshift for the foreseeable future without the aid of advanced RFI mitigation techniques, or without sufficiently deep data to throw away the short baselines.  Nonetheless, the data presented here provide a first glimpse into the future of complementary galaxy evolution studies with the Square Kilometre Array (SKA) and ALMA.  As \hi\ studies start to catch up with existing surveys of molecular gas (e.g.,~PHIBSS and EGNoG) by detecting atomic gas in the most massive galaxies at redshifts out to $z=1$, the challenge will be to push our understanding to lower stellar mass galaxies.  We have shown that stacking of individual low-stellar mass objects that are un-detected by ALMA can provide an average estimate for their molecular gas content, opening the possibility for understanding variations in $\alpha_{CO}$ for low mass galaxies at higher redshift.

\begin{acknowledgements}
KMH acknowledges financial support from the grant CEX2021-001131-S funded by MCIN/AEI/ 10.13039/501100011033 from the coordination of the participation in SKA-SPAIN funded by the Ministry of Science and Innovation (MCIN); from grant PID2021-123930OB-C21 funded by MCIN/AEI/ 10.13039/501100011033 by “ERDF A way of making Europe” and by the "European Union."  JBB was a Jansky Fellow of the National Radio Astronomy Observatory.  CHILES was supported by NSF collaborative grants AST-1413102, AST-1412843, AST-1413099 and AST-1412503. DJP is supported through the South African Research Chairs Initiative of the Department of Science and Technology and National Research Foundation and acknowledges partial support from NSF CAREER grant AST-1149491 and NSF grant AST-1412578.  
The authors acknowledge the computational resources provided by the WVU Research Computing Spruce Knob HPC cluster, which is funded in part by NSF EPS-1003907.

This paper makes use of the following ALMA data: ADS/JAO.ALMA\#2018.1.01852.S and ADS/JAO.ALMA\#2019.1.01615.S. ALMA is a partnership of ESO (representing its member states), NSF (USA) and NINS (Japan), together with NRC (Canada), MOST and ASIAA (Taiwan), and KASI (Republic of Korea), in cooperation with the Republic of Chile. The Joint ALMA Observatory is operated by ESO, AUI/NRAO and NAOJ. The National Radio Astronomy Observatory is a facility of the National Science Foundation operated under cooperative agreement by Associated Universities, Inc.

The Legacy Surveys consist of three individual and complementary projects: the Dark Energy Camera Legacy Survey (DECaLS; Proposal ID \#2014B-0404; PIs: David Schlegel and Arjun Dey), the Beijing-Arizona Sky Survey (BASS; NOAO Prop. ID \#2015A-0801; PIs: Zhou Xu and Xiaohui Fan), and the Mayall z-band Legacy Survey (MzLS; Prop. ID \#2016A-0453; PI: Arjun Dey). DECaLS, BASS and MzLS together include data obtained, respectively, at the Blanco telescope, Cerro Tololo Inter-American Observatory, NSF’s NOIRLab; the Bok telescope, Steward Observatory, University of Arizona; and the Mayall telescope, Kitt Peak National Observatory, NOIRLab. Pipeline processing and analyses of the data were supported by NOIRLab and the Lawrence Berkeley National Laboratory (LBNL). The Legacy Surveys project is honored to be permitted to conduct astronomical research on Iolkam Du’ag (Kitt Peak), a mountain with particular significance to the Tohono O’odham Nation.

NOIRLab is operated by the Association of Universities for Research in Astronomy (AURA) under a cooperative agreement with the National Science Foundation. LBNL is managed by the Regents of the University of California under contract to the U.S. Department of Energy.

This project used data obtained with the Dark Energy Camera (DECam), which was constructed by the Dark Energy Survey (DES) collaboration. Funding for the DES Projects has been provided by the U.S. Department of Energy, the U.S. National Science Foundation, the Ministry of Science and Education of Spain, the Science and Technology Facilities Council of the United Kingdom, the Higher Education Funding Council for England, the National Center for Supercomputing Applications at the University of Illinois at Urbana-Champaign, the Kavli Institute of Cosmological Physics at the University of Chicago, Center for Cosmology and Astro-Particle Physics at the Ohio State University, the Mitchell Institute for Fundamental Physics and Astronomy at Texas A\&M University, Financiadora de Estudos e Projetos, Fundacao Carlos Chagas Filho de Amparo, Financiadora de Estudos e Projetos, Fundacao Carlos Chagas Filho de Amparo a Pesquisa do Estado do Rio de Janeiro, Conselho Nacional de Desenvolvimento Cientifico e Tecnologico and the Ministerio da Ciencia, Tecnologia e Inovacao, the Deutsche Forschungsgemeinschaft and the Collaborating Institutions in the Dark Energy Survey. The Collaborating Institutions are Argonne National Laboratory, the University of California at Santa Cruz, the University of Cambridge, Centro de Investigaciones Energeticas, Medioambientales y Tecnologicas-Madrid, the University of Chicago, University College London, the DES-Brazil Consortium, the University of Edinburgh, the Eidgenossische Technische Hochschule (ETH) Zurich, Fermi National Accelerator Laboratory, the University of Illinois at Urbana-Champaign, the Institut de Ciencies de l'Espai (IEEC/CSIC), the Institut de Fisica d'Altes Energies, Lawrence Berkeley National Laboratory, the Ludwig Maximilians Universitat Munchen and the associated Excellence Cluster Universe, the University of Michigan, NSF's NOIRLab, the University of Nottingham, the Ohio State University, the University of Pennsylvania, the University of Portsmouth, SLAC National Accelerator Laboratory, Stanford University, the University of Sussex, and Texas A\&M University.

BASS is a key project of the Telescope Access Program (TAP), which has been funded by the National Astronomical Observatories of China, the Chinese Academy of Sciences (the Strategic Priority Research Program “The Emergence of Cosmological Structures” Grant \# XDB09000000), and the Special Fund for Astronomy from the Ministry of Finance. The BASS is also supported by the External Cooperation Program of Chinese Academy of Sciences (Grant \# 114A11KYSB20160057), and Chinese National Natural Science Foundation (Grant \# 12120101003, \# 11433005).

The Legacy Survey team makes use of data products from the Near-Earth Object Wide-field Infrared Survey Explorer (NEOWISE), which is a project of the Jet Propulsion Laboratory/California Institute of Technology. NEOWISE is funded by the National Aeronautics and Space Administration.

The Legacy Surveys imaging of the DESI footprint is supported by the Director, Office of Science, Office of High Energy Physics of the U.S. Department of Energy under Contract No. DE-AC02-05CH1123, by the National Energy Research Scientific Computing Center, a DOE Office of Science User Facility under the same contract; and by the U.S. National Science Foundation, Division of Astronomical Sciences under Contract No. AST-0950945 to NOAO.
\end{acknowledgements}

\bibliographystyle{aa}
\bibliography{aa56121-25.bib}

\begin{appendix} 

\section{Serendipitous detection of CO (2-1) line in the $z=1.290$ ULIRG J100111.86+023217.8}
\label{sect:serendip}

\begin{figure}
    \centering
    \includegraphics[width=0.503\linewidth]{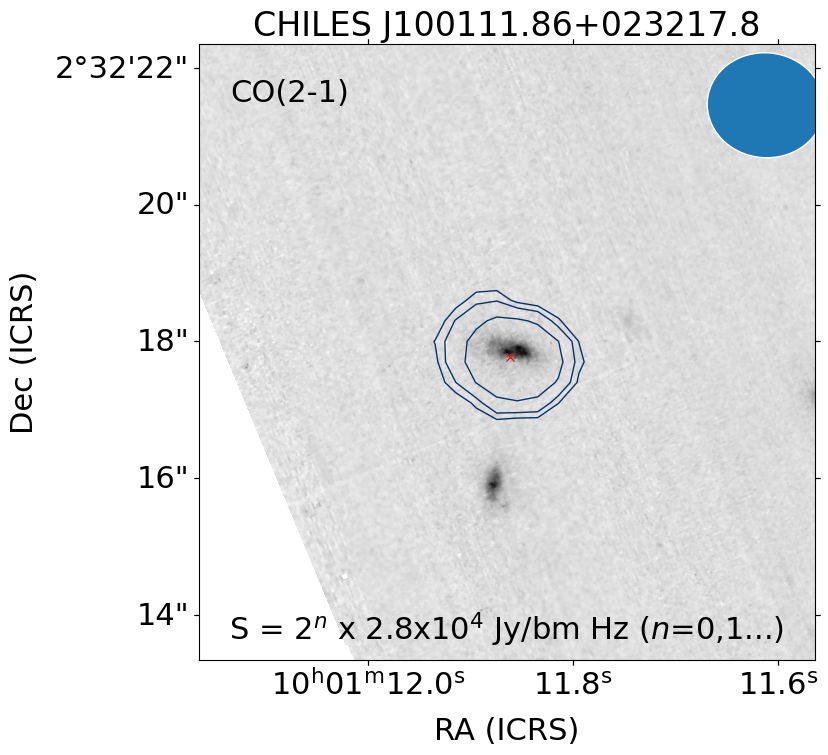}
    \includegraphics[width=0.475\linewidth,trim=141 0 0 0 ,clip]{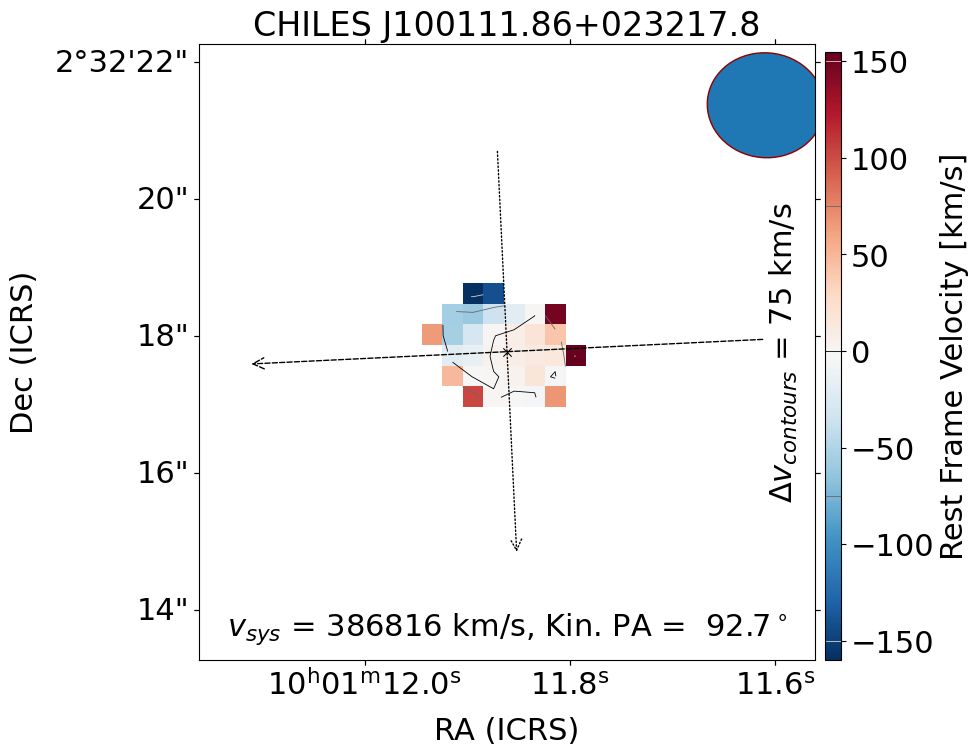}
    \includegraphics[width=1.0\linewidth,trim=0 0 0 30 ,clip]{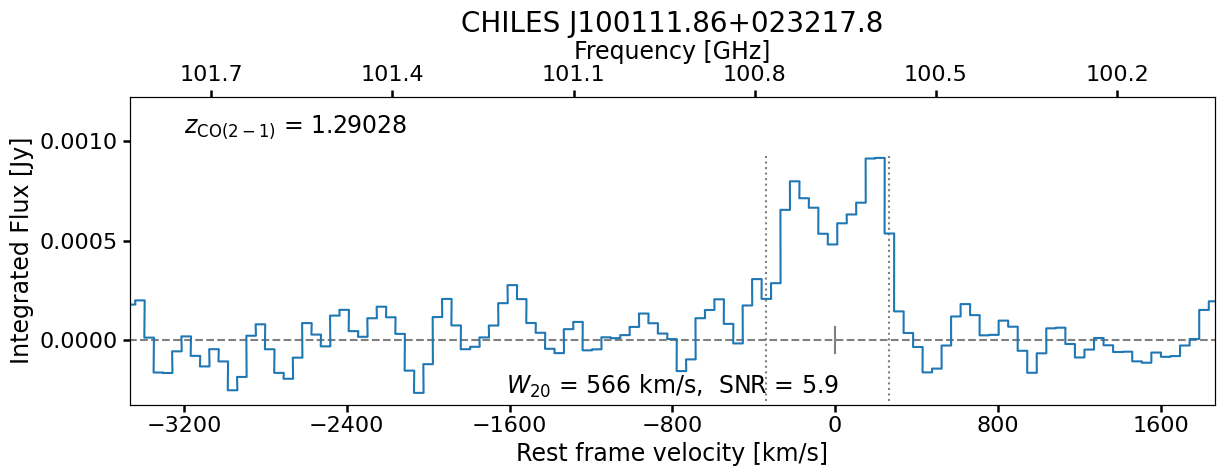}
    \caption{Properties of the serendipitous detection.  Top left: JWST f150w image with CO(2-1) contours. Top right: Intensity weighted velocity map showing the kinematic major and minor axes.  In both cases, the ALMA beam is shown in the top right.  Bottom: Aperture spectrum over the CO detection.  The gray lines denote the spectral extent of the source mask.}
    \label{fig:serendip}
\end{figure}

Figure \ref{fig:serendip} shows the properties of the serendipitous detection found in spectral window (spw) 19 of the first COSMOS 1403950 observation.  The object appears in the COSMOS2020 catalog with ID 1178832, and in COSMOS2015 with ID 815955 \citep{Weaver22}. In the COSMOS2020 catalog, the source is assigned a photometric redshift of 1.2867 \citep{Weaver22}.  However, as part of a redshift survey of \textit{Herschel}-selected far IR starbursting galaxies, the source was found to have an optical spectroscopic redshift of $z=1.284$ from Keck  \citep{Casey12}.  In addition, the source has been studied in detail by \citet{Ling22} who find a photo-z from COSMOS2010 of 1.47 with $\log(L_{IR}/L_{\odot}) = 12.34$ and a stellar mass of $\log(M_*/M_{\odot})=10.64$, which make it an ultra luminous infrared galaxy (ULIRG).  

In the CHILES Con Pol images, we find an unresolved point source at the exact location of the CO detection, with a peak 1.4 GHz flux density of $58.7\pm4.4$ $\mu$Jy.  If the radio emission is due to star formation, this corresponds to a star formation rate of 2.39~\msun\ yr$^{-1}$ at our assumed redshift.

If we assume the line is CO(2-1), then we derive a submm spectroscopic redshift of 1.290, consistent with the COSMOS2020 photometric value, and the Keck optical spectroscopic redshift.  To calculate the molecular mass for higher order transitions, we can generalize Equation \ref{eqn:mh2vel} to the following, analogous to Equation \ref{eqn:mh2}:
\begin{equation}
    M_{H_2} = \alpha_{CO} \times 9.743\times10^{3}\, S_{CO}\, \nu_{rest}^{-3}\, D_L^2 \, r_{xy}^{-1}.
\label{eqn:mh2b}
\end{equation}
where $r_{xy}$ is the line ratio for the appropriate CO transition to CO(1-0).  In this case, if we assume common values for ULIRGs, $\alpha_{CO}^{ULIRG} = 0.8$ \citep{Downs98}, and $r_{21}=0.91$ \citep{Papadopoulos12}, we estimate a molecular gas mass of $\log(M_{H_2}/M_{\odot}) = 9.81$.  The $W_{50}$ ($W_{20}$) line width measured by SoFiA-2 is 511 (566)~\kms. 

The Bayesian method of SED fitting with CIGALE \citep{Boquien19,Yang22} for the combined photometry data from the COSMOS2020 catalog and new JWST photometry yields a stellar mass of ($2.96\pm0.17)\times10^{10}$~\msun\ and SFR of $154\pm8$~\msun\ yr$^{-1}$ for a fixed redshift of $z_{CO}=1.290$.
Both of these quantities are in excellent agreement with the earlier estimate by \citet{Ling22} after correcing for the redshift.  A more detailed discussion of the physical properties of this galaxy will be presented in a separate paper.

\section{CO(1-0) and \HI\ atlases for ALMA CO detections}
\label{sect:sip_co}

\begin{figure*}
    \centering
    \includegraphics[angle=90,width=0.45\linewidth]{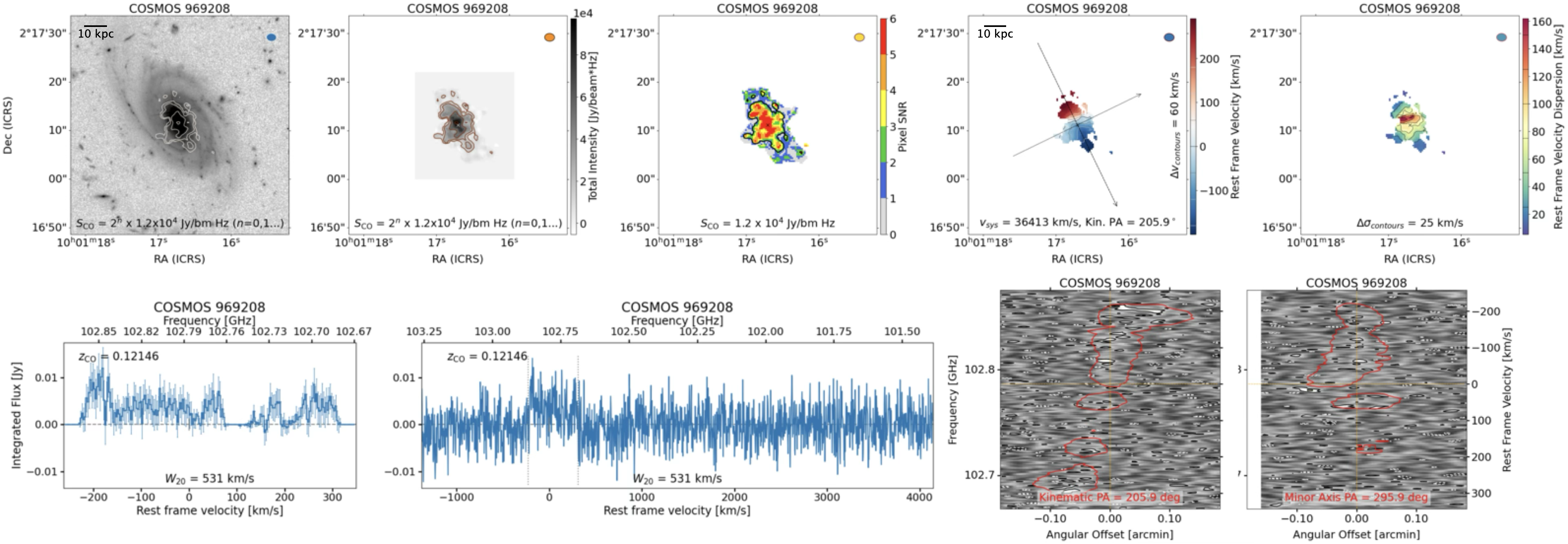}
    \hfill
    \includegraphics[angle=90,width=0.45\textwidth]{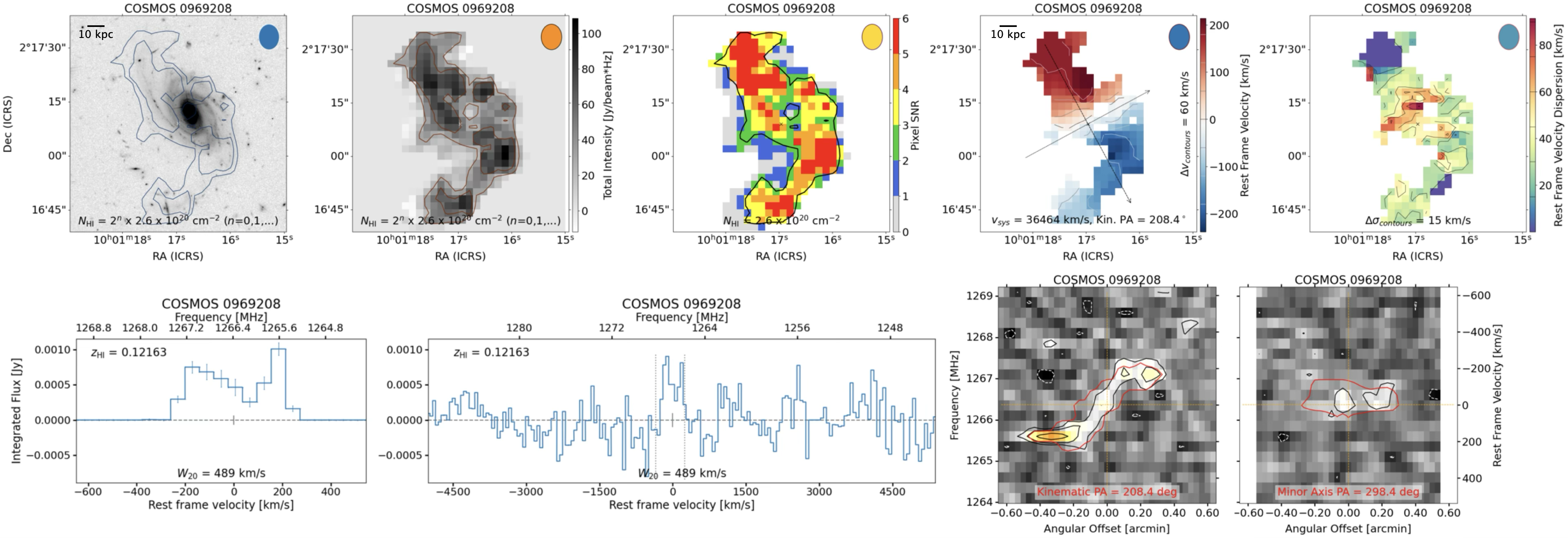}
    \caption{Atlas page of CO(1-0) (left) and \hi\ (right) for COSMOS 0969208.  See Appendix text for description of panels.}
    \label{fig:co_atlas0}
\end{figure*}

\begin{figure*}
    \centering
    \includegraphics[angle=90,width=0.45\linewidth]{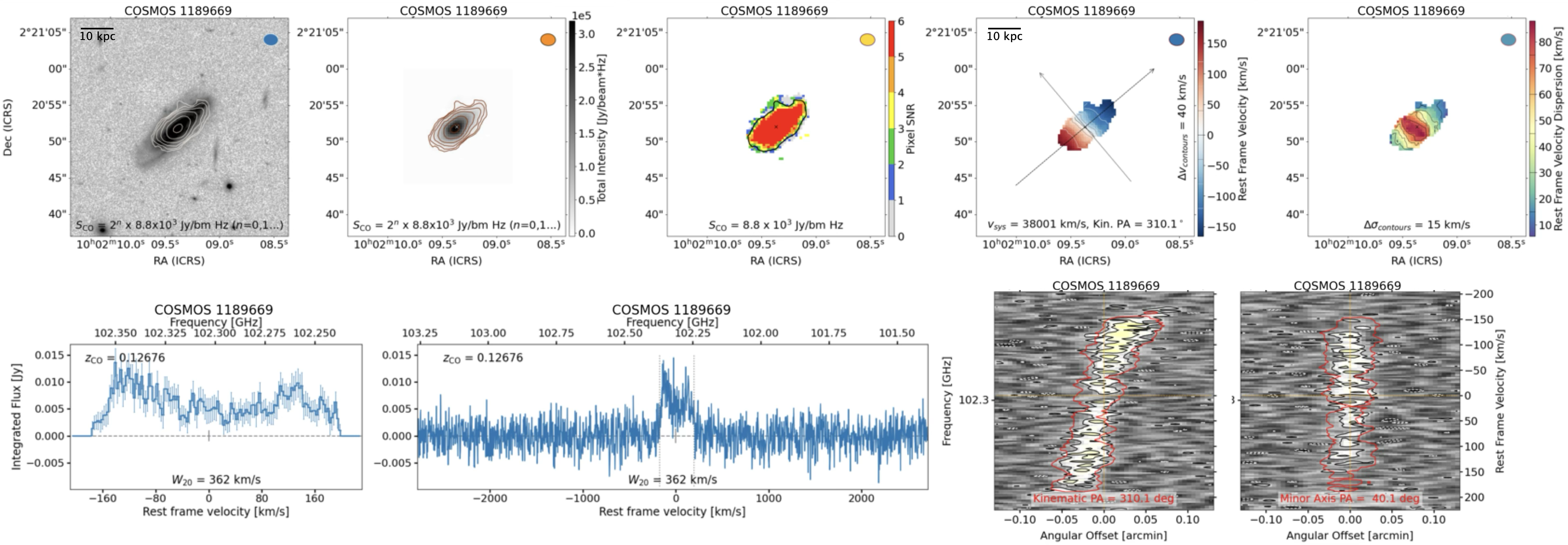}
    \hfill
    \includegraphics[angle=90,width=0.45\textwidth]{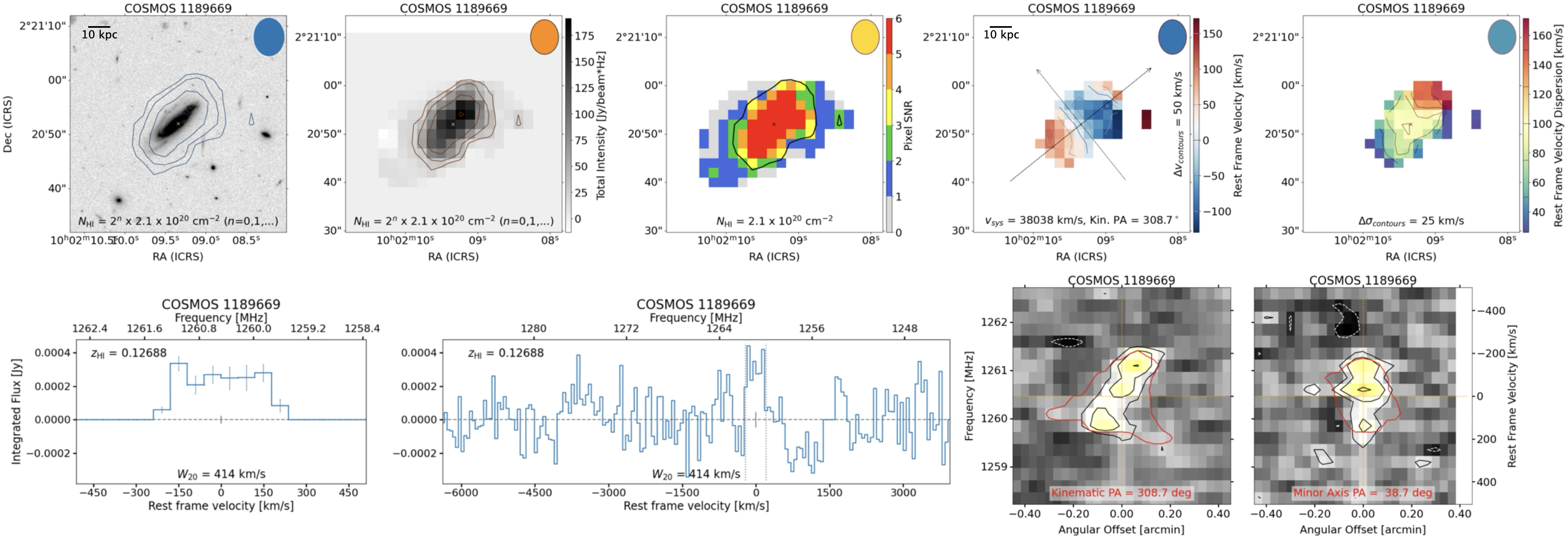}
    \caption{Atlas page of CO(1-0) (left) and \hi\ (right) for COSMOS 1189669.  See Appendix text for description of panels.}
    \label{fig:co_atlas1}
\end{figure*}

\begin{figure*}
    \centering    
    \hfill
    \includegraphics[angle=90,width=0.45\textwidth]{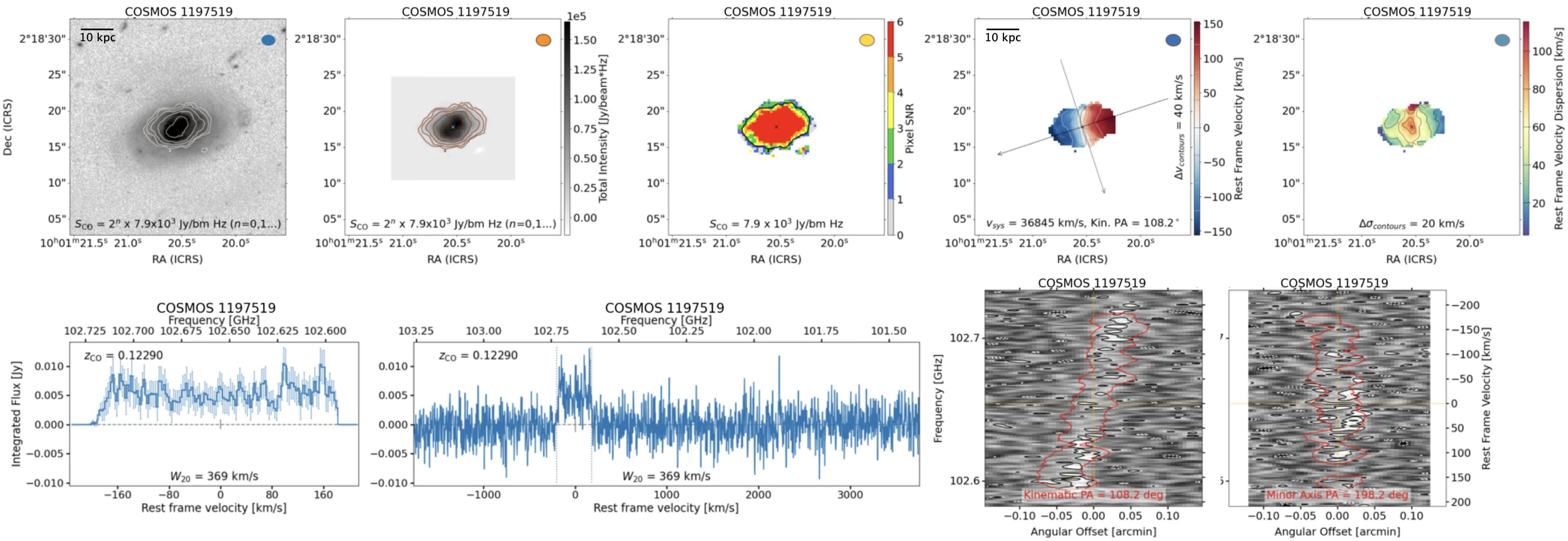} \hfill
    \includegraphics[angle=90,width=0.45\textwidth]{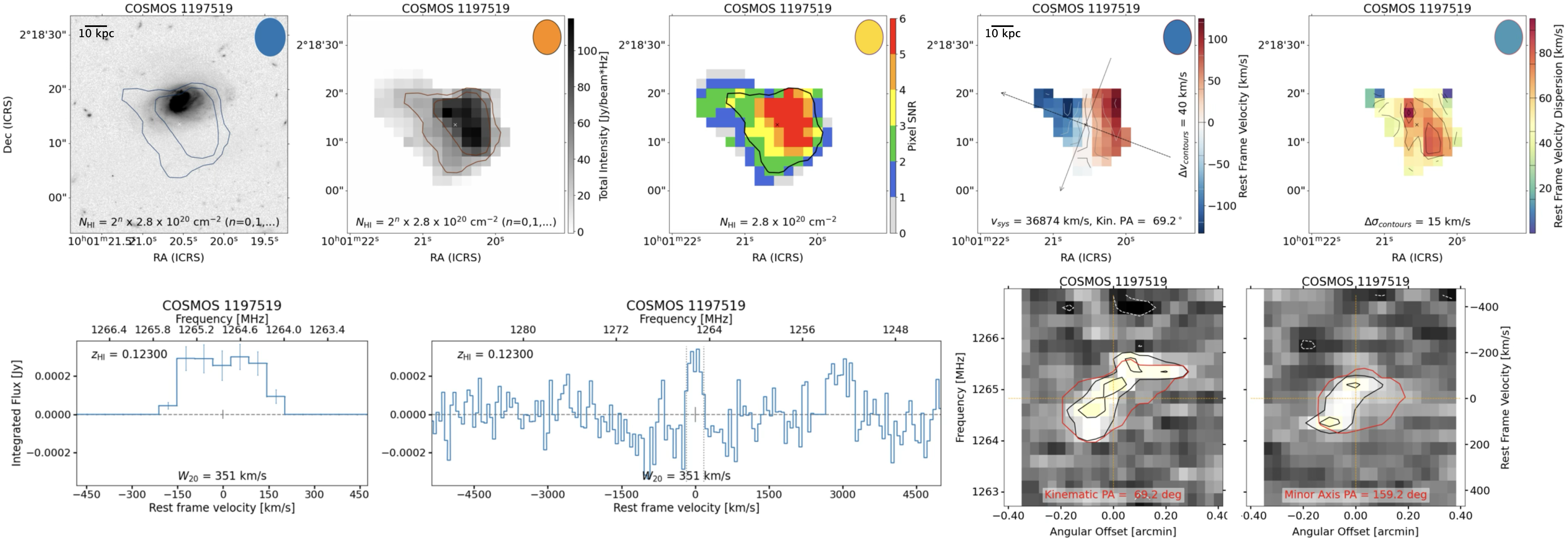}
    \caption{Atlas page of CO(1-0) (left) and \hi\ (right) for COSMOS 1197519. See Appendix text for description.
    }
    \label{fig:co_atlas2}
\end{figure*}

\begin{figure*}
    \centering    
    \hfill
    \includegraphics[angle=90,width=0.45\textwidth]{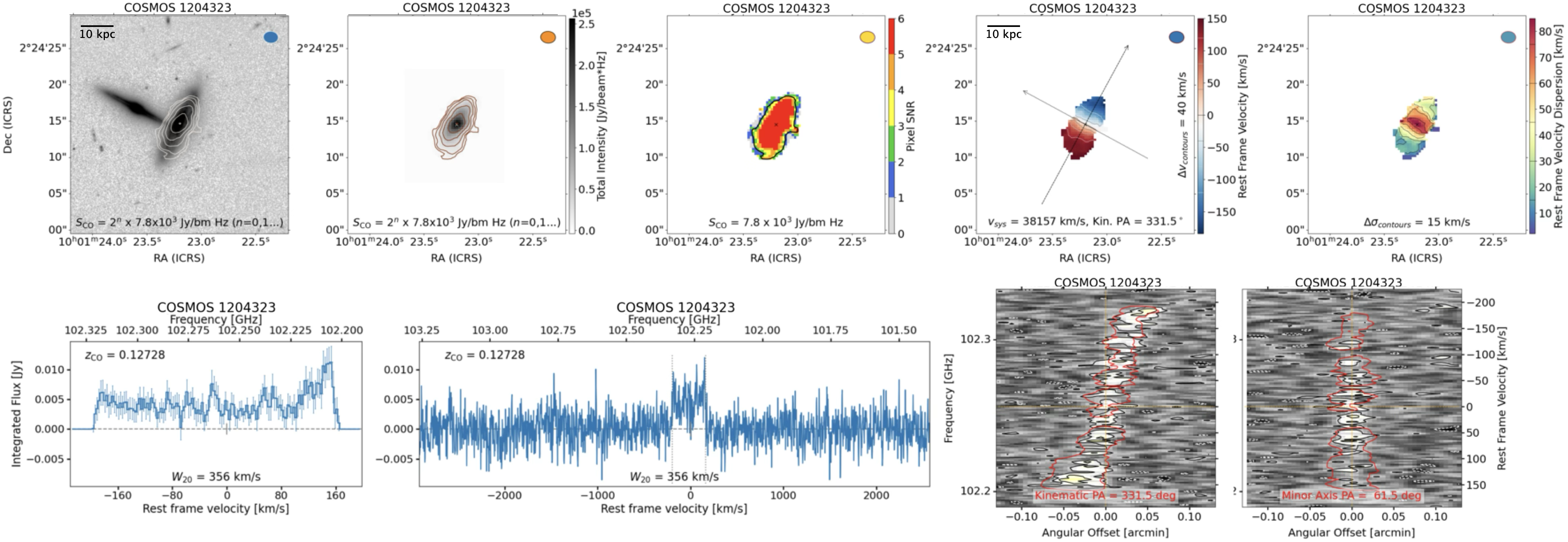} \hfill
    \includegraphics[angle=90,width=0.45\textwidth]{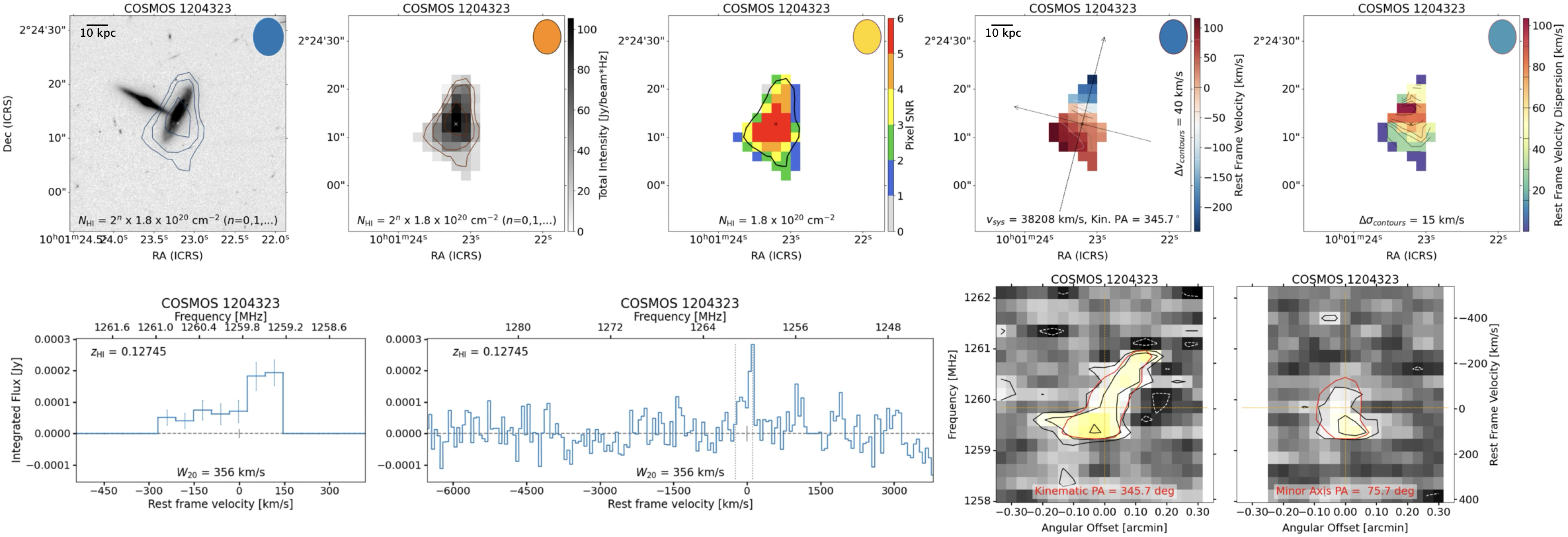}
    \caption{Atlas page of CO(1-0) (left) and \hi\ (right) for COSMOS 1204323. See Appendix text for description.
    }
    \label{fig:co_atlas3}
\end{figure*}

\begin{figure*}
    \centering
    \hfill
    \includegraphics[angle=90,width=0.45\textwidth]{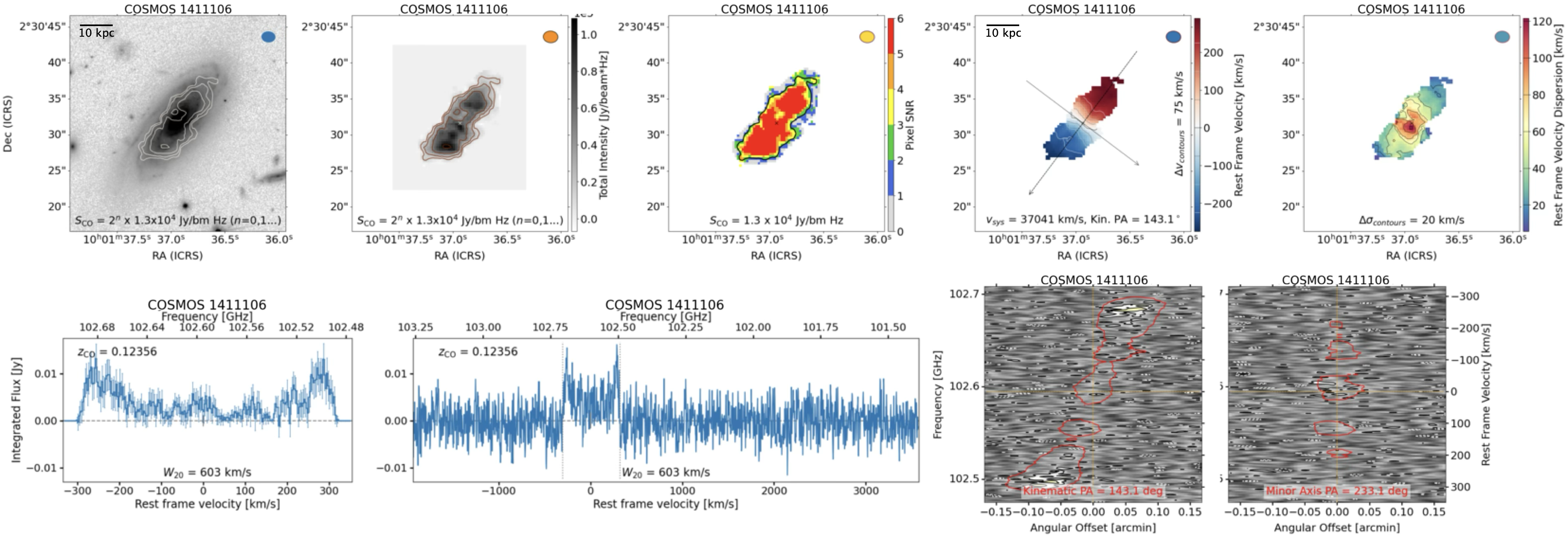} \hfill
    \includegraphics[angle=90,width=0.45\textwidth]{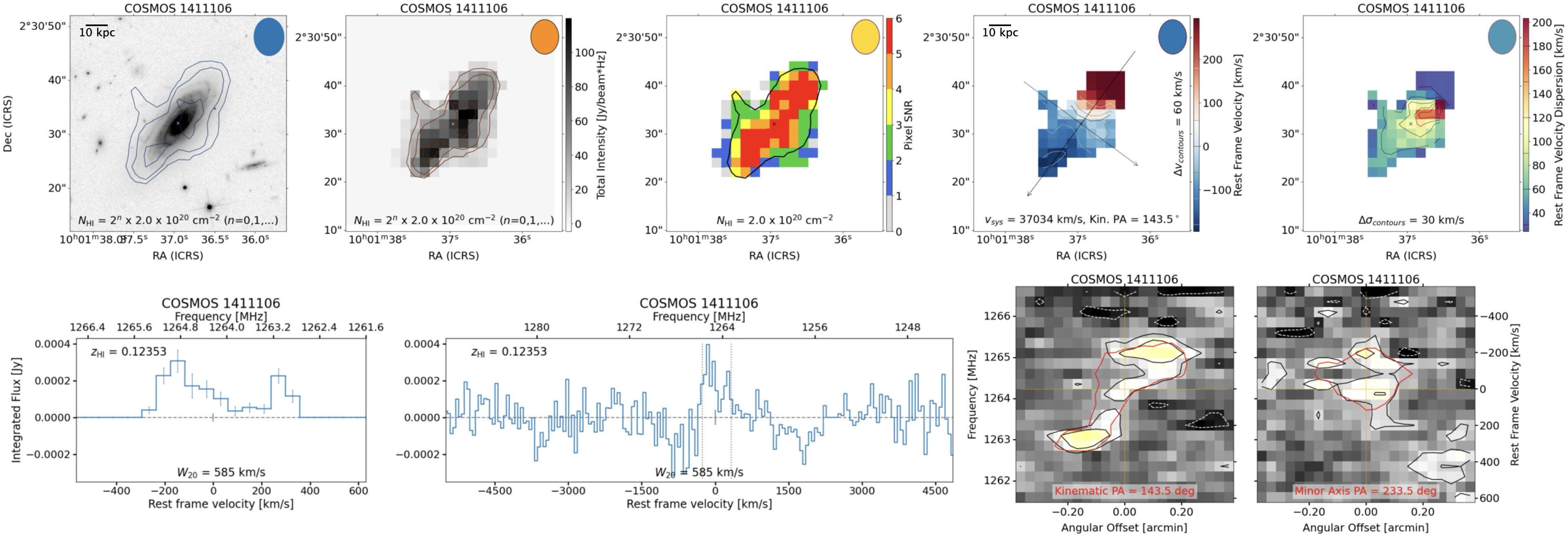}
    \caption{Atlas page of CO(1-0) (left) and \hi\ (right) for COSMOS 1411106. See Appendix text for description.
    }
    \label{fig:co_atlas4}
\end{figure*}

Figures \ref{fig:co_atlas0}-\ref{fig:co_atlas4} show the CO(1-0) and \hi\ atlases for the five ALMA detected galaxies.  After rotating the figures 90 degrees to the right, the figure panels are as follows.  Top row from left to right: total intensity contours on HST/ACS F814W image; total intensity contours on CO (\hi) grayscale; pixel SNR map; intensity-weighted velocity map with kinematic position angles; velocity dispersion map. Bottom: masked CO (\hi) spectrum; CO (\hi) aperture spectrum; position-velocity slice along the kinematic major axis; pv-slice along the minor axis.  Figures were generated by the SoFiA Image Pipeline (SIP)\footnote{\url{https://github.com/kmhess/SoFiA-image-pipeline}} software \citep{Hess22}.

\section{\HI\ atlases}
\label{sect:sip_hi}

\begin{figure*}
    \centering
    \hfill
    \includegraphics[angle=90,width=0.45\textwidth]{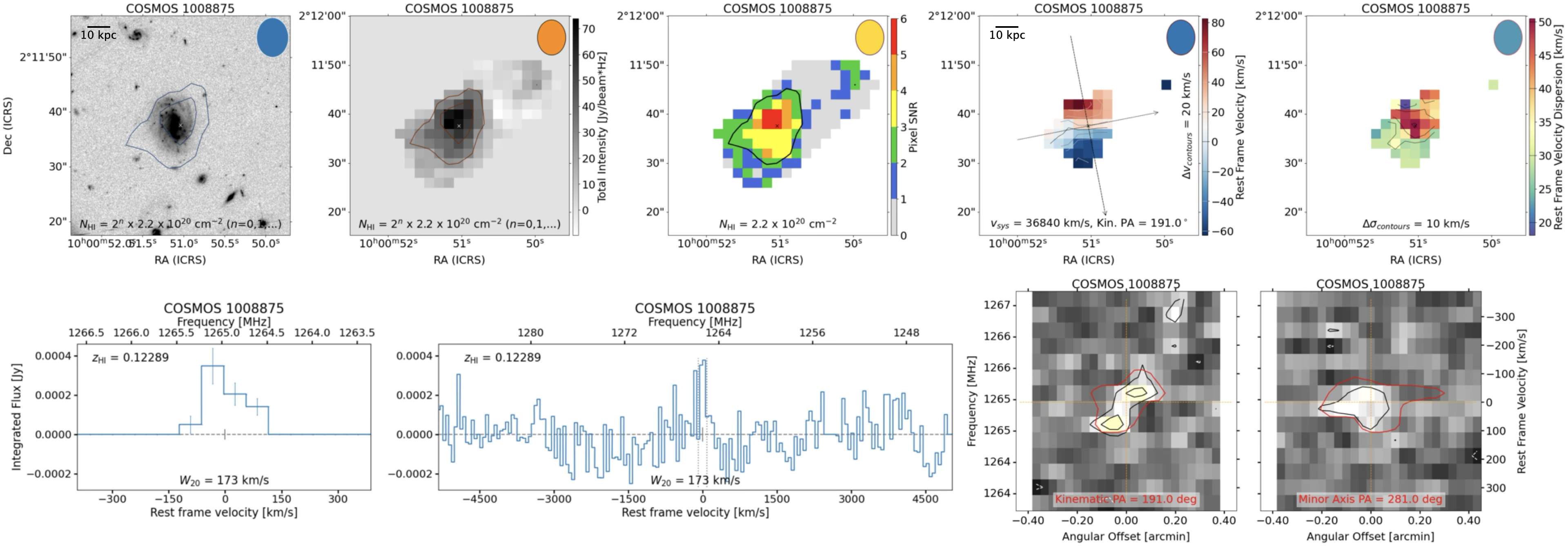} \hfill
    \includegraphics[angle=90,width=0.45\textwidth]{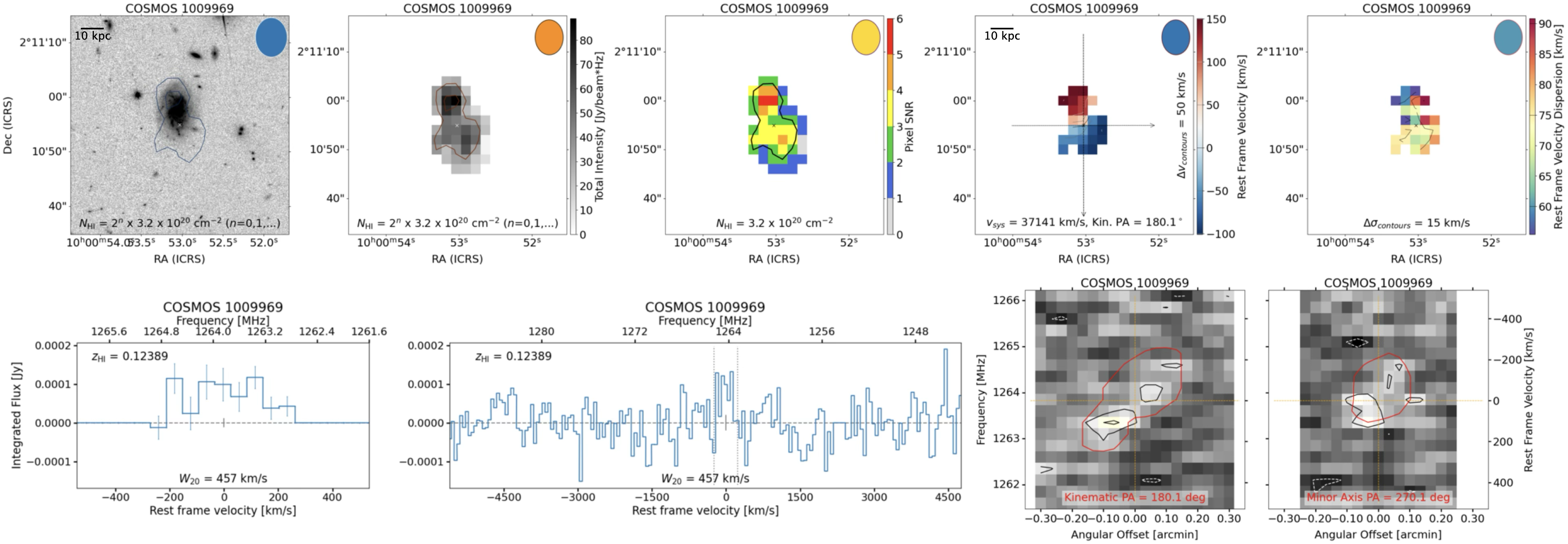}
    \caption{Atlas pages of \hi\ detection for COSMOS 1008875 and COSMOS 1009969. See Appendix text for description.
    }
    \label{fig:hi_atlas1}
\end{figure*}

\begin{figure*}
    \centering
    \hfill
    \includegraphics[angle=90,width=0.45\textwidth]{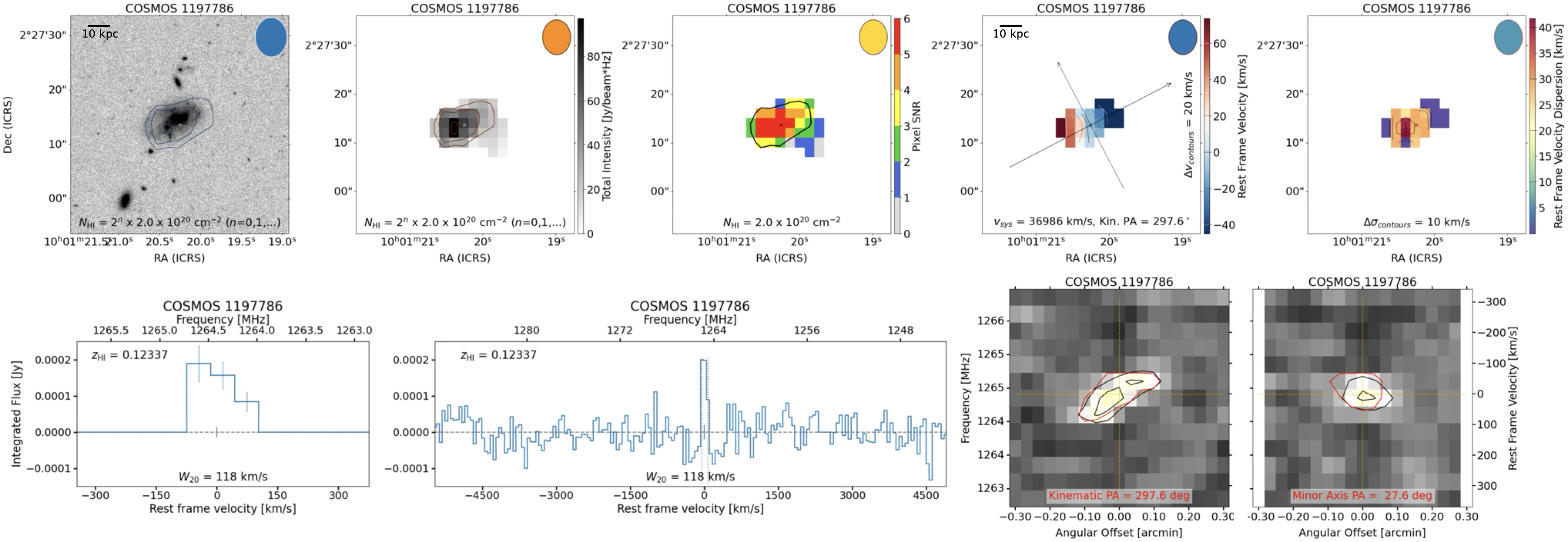} \hfill
    \includegraphics[angle=90,width=0.45\textwidth]{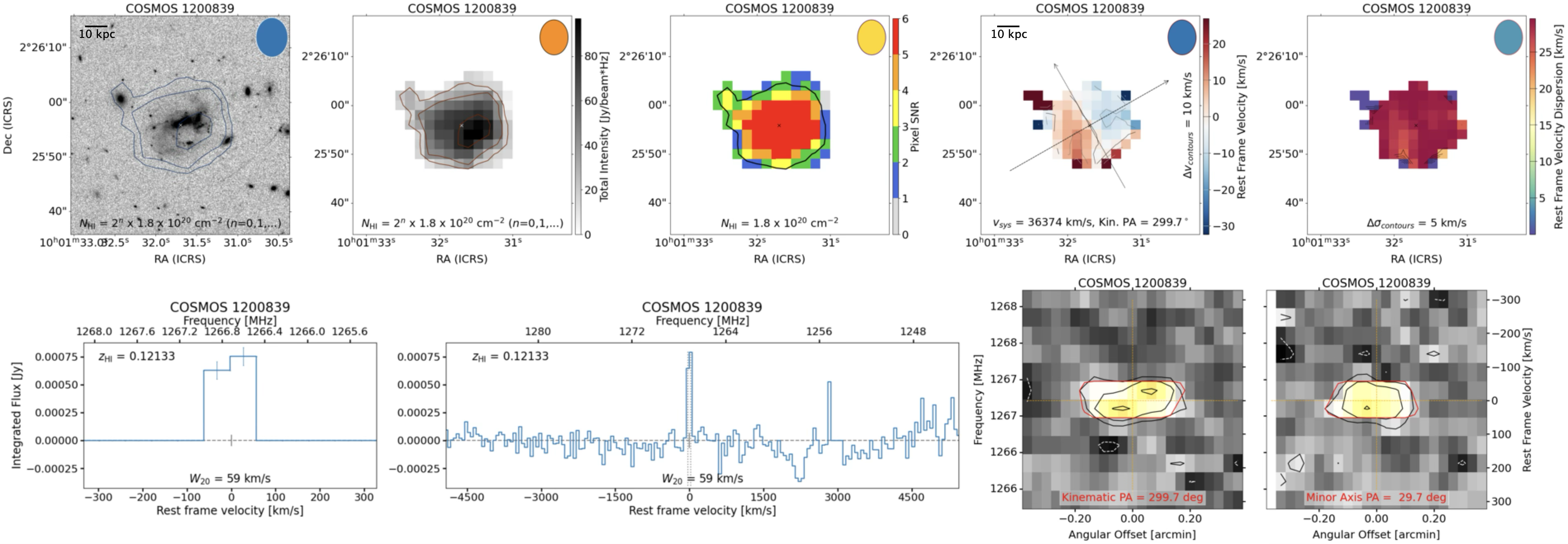}
    \caption{Atlas pages of \hi\ detection for COSMOS 1197786 and COSMOS 1200839. See Appendix text for description.}
    \label{fig:hi_atlas2}
\end{figure*}

\begin{figure*}
    \centering
    \hfill
    \includegraphics[angle=90,width=0.45\textwidth]{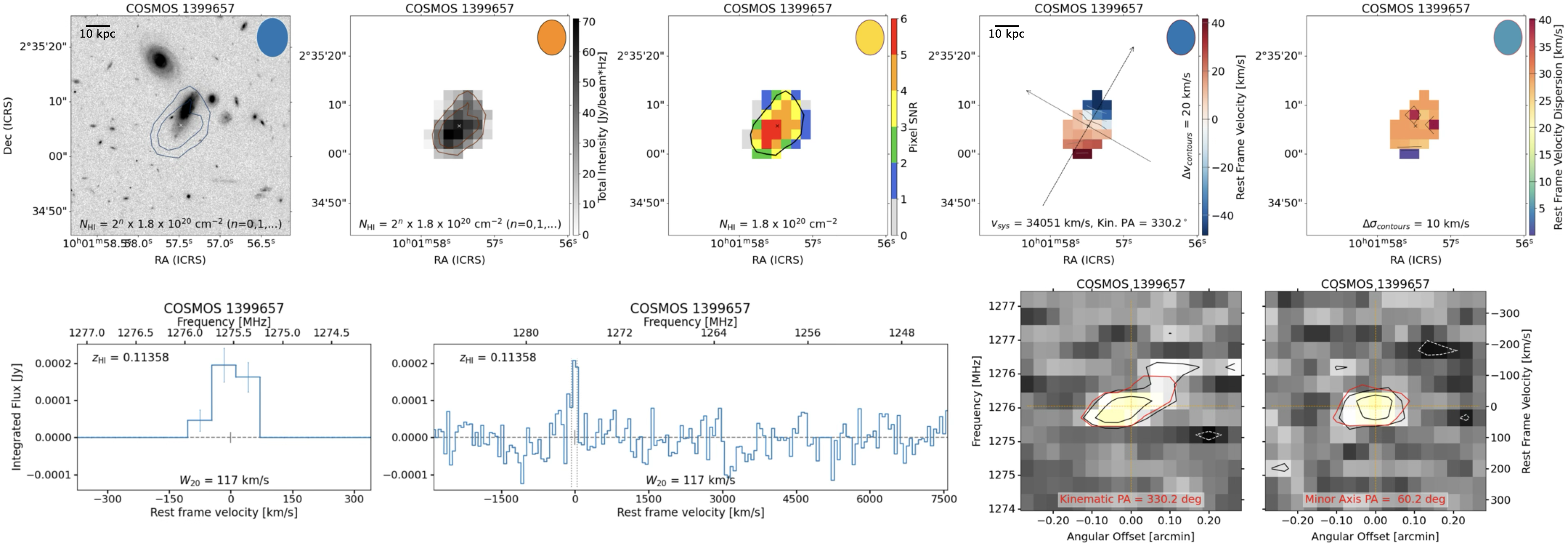} \hfill
    \includegraphics[angle=90,width=0.45\textwidth]{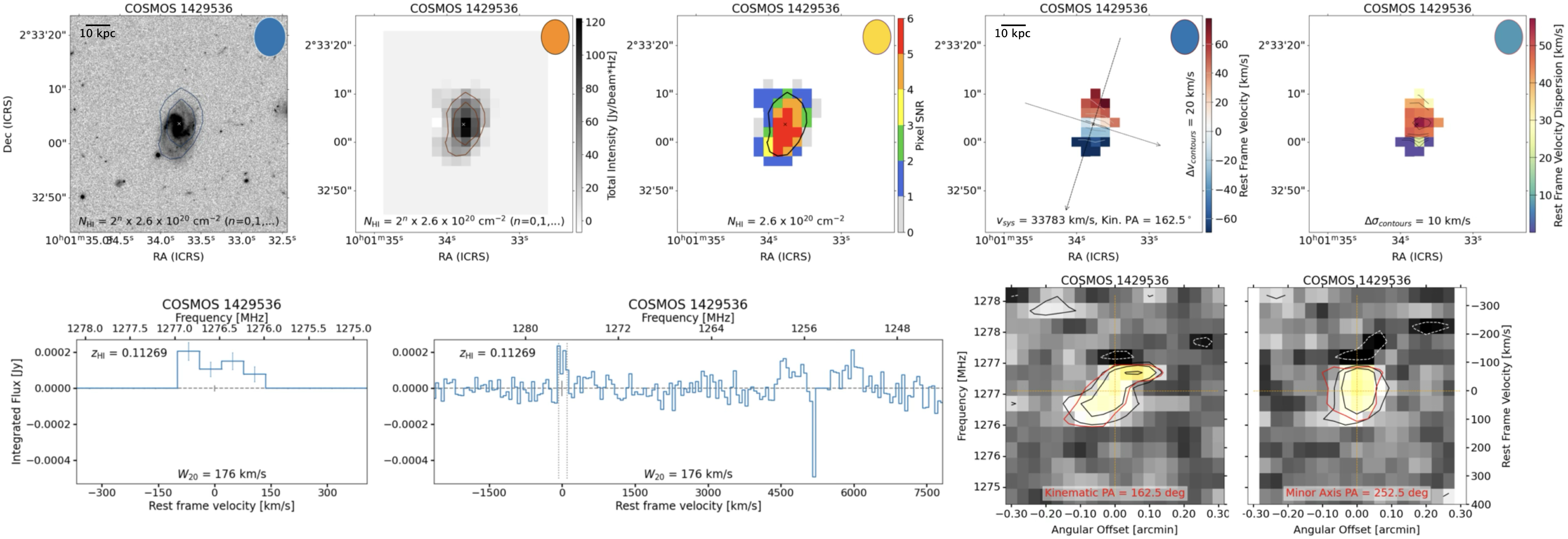}
    \caption{Atlas pages of \hi\ detection for COSMOS 1399657 and COSMOS 1429536. See Appendix text for description.}
    \label{fig:hi_atlas3}
\end{figure*}

\begin{figure*}
    \centering
    \hfill
    \includegraphics[angle=90,width=0.45\textwidth]{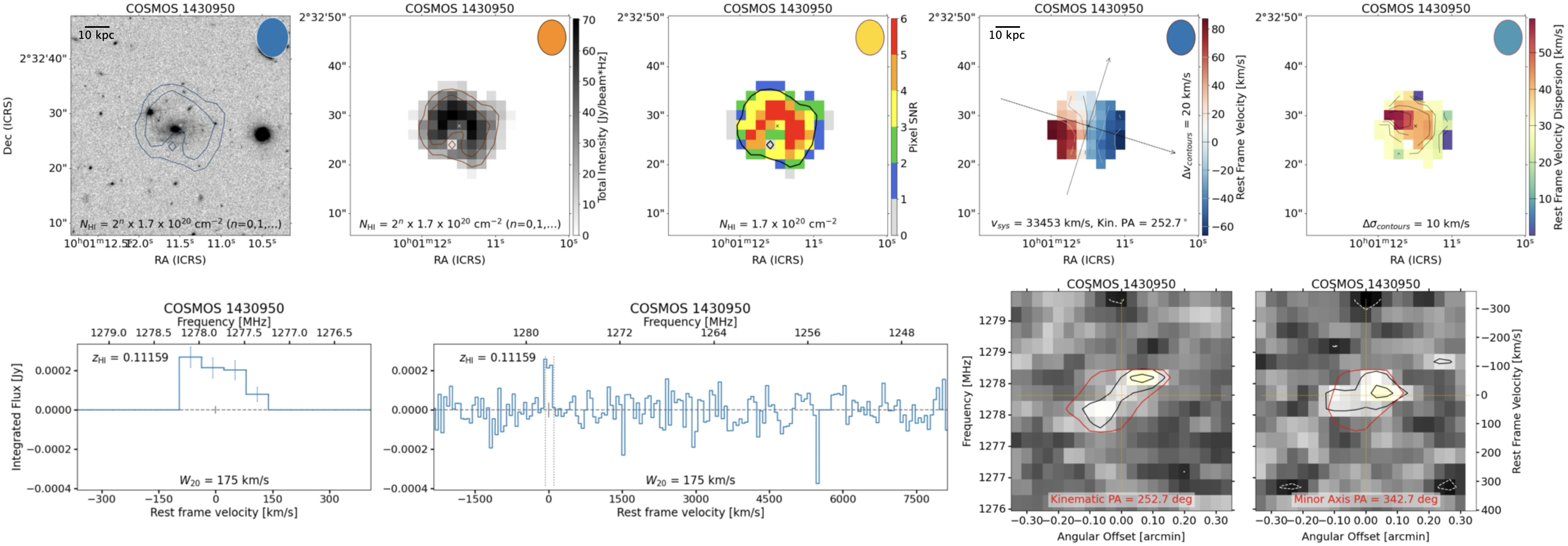} \hfill
    \includegraphics[angle=90,width=0.45\textwidth]{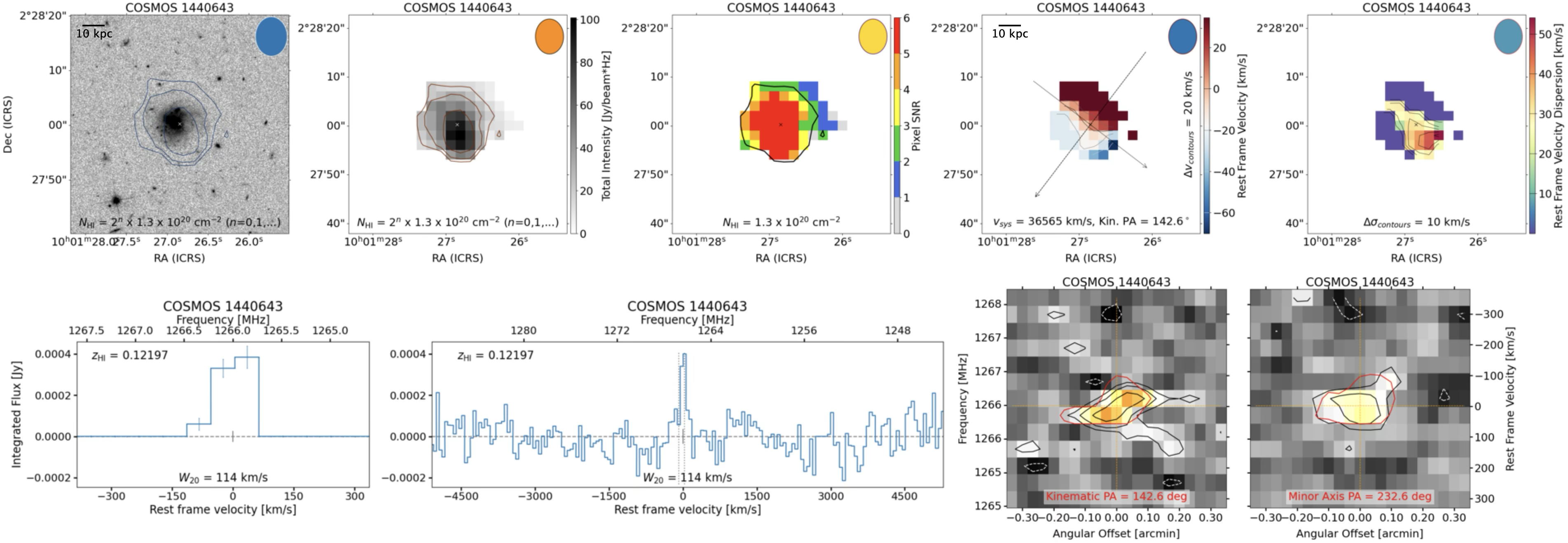}
    \caption{Atlas pages of \hi\ detection for COSMOS 1430950 and COSMOS 1440643. See Appendix text for description.}
    \label{fig:hi_atlas4}
\end{figure*}

\begin{figure}
    \centering
    \includegraphics[angle=90,width=0.45\textwidth]{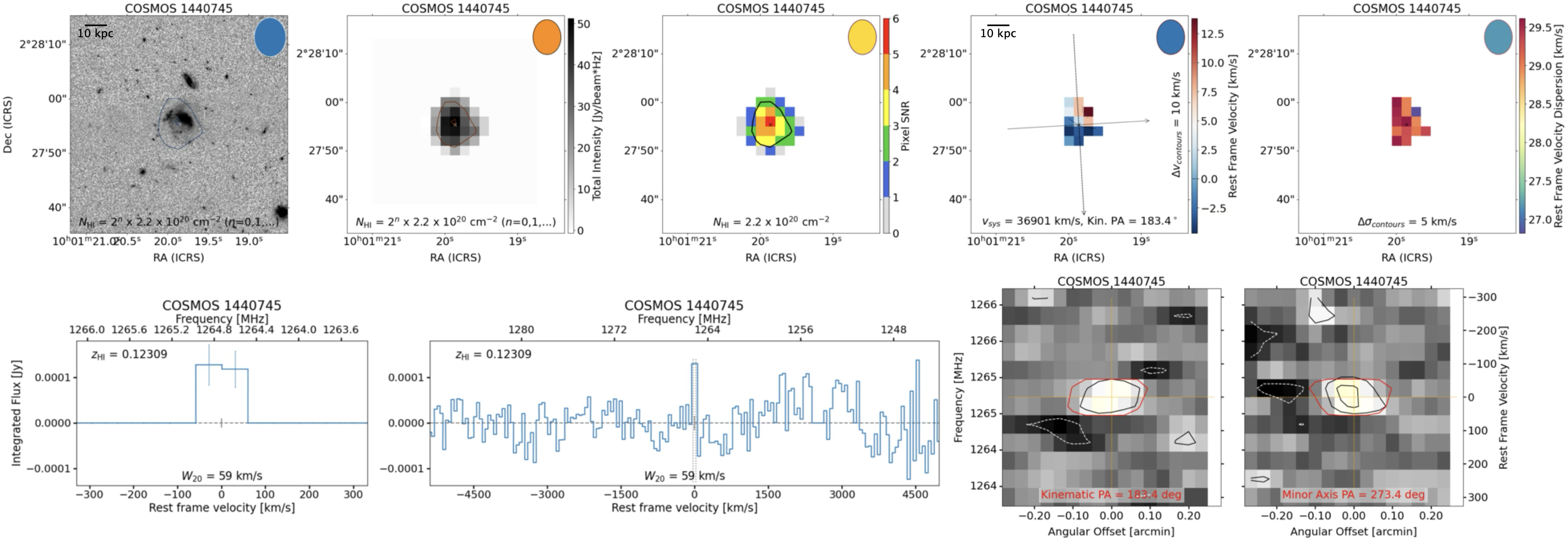}
    \caption{Atlas pages of \hi\ detection for COSMOS 1440745. See Appendix text for description.}
    \label{fig:hi_atlas5}
\end{figure}

Figures \ref{fig:hi_atlas1}-\ref{fig:hi_atlas5} show the \HI\ atlases for the nine CHILES detections around $z=0.12$ which were not directly detected by ALMA in CO(1-0).  The figure panels are as follows. Top row from left to right: \hi\ total intensity contours on HST/ACS F814W image; \HI\ contours on \HI\ grayscale; pixel SNR map; intensity-weighted velocity map with kinematic position angles. Bottom row: masked \HI\ spectrum; \HI\ aperture spectrum; position-velocity slice along the kinematic major axis; pv-slice along the minor axis.  Figures were generated by the SIP software \citep{Hess22}.

\end{appendix}

\end{document}